\documentclass[11pt]{article}
\usepackage{iap2000,epsfig}

\def\Journal#1#2#3#4{{#1} {\bf #2}, #3 (#4)}
\def\PRL{\em Phys. Rev. Lett.}
\def\APJ{\em ApJ}
\def\APJS{\em ApJS}
\def\AJ{\em AJ}
\def\AA{\em AA}
\def\AAS{\em AAS}
\def\AAR{\em AA Rev}
\def\ARAA{\em ARAA}
\def\AN{\em Astr. Nachr.}
\def\MN{\em MNRAS}
\def\PA{\em PASP}
\def\PAJ{\em PASJ}
\def\NAT{\em Nature}
\def\JHA{\em J. for History Astr.}
\def\SKY{\em Sky and Telescope}
\def\SA{\em Sov. Astron.}
\def\SAL{\em Sov. Astron. Lett.}

\def\dV{de~Vaucouleurs}

\def\be{\begin{equation}}
\def\ee{\end{equation}}
\def\bea{\begin{eqnarray}}
\def\eea{\end{eqnarray}}

\begin{document}
\title{FROM MESSIER TO ABELL: \\ 200 YEARS OF SCIENCE WITH GALAXY CLUSTERS
}

\author{ Andrea BIVIANO }

\address{Osservatorio Astronomico di Trieste \\
via G.B. Tiepolo 11 -- I-34131 Trieste, Italy \\
biviano@oat.ts.astro.it}

\maketitle

\abstracts{ }

\section{Introduction}\label{s-intro}
The history of the scientific investigation of galaxy clusters starts
with the XVIII century, when Charles Messier and F. Wilhelm Herschel
independently produced the first catalogues of nebul\ae, and noticed
remarkable concentrations of nebul\ae~ on the sky.  Many astronomers
of the XIX and early XX century investigated the distribution of
nebul\ae~ in order to understand their relation to the local {\em
``sidereal system''}, the Milky Way. The question they were trying to
answer was whether or not the nebul\ae~ are external to our own
galaxy. The answer came at the beginning of the XX century, mainly
through the works of V.M.~Slipher and E.~Hubble (see, e.g.,
Smith\cite{smi79a}).

The extragalactic nature of nebul\ae~ being established, astronomers
started to consider clusters of galaxies as physical systems. The
issue of how clusters form attracted the attention of
K.~Lundmark\cite{lun27} as early as in 1927. Six years later,
F.~Zwicky\cite{zwi33} first estimated the mass of a galaxy cluster,
thus establishing the need for dark matter. The role of clusters as
laboratories for studying the evolution of galaxies was also soon
realized (notably with the collisional stripping theory of Spitzer \&
Baade\cite{spi51}).

In the 50's the investigation of galaxy clusters started to cover all
aspects, from the distribution and properties of galaxies in clusters,
to the existence of sub- and super-clustering, from the origin and
evolution of clusters, to their dynamical status, and the nature of
dark matter (or {\em ``positive energy''}, see e.g.,
Ambartsumian\cite{amb61}).  As a matter of fact, the topic expanded so
much that in 1959 a new separate section specifically devoted to
galaxy clusters -- {\sl Galaxienhaufen} -- appeared in the {\sl
Astronomischer Jahresbericht.}  Galaxy clusters had become one of the
main research topics in extragalactic astrophysics.

In this historical review I have tried to cover all aspects of
astrophysics research on galaxy clusters, spanning a temporal range of
exactly 200 years, from 1784 to 1983.  In 1784, Charles
Messier\cite{mes84} was the first to write about a cluster of
galaxies, Virgo, in his {\sl Catalogue des n\'ebuleuses et des amas
d'\'etoiles que l'on d\'ecouvre parmi les \'etoiles fixes, sur
l'horizon de Paris.} In 1983, on October 7$^{th}$, George O. Abell,
the eponymous of nearby rich clusters of galaxies, prematurely died at
the age of 56. A practical reason for stopping this review with 1983,
is that the exponential increase of publications makes it increasingly
difficult for the historian to keep pace with the new scientific
results.

This review is divided into four main topics:
\begin{enumerate}
\item {\sc The distribution of clusters}, including:
\begin{itemize}
\item the discovery of clusters
\item cluster catalogues
\item the large scale structure (superclusters)
\item distribution functions of cluster properties
\end{itemize}
\item {\sc The cluster components}, including:
\begin{itemize}
\item the properties and distribution of cluster galaxies
\item the properties and distribution of intracluster (IC hereafter) hot gas
\item cluster radio-sources
\end{itemize}
\item {\sc The cluster structure}, including:
\begin{itemize}
\item the dynamical status of clusters (stability and subclustering)
\item cluster masses
\item cluster luminosities (the luminosity function)
\item the nature of the {\sl missing mass}
\end{itemize}
\item {\sc The evolution of clusters}, including:
\begin{itemize}
\item the evolution of clustering
\item the evolution of galaxies in clusters
\item the evolution of the IC gas
\item cooling flows and the evolution of cD galaxies
\end{itemize}
\end{enumerate}

I consider here both theoretical and observational aspects.  However,
I rarely mention technical aspects, such as the development of new
telescopes and instruments, which were certainly very relevant to our
understanding of galaxy clusters. In this respect, this review traces
the history of the scientific thought, rather than the history of
science.

For the sake of homogeneity, all quantities that are $H_0$-dependent,
have been re-scaled to the same value the Hubble constant, $H_0 =
75$~km~s$^{-1}$~Mpc$^{-1}$.
 
\section{The distribution of clusters}
\subsection{Early days}
The first written reference to a cluster of galaxies is probably that
of the French astronomer Charles Messier\cite{mes84} in 1784. In his
{\sl Catalogue des n\'ebuleuses et des amas d'\'etoiles que l'on
d\'ecouvre parmi les \'etoiles fixes, sur l'horizon de Paris}, he
listed 103 nebul\ae, 30 of which we now identify as
galaxies\footnote{Of the 30 extragalactic objects in Messier's
catalogue, only 13 are listed in the Virgo Cluster Catalogue of
Binggeli et al.\cite{bin85}.}. Messier already noticed the exceptional
concentration of nebul\ae~ in the Virgo constellation. However,
Messier's interest in nebul\ae~ was very marginal. He seeked to define
the positions of nebul\ae~ in order not to misidentify them with new
comets\footnote{Charles Messier was nicknamed {\em ``le furet des
com\`etes''} by Louis XV.}.

\begin{figure}
\begin{center}
\psfig{figure=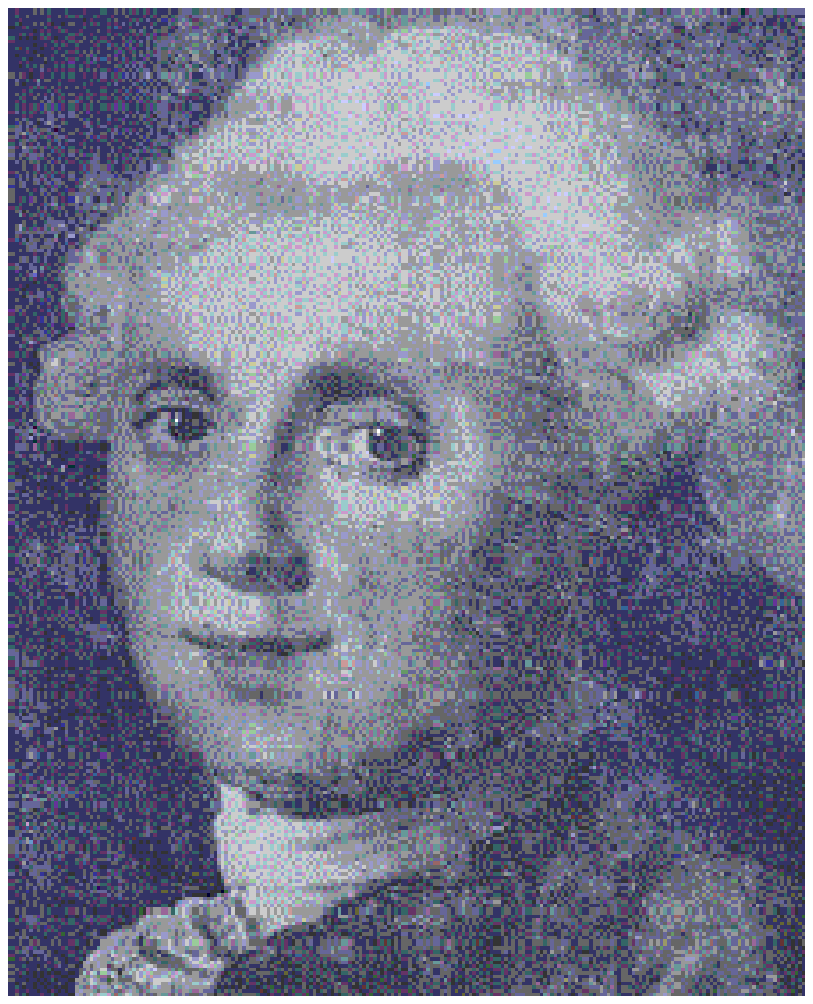,height=7cm,angle=0}
\psfig{figure=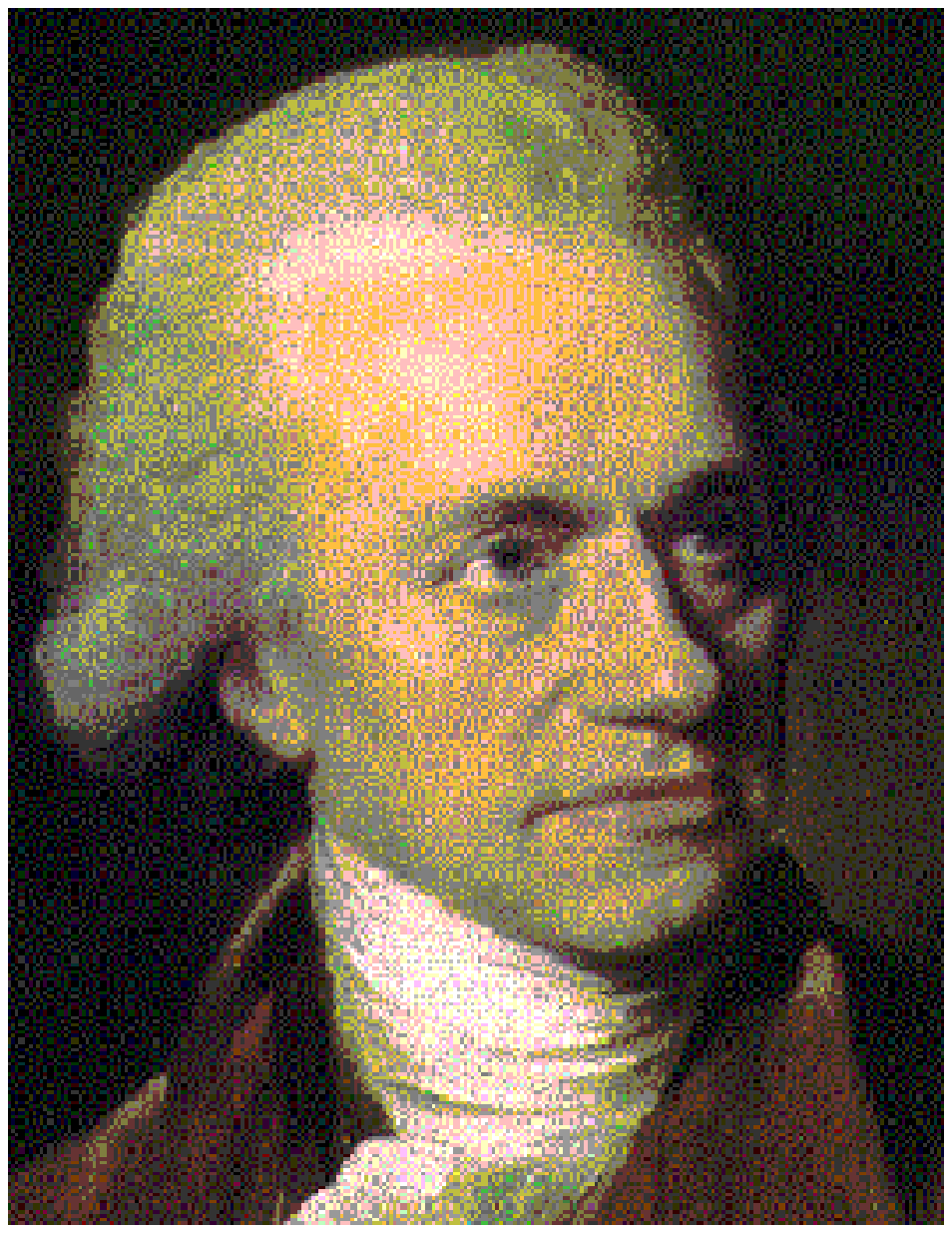,height=7cm,angle=0}
\end{center}
\caption{Portraits of C. Messier (left) and F. Wilhelm Herschel.}
\end{figure}

F. Wilhelm Herschel had a quite different approach to the
investigation of nebul\ae. German born, he escaped from Hanover and
reached England during the War of the Seven Years.  A musician, he
became interested in astronomy after reading a popular book. After the
first successful discoveries\footnote{W. Herschel became very famous
after his discovery of Uranus in 1781.} with his self-made telescopes,
the king of England granted him the money to build the largest
telescope of his times, a 1.47~m aperture, 12.2~m focal length
refractor. Herschel was interested in what we would now call
the Large Scale Structure of the Universe. In 1785 he published {\sl
On the Construction of the Heavens}\cite{her85}, where he suggested
that the {\em ``sidereal system we inhabit''} is a nebula, common in
appearance to many others, which therefore must be external to our
own. Most relevant here is W.~Herschel's description of the
Coma cluster of galaxies:
\begin{quote}
{\em ``that remarkable collection of many hundreds if nebul\ae~ which are
to be seen in what I have called the nebulous stratum of Coma Berenices''}
\end{quote}
In the same paper, W.~Herschel mentioned her sister's discovery of the
second small companion of M~31, NGC~205. With M~32, these three galaxies
make a triplet similar to that composed by the Milky~Way and the two
Magellanic clouds. The other giant galaxy in the Local Group, M~33 was
listed in Messier's catalogue. So, 7 members of the Local Group of
galaxies were already known at that time. Their distances being
unknown, it was only in 1936 that E.~Hubble\cite{hub36} pointed out
that these galaxies (and a few more) belong to the same system, which
he named {\em ``The Local Group''} (see, e.g., van~den~Bergh\cite{vdb99a}).

In the course of his life, W.~Herschel\cite{her11} classified some
2500 nebul\ae~ and recognized several other nearby clusters and groups
of galaxies, such as Leo, Ursa Major, Hydra, NGC4169, etc. His work
was continued by his son, John~F.W.~Herschel.  J.~Herschel surveyed
the southern sky from Cape of Good Hope, and catalogued over 6000
nebul\ae~ that in 1864 he collected in his {\sl General Catalogue of
Nebul\ae~ and Clusters of Stars.} During the first part of the XIX
century, J.~Herschel noted that the northern hemisphere has an excess
of nebul\ae~ with respect to the southern hemisphere, and he recognized
several concentrations of nebul\ae~ (in Pisces and Fornax, in
particular). He already hinted at the existence of the Local
Supercluster, with the Virgo concentration {\em ``being regarded as
the main body of this system'',} and our own Galaxy {\em ``placed
somewhat beyond the borders of its densest portion, yet involved among
its outlying members''} (see, e.g., Flin\cite{fli88}).

In J.~Herschel's times, d'Arrest\cite{dar65} and Proctor\cite{pro72}
published new positions and finding charts of nebul\ae~ in the Coma
and Virgo clusters, Stephan\cite{ste77} discovered the famous galaxy
quintet, and Dreyer\cite{dre88b} published his {\sl New General
Catalogue}. Complemented by the {\sl Index Catalogues,} the {\sl NGC}
listed roughly 13000 nebul\ae~ in 1908.

At the beginning of the new century, the extensive photographic work
of Max Wolf\cite{wol01,wol02,wol05} led to a detailed description of
the Coma and Perseus clusters. In 1918 Curtis\cite{cur18a} added more
nebul\ae~ to Wolf's list, reaching a total of 300 nebul\ae~ in the
Coma cluster.

In the early years of the XX century, intensive photographic
observations of nebul\ae~ were done mostly with the aim of
establishing whether they were external to our own galaxy or not.  The
{\sl Great Debate} on the nature of nebul\ae~ between Shapley and
Curtis, took place on April, 26$^{th}$ 1920, with no clear winner. Not
only were astronomers trying to determine the distribution of
nebul\ae~ with respect to the galactic plane, they were also trying to
count them! Curtis\cite{cur18b}' estimate of 722,000 nebul\ae~ in
1918, was revised to 60 millions by Hubble\cite{hub36} in 1936.

In 1904 Easton\cite{eas04} noted an asymmetry in the distribution of
the nebul\ae~ with respect to the galactic plane, with an excess of
nebul\ae~ in the northern hemisphere. Nineteen years later, this
asymmetry was re-discovered by Reynolds\cite{rey23a,rey23b} who noted
that
\begin{quote}
{\em ``many of the spirals 10' diameter and upwards lie along 100$^{\circ}$,
and form part of a well-marked band of nebul\ae~ passing over the
north galactic pole, which comes out conspicuously if the spirals
ranging down to 2' diameter are plotted together.''}
\end{quote}
A clear reference to the Local Supercluster! In the same years,
C.~Wirtz, using Dreyer's catalogues and Curtis'
surveys, called the attention to several conspicuous well-defined
centers of clustering (see, e.g., Abell\cite{abe75}).

In the early twenties, Edwin Hubble discovered cepheids in
M31, and definitely established the extragalactic nature of
nebul\ae. A few years later he published his work\cite{hub29} on the
velocity-distance relation for extragalactic nebul\ae. Extending this
relation to higher redshifts became the main driver for Hubble \&
Humason's great observational work on extragalactic
nebul\ae\cite{hub31}.  In 1934 and 1936 Milton
Humason\cite{hum34,hum36} measured velocities of 39,200 km/s and
42,000 km/s for galaxies in the Bo\"otis and Ursa Major II clusters,
making them the most distant clusters known at that time.

\begin{figure}
\begin{center}
\psfig{figure=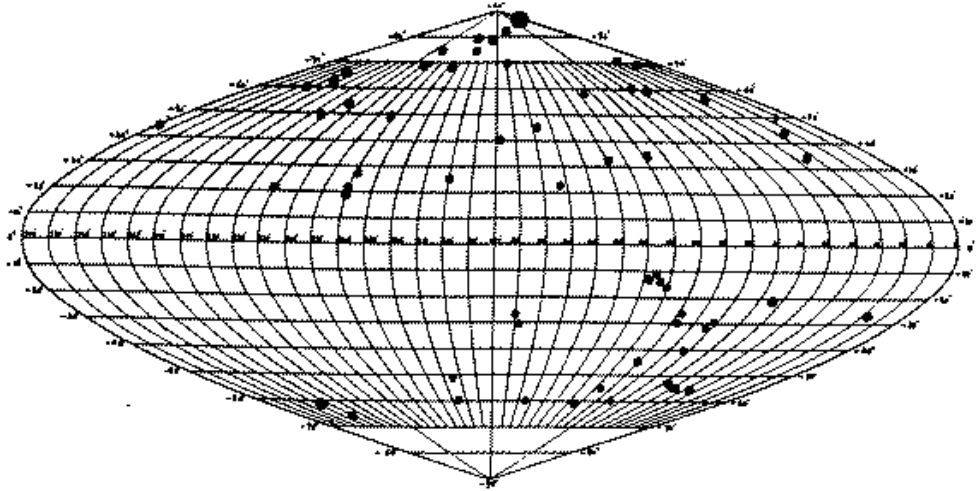,height=7cm,angle=0}
\end{center}
\caption{Galactic distribution of the clusters of anagalactic
nebul\ae. From Lundmark (1927).
\label{fig-kl55cls}}
\end{figure}

More galaxy systems were discovered in those years: Cancer, Hercules,
Leo, and notably the {\em ``Centaurus cloud''}, today's {\sl Shapley
concentration} (see, e.g., Bardelli et al.\cite{bar94}).
Shapley\cite{sha30} correctly estimated it to be 14 times more distant
than Virgo, and 10 times as rich in nebul\ae.  All these discoveries
were serendipitous; as an example, the Perseus-Pisces stratum was
noted by Tombaugh\cite{tom37} as {\em ``a by-product of the extensive
trans-Neptunian planet search''} which eventually led to the discovery
of Pluto. Knut~Lundmark\cite{lun27} plotted the sky distribution of 55
clusters of {\em ``anagalactic nebul\ae''} -- see
Fig.~\ref{fig-kl55cls}.  Coordinates of these clusters were not
listed, but it is likely that many of them were {\sl groups} rather
than {\sl clusters}. Lundmark noted {\em ``the most characteristic
feature in the charts of the nebular distribution is the clustering
tendency'',} a tendency confirmed in the Harvard survey\cite{sha32}.
While presenting results from this survey, Shapley\cite{sha33}
provided a list of 25 clusters and suggested the existence of {\em
``metagalactic clouds''} (today's superclusters), such as those in
Coma, Centaurus and Hercules\cite{sha34}.  E.F. Carpenter\cite{car38}
described clusters as the extremes of a continuous non-uniform spatial
distribution of galaxies, thus anticipating the works of Neyman \&
Scott\cite{ney52} and Peebles\cite{pee74}.

In contrast to the growing dominant opinion, in 1936
Hubble\cite{hub36} described the distribution of nebul\ae~ as {\em
``moderately uniform''} and noted that {\em ``no organization on a
scale larger than the great clusters''} was definitely known. However,
he recognized our own Galaxy as a member of a galaxy system, which he
named {\em ``The Local Group''}. Zwicky\cite{zwi38} noted that the
local group may well be part of the Virgo galaxy system, that
Holmberg\cite{hol37} described as a {\em ``Metagalactic cloud''} of
$\sim 100$ Mpc size.

\subsection{Surveys and catalogues}
After the Second World War, the Lick and Palomar sky surveys and the
spectroscopic observations of Humason, Mayall \& Sandage\cite{hum56}
provided the essential data-base for the analysis of the distribution
of galaxies. The 1956 paper of Humason et al.\cite{hum56} collected
the results of twenty years of spectroscopic observations, providing more than
800 redshifts of galaxies, of which 75 in Virgo, 23 in Coma, and a
few dozens in several other clusters. They noted that there was {\em
``increasing evidence''} for a general clustering phenomenon, and
dismissed Hubble's view of a uniform galaxy distribution with a few
sporadic isolated clusters.

\begin{figure}
\begin{center}
\psfig{figure=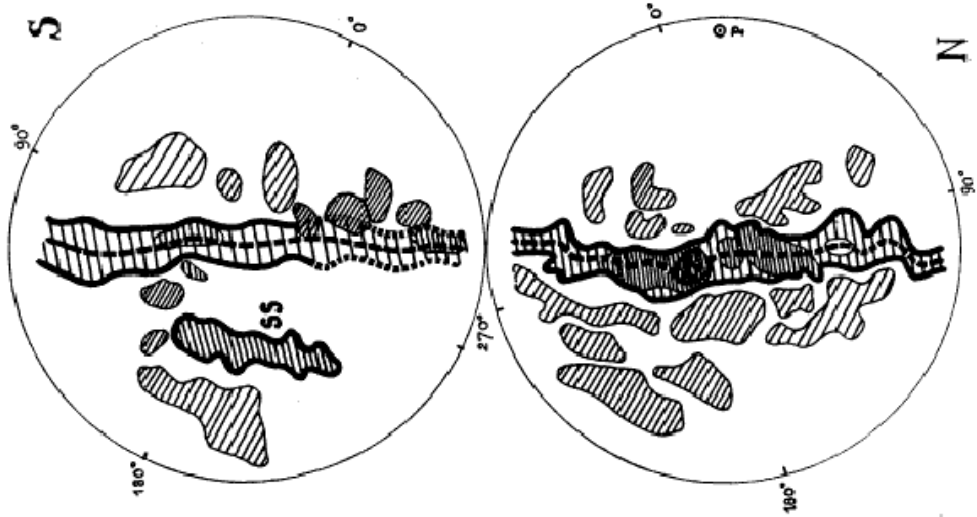,height=13cm,angle=90}
\end{center}
\caption{Clustering of the nebul\ae~ in the southern and northern emisphere
giving evidence of the Local Supercluster. The density of the
shading gives in a qualitative way an idea of the nebular density 
-- from \dV~ (1953).
\label{fig-devlsg}}
\end{figure}

The evidence for the {\em ``Local Supergalaxy''} and for many other
superclusters grew stronger mainly through the works of
\dV\cite{dev53,dev58} -- see Fig.~\ref{fig-devlsg} --
Shane \& Wirtanen\cite{sha54},
van den Bergh\cite{vdb60c}, and Abell\cite{abe61}.  Only
Zwicky\cite{zwi52a} continued to deny the existence of
superclusters. Zwicky\cite{zwi57a} thought that the apparent
non-uniform distribution of clusters was due to the obscuration
effects of inter-galactic and IC dust.  He eventually discovered a
supercluster himself\cite{zwi63} (no.20 in Zucca et al.\cite{zuc93}'s
catalogue), but refused to call it a supercluster. Zwicky's point of
view was however very different from Hubble's. Zwicky thought galaxy
clusters to be much larger than usually accepted, almost reaching to
the sizes of superclusters.  Clusters, he wrote in 1952, {\em ``fill
the universe just as the bubbles fill a volume of suds''.} For these
reasons, Abell\cite{abe65} thought that Zwicky's opposition to the
idea of superclusters was purely semantic.

In a series of papers, Neyman, Scott, Shane \&
Swanson\cite{ney52,ney53,ney54,sco54} addressed the issue of galaxy
clustering by applying mathematical models to the Lick galaxy counts
of Shane \& Wirtanen\cite{sha54}, and were the first to compare the
observed galaxy distribution to synthetic images of the
Universe\cite{sco54}.

The introduction of new techniques and new ideas pushed the search for
clusters to higher redshifts.  Baum\cite{bau58} pointed out that
clusters at redshifts $\sim 0.5$ could be most easily detected by
moving redwards the observing waveband. Minkowski\cite{min60}
speculated that collisions between galaxies could produce
radio-emission; since collisions should be frequent in dense
environments, he suggested that clusters could be found around
radio-galaxies. In 1960 he applied this idea to the region around
3C295, and found a system of galaxies at a redshift $\simeq
0.44$--0.46.  3C295 held the record of the highest redshift cluster
for a long time\footnote{3C295 later became one of the two clusters
where Butcher \& Oemler\cite{but78a} found evidence for an increased
fraction of blue galaxies.}.

Meanwhile, the search for nearby galaxy clusters had become
systematic. The time of serendipitous discoveries was long gone, and
in 1957 Herzog, Wild \& Zwicky\cite{her57} announced the construction
of a {\sl Catalogue of Galaxies and Clusters of Galaxies}\cite{zwi61},
that upon completion would contain $\sim 10000$ clusters. Their
announcement came just one year before the publication of Abell's
catalogue\cite{abe58}, but the final {\sl CGCG} was to be published
only in 1967.

Abell's paper, {\sl The distribution of rich clusters of galaxies}, is
a milestone in the history of science with galaxy
clusters\footnote{Abell's paper was just {\em ``a portion of a thesis
submitted in partial fulfillment of the requirements for the
Ph.D. degree''} -- though requirements, no doubt!}. The very fact that
{\sl Abell cluster} has become a synonymous with {\sl rich cluster}
tells us a lot about the importance of this paper.

Abell's 2712 clusters were selected on red {\sl POSS} plates because
he realized the advantage of the red band over the blue band for the
identification of distant clusters. Abell's radius was subjectively
chosen by looking at the projected overdensities of clusters, and yet
is close to the cluster gravitational radius\cite{biv01}. Abell's
subjective selection criteria were extremely well chosen, and even the
background subtraction was quite accurate.

Abell's paper was much more than a catalogue of clusters. He was the
first to show that the distribution of cluster richnesses -- which is
broadly related to the mass distribution -- is very steep. He knew
that his cluster sample was incomplete at the low richness end, and
for this reason he defined a statistical subsample of the richest 1682
clusters.  As a matter of fact, he wrote
\begin{quote}
{\em ``during the course of the plate inspections, many thousands of clusters 
and groups of galaxies were recognized which were not catalogued because
they obviously were not sufficiently rich to insure their essentially
complete identification. Thus neither the statistical sample of clusters
nor a subjective impression indicates a maximum in the $N(n)$ versus
$n$ relation.''}
\end{quote}
We better remember this statement when commenting upon the results of
modern optical cluster surveys\cite{lum92,pos96} (see also {\sc Lobo},
these proceedings).

The publication of Abell's catalogue opened a new era in the
investigation of galaxy clusters. All of a sudden, researchers had a
catalogue of clusters, and they could start look at them as a
population, rather than as individual objects. The first volume of
Zwicky et al.\cite{zwi61}'s {\sl Catalogue of Galaxies and Clusters of
Galaxies} was published only a few years later, but it did not exert
such a large influence on the study of clusters. The main problem with
the {\sl CGCG}, as immediately pointed out by Abell\cite{abe62}, was
that the sizes of Zwicky's clusters were distance-dependent, since
they were defined within the isopleth contour that represents twice
the field density. The {\sl CGCG} could then not be used as a
statistical homogeneous cluster catalogue, and most researchers
preferred to base their analysis on Abell's catalogue (and they still
do).

The first critical examination of Abell's and Zwicky's catalogues was
done by Reaves\cite{rea74}. Abell's statistical subsample was shown to
be $\sim 85$~\% complete, while the completeness of the full Abell
catalogue is only $\sim 40$~\%, similar to that of the Zwicky
catalogue. Reaves' estimates were based on how frequently a given
cluster detected on one plate was missing on another plate where it
should have been seen. His conclusions are quite close to those
obtained by Lucey\cite{luc83} and Briel \& Henry\cite{bri93} several
years after.

In the following years, there was an increase and an improvement in the
classification of clusters, along these five main
research lines:
\begin{itemize}
\item{Finer classifications:} Bautz \& Morgan\cite{bau70} and Rood \&
Sastry\cite{roo71} invented finer cluster classification schemes, to
supersede the traditional regular--irregular cluster classification.
Oemler\cite{oem74} classified clusters according to their galaxy
morphological content, and suggested a relationship
between a cluster compactness and its galaxy morphological mix.
\item{Redshift determinations:} 
Noonan\cite{noo73,noo81} published lists of cluster
redshifts (138 in 1973, and four times as many in 1981).
\item{Southern clusters:}
Klemola\cite{kle69}, Snow\cite{sno70}, Rose\cite{ros76}, Duus
\& Newell\cite{duu77} provided lists of hundreds of clusters in the
southern hemisphere.
\item{Poor galaxy systems:} \dV\cite{dev75} published a
list of 55 groups of galaxies, based on his {\sl Reference
Catalogue}\cite{dev64}. Another list of 174 groups was published
by Holmberg\cite{hol74}. Shakbazyan \& Petrosyan\cite{sha74} published
a catalogue of {\sl Compact groups of compact galaxies,} followed by
Rose's catalogue of compact groups in 1977\cite{ros77}. Turner \&
Gott\cite{tur76a} provided the first complete catalogue of galaxy groups.
Morgan et al.\cite{mor75} and Albert et al.\cite{alb77} identified
poor clusters dominated by giant elliptical at their centre.
\item{Automated search for clusters:} in 1976 MacGillivray et
al.\cite{mac76} inaugurated the automated search for galaxy clusters.
Clusters were identified in galaxy catalogues built using the {\sl COSMOS}
automatic plate-measuring machine.
\end{itemize}
In 1973, Karachentseva\cite{kar73} published a {\sl Catalogue of
isolated galaxies.} Clustered galaxies have become the rule, isolated
galaxies the exception, to such a point that two years later
\dV~ could ask: {\em ``Are there isolated galaxies?''}

\begin{figure}
\begin{center}
\psfig{figure=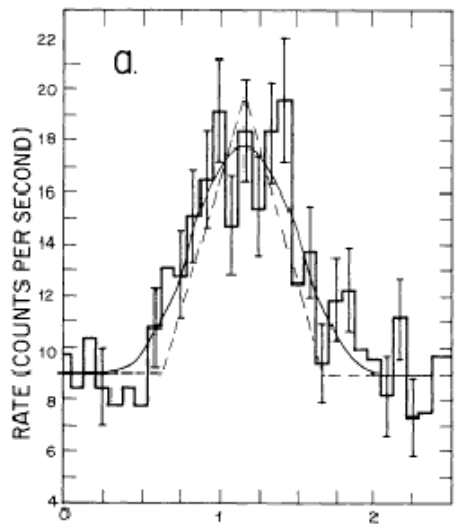,height=7cm,angle=0}
\end{center}
\caption{Counting rates per degrees (relative azimuth on x-axis) in
the Coma cluster. The solid line indicates a fit with an extended source,
the dashed line the expected response to a point source.
From Gursky et al. (1971).
\label{fig-hgcomax}}
\end{figure}

In 1971 Meekins et al.\cite{mee71} and Gursky et al.\cite{gur71}
detected extended X-ray emission from the Coma cluster (see
Fig.~fig-hgcomax). Little by little, optical catalogues of galaxy
clusters would give way to X-ray catalogues. Initially there were just
lists of optical counterparts for a few X-ray sources (e.g. Melnick \&
Quintana\cite{mel75}), but soon after extensive X-ray surveys of
hundreds of Abell clusters were published (see, e.g., Ulmer et
al.\cite{ulm81}).

\subsection{Superclusters and voids}
In his milestone paper, Abell\cite{abe58} also demonstrated the
existence of {\em ``clusters of clusters''} in 3 dimensions.  Abell
used his magnitude-based cluster distance estimates to establish that
the average size of superclusters is $\simeq 60$ Mpc. He rejected
Zwicky's hypothesis of IC dust by showing that regions of the sky
devoid of intermediate-distance clusters were nevertheless occupied by
even more distant clusters. Ten years after, Reaves\cite{rea68} was
able to set an upper limit of $0.1$ magnitudes to the extinction by IC
dust, based on the colour vs. redshift relation for galaxies in
cluster fields. Despite Abell's and Reaves' results, Bogart \&
Wagoner\cite{bog73} in 1973 still invoked IC dust as the origin of an
apparent cluster--cluster anti-correlation.

In 1962 Abell\cite{abe62} published the first list of (seventeen)
superclusters. He noted that the existence of superclusters was to be
taken into account when estimating the probability of chance
projection effects in a cluster catalogue, thus anticipating the ideas
of Lucey\cite{luc83}. A few years later, Abell \& Seligman\cite{abe67}
showed that superclusters could be easily identified even in Zwicky's
{\sl CGCG}\cite{zwi61}.

A step further towards establishing the reality and properties of the
Local Supercluster, was done by \dV\cite{dev75}. He considered the
distribution of 55 nearby groups. By noting that 85~\% of all nearby
galaxies are in groups, he suggested that superclusters may well
overlap and fill all the space available. He correctly argued that the
observational samples had not yet reached to the distance of
homogeneity, thus making it meaningless any attempt to estimate the
mean density of the Universe. The concept of the Large Scale Structure
of the Universe was taking his first steps.

Despite this observational progress, the reality of superclusters
remained an open issue. Peebles and collaborators published papers
arguing both against\cite{yup69} and in favour of the
existence\cite{hau73} of superclusters. Peebles' final word came in
1974, with the development of a mathematical tool that was to stay
with cosmologists ever since: the covariance function\cite{pee74}.  By
showing that the covariance function is a simple power law over a very
large distance range, he concluded that there was no physical division
between groups and clusters, nor between clusters and superclusters.

Zwicky continued to reject all evidences in favour of the existence of
superclusters. He thought that IC dust could account for
irregularities of the clusters distribution. Zwicky's hypothesis was
finally falsified by Reaves\cite{rea74} in 1974. Reaves showed that
intermediate-distance clusters are less often seen behind nearby
clusters than very distant ones. Correctly, he attributed this to the
difficulty of distinguishing clusters in projection when they are not
well separated along the line of sight, and the two cluster luminosity
functions peak at a similar magnitude. 

Fritz Zwicky did not live long enough to read Reaves' paper.  He died
on Feb. 8$^{th}$ 1974, just a few days before his 76$^{th}$ birthday.

\begin{figure}
\begin{center}
\psfig{figure=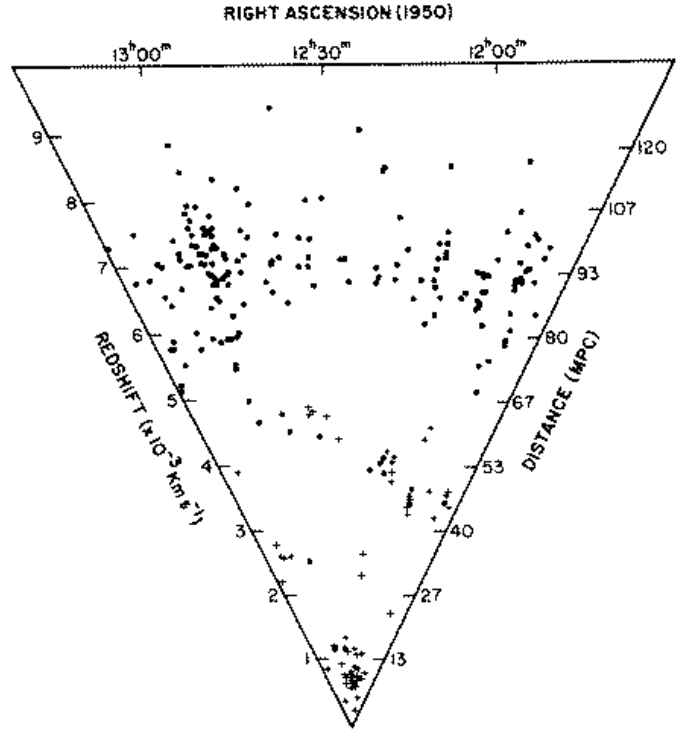,height=10cm,angle=0}
\end{center}
\caption{The wedge diagram of the Coma supercluster; crosses indicate
galaxies that would be too faint to be detected if they were at the
distance of the Coma cluster -- from Gregory \& Thompson (1978)
\label{fig-gtcomasc}}
\end{figure}

After Zwicky's death the reality of superclusters was no longer
questioned.  A major breakthrough in this topic came with the
extensive redshift surveys of Chincarini, Gregory, Rood, Tarenghi,
Thompson \& Tifft\cite{tif76,gre78,tho78,tar78a,chi79,tar80,gre81},
that drew the 3-dimensional structures of the Coma -- see
Fig.\ref{fig-gtcomasc} --, Hercules, Hydra-Centaurus, Perseus and
Pisces superclusters. Cluster-connecting filaments and voids were
identified. The emerging picture was thus summarized by
Abell\cite{abe77}:
\begin{quote}
{\em ``The picture that suggests itself is that of a large
inhomogeneity or region of space containing galaxies, groups, and
clusters, in which what is commonly called the Coma cluster is simply
a dense concentration, rather like an urban center in a large
metropolitan area''}
\end{quote}
In 1978 J\^oeveer et al.\cite{joe78} described Perseus and other
eight superclusters, and noted that the majority of clusters of
galaxies form chains. Einasto et al.\cite{ein80} pointed out that the
large scale structure of the Universe resembles cells, with galaxies
and galaxy clusters concentrated towards cell walls, whereas the
spatial density of galaxies inside cells is very low. In 1981
Kirshner et al.\cite{kir81} found the million Mpc$^3$ Bo\"otes void,
that Bahcall \& Soneira\cite{bah82} showed to be associated with the
Hercules supercluster and the CorBor extension.

Numerical simulations were keeping abreast of observations: in 1979
Aarseth et al.\cite{aar79} were able to produce 3-dimensional plots of
the galaxy distribution where the recently discovered huge voids were
quite evident\footnote{In the discussion following Aarseth's
talk\cite{aar78}, Peebles referred to Aarseth's plots as {\em
``propaganda films''} and deemed it {\em ``very dangerous to compare
them too closely to the real Universe''.}}.

\subsection{Clusters and the Large Scale Structure of the Universe}
\begin{figure}
\begin{center}
\psfig{figure=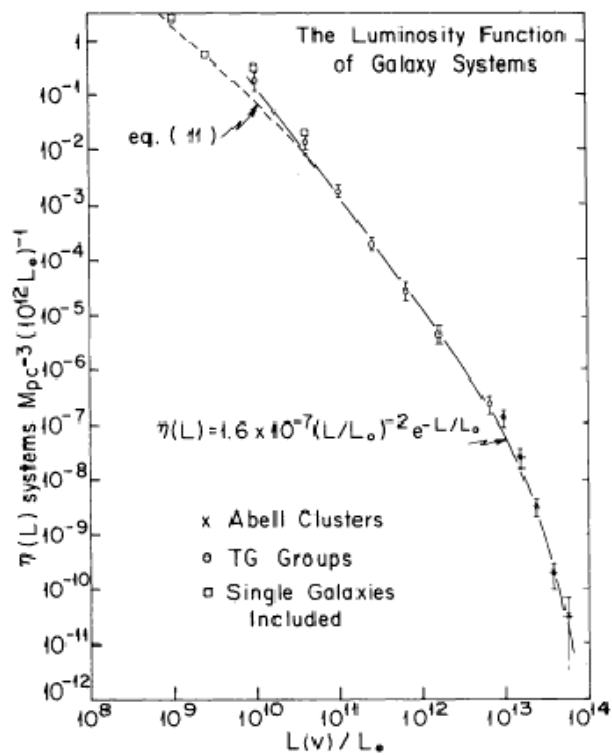,height=10cm,angle=0}
\end{center}
\caption{The luminosity function of all galaxy systems. The solid line
represents the best fitting curve. From Bahcall (1979a).
\label{fig-nblf}}
\end{figure}

The huge observational effort of the seventies made it possible to
evaluate the distribution functions of cluster properties. At the end
of the 70's Chincarini\cite{chi78} established the relation between
cluster luminosities and their richness classes.  One year later,
based on similar relations, Neta~Bahcall\cite{bah79a,bah79b} produced
the first optical -- see Fig.~\ref{fig-nblf} --
and X-ray luminosity functions of galaxy systems,
ranging six decades in luminosity. Subsequent studies, based on larger
data-sets, confirmed the validity of Bahcall's determinations (see,
e.g., McKee et al.\cite{mck80}, Hintzen et al.\cite{hin80} and
Abramopoulous \& Ku\cite{abr83}). A preliminary attempt to produce the
virial mass function of clusters was done by Struble \&
Bludman\cite{str79a}, but their sample was incomplete and biased at
the low-mass end. The first unbiased estimates of the cluster mass
function\cite{bah93,biv93} would only come in 1993, 14 years later.

In 1982 Davis et al.\cite{dav82} produced the first wide-angle galaxy
redshift survey, not dominated by the Local Supercluster.  The authors
hoped that their survey {\em ``would begin to approximate a fair
sample volume of the universe''.} Maybe the first CfA survey was no so
{\em ``fair''} after all, but Davis et al.\cite{dav82}'s description
of the galaxy distribution was fairly correct. The galaxy
distribution, they wrote, {\em ``is frothy, characterized by large
filamentary superclusters of up to 45 Mpc in extent, and corresponding
large holes devoid of galaxies''}. 

A major output of the first CfA survey was Huchra \&
Geller\cite{huc82}'s catalogue of groups of galaxies. For the first
time, groups were identified in 3-dimensions, as volume-density
enhancements in the distribution of galaxies. Of the 176 catalogued
groups, 74 were identified for the first time\cite{gel83}. In those
years, another famous catalogue of groups was created,
Hickson\cite{hic82}'s catalogue of 100 compact groups.

\begin{figure}
\begin{center}
\psfig{figure=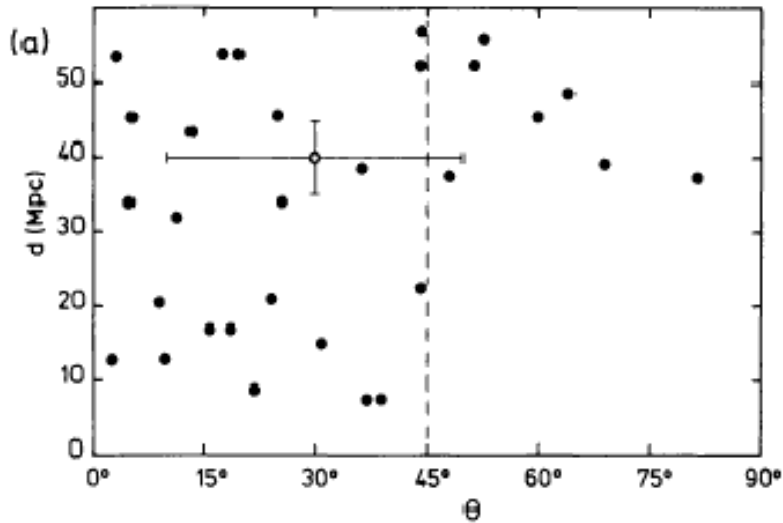,height=7cm,angle=0}
\end{center}
\caption{The difference between the cluster
position angle and the position angle defined by the
direction to the closest neighbouring cluster (x-axis), vs.
the spatial distance to the closest neighbour (y-axis).
From Binggeli (1982).
\label{fig-bbalign}}
\end{figure}

Meanwhile, astronomers started to use galaxy clusters as tracers of
the Large Scale Structure of the Universe. Binggeli\cite{bin82}
showed the existence of cluster alignments on scales up to 45 Mpc
-- see Fig.fig-bbalign. The
cluster correlation function was computed by Bahcall \&
Soneira\cite{bah83} and Klypin \& Kopylov\cite{kly83}, and shown to
extend to 200 Mpc. Other useful tracers of the Large Scale Structure
were found to be voids (Sharp\cite{sha81}) and Lyman-$\alpha$
absorbers, which Oort\cite{oor81} used for the first time to shed
light on the clustering at very high redshift ($z > 2$).

In 1983 Abell\cite{abe83a} revised the properties of superclusters and
suggested that they constitute the end of the clustering hierarchy,
since their separations are comparable to their sizes, so that
superclusters are interconnected. Shortly before his death, occurred
on October 7$^{th}$ 1983, Abell\cite{abe83b} (together with Corwin)
announced the preparation of the southern extension of his catalogue,
a work that would keep busy his collaborators for six more
years\cite{abe89}. Abell's original catalogue was however to remain
unsurpassed for the quality of the cluster richness estimates (see
Girardi et al.\cite{gir93}).

\section{The cluster components}
\subsection{The morphology-density relation}
It was probably Harold Shapley\cite{sha26} in 1926 the first to
explicitly refer to the different galaxy content of the Virgo and the
Coma cluster, Coma being dominated by {\em ``spheroidal''} galaxy
types\footnote{It was only in 1923 that Reynolds\cite{rey23b} pointed
out the existence of many {\em ``globular or ovoid''} nebul\ae,
distinctly different from spirals.}.  However, Shapley thought that
with increasing resolution many apparently featureless spheroidals
would turn out to be real spirals.  Ten years after, in {\sl The Realm
of the Nebul\ae}, Hubble first hinted at the existence of a
morphology--density relation:
\begin{quote}
{\em ``There are some indications of a correlation between characteristic type
and compactness, the density of the cluster diminishing as the most frequent
type advances along the sequence of classification''} 
\end{quote}
Hubble also noted the {\em ``dominance of late typed among isolated
nebul\ae~ in the general field''.} The morphology-density relation was
immediately regarded as fundamental, to such a point that
Tombaugh\cite{tom37}, in 1937, thought that a galaxy overdensity
dominated by spirals could not be a real cluster. In the same year,
Tombaugh noted that cluster ellipticals are more centrally
concentrated than cluster spirals. In 1942 Zwicky\cite{zwi42a} showed
that S0s in Virgo are distributed like ellipticals and unlike spirals.

In 1960 van~den~Bergh\cite{vdb60a} first noted the existence of a
correlation between morphology and {\sl local} galaxy density. By
examining the Ursa Major and Virgo clusters, he noted that
\begin{quote}
{\em ``there is some indication that the nebular population type
is related to the surface density of galaxies''}
\end{quote}
In those years, \dV\cite{dev61a,dev61b} (see also Abell\cite{abe62})
suggested that spirals and ellipticals in Virgo have different
distributions simply because they belong to different clusters. The
morphology-density relation was thus reduced to a mere projection
effect. An even more extreme view was taken by Neyman et
al.\cite{ney62} who maintained that the observed scarcity of spirals
in clusters with respect to the field could be understood as {\em ``a
difference in the difficulty of observations''}!

In 1965 an extreme case of morphological segregation was discovered.
Morgan \& Lesh\cite{mor65} noted that many clusters are centrally
dominated by {\em ``supergiant galaxies'',} that they called
cDs. These galaxies were shown to live in the densest cluster
environment only. Not only are cDs lacking in the field, but also in
poor clusters and groups. In fact, the central dominant galaxies of
the poor clusters classified by Morgan et al.\cite{mor75}, were later
shown to lack the characteristic extended envelope of cDs (Thuan \&
Romanishin\cite{thu81}).

In the 70's the number of available galaxy redshifts increased
considerably, finally allowing a more reliable identification of
cluster members. Rood et al.\cite{roo72} were then able to identify 16
spirals as members of the Coma cluster.  The idea that rich clusters
are dominated by ellipticals and S0s was so firmly established that
Rood et al.'s was considered a {\em ``striking''} result.

In 1974 Oemler\cite{oem74} published his seminal paper {\sl The
systematic properties of clusters of galaxies. I. Photometry of 15
clusters.}  He noted that the morphological segregation in clusters
depends on the cluster content. The morphology-density relation was
interpreted as a relation between the morphological content of a
cluster and its compactness. Oemler constructed galaxy number density
profiles by type, and noticed a decreasing space density of spirals
towards the cluster centres, except in spiral-rich clusters. He also
noticed that spirals in cD-clusters have a shallower density
profile than ellipticals at large radii. However, he could not notice
any difference between the density profiles of S0s and ellipticals.

A year later, Gregory\cite{gre75} showed that the fraction of spirals
indeed increases with the distance from the Coma cluster centre. He
wrote:
\begin{quote}
{\em ``The increase in relative numbers of spiral and irregular
galaxies with radial distance seems incontestable. The effect
is so strong as to be obvious to the eye on a casual inspection
of the Sky Survey''}
\end{quote}
Melnick \& Sargent\cite{mel77} confirmed Gregory's finding in
other six X-ray bright clusters.

This tendency for ellipticals to be more clustered than spirals was
shown by Davis \& Geller\cite{dav76} not to be restricted to
clusters. They applied the 2-point correlation function to the Uppsala
catalogue to show that morphological segregation exists on scales up
to 6 Mpc. Four years earlier, in 1972, Takase\cite{tak72} had already
pointed out a colour segregation of galaxies on the scale of the Local
Supercluster.

\begin{figure}
\begin{center}
\psfig{figure=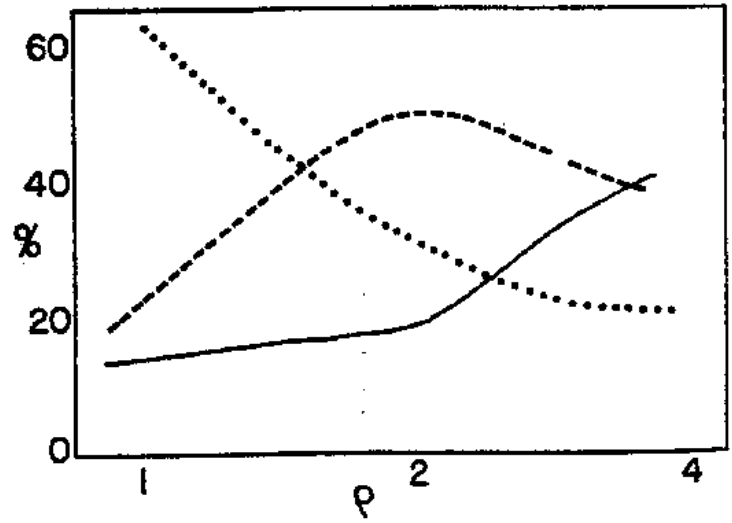,height=7cm,angle=0}
\end{center}
\caption{The variation of galaxy population with the mean density
of clusters. Solid-line: ellipticals; dashed-line: S0s;
dotted-line: spirals. From Oemler (1977).
\label{fig-aomorph}}
\end{figure}

\begin{figure}
\begin{center}
\psfig{figure=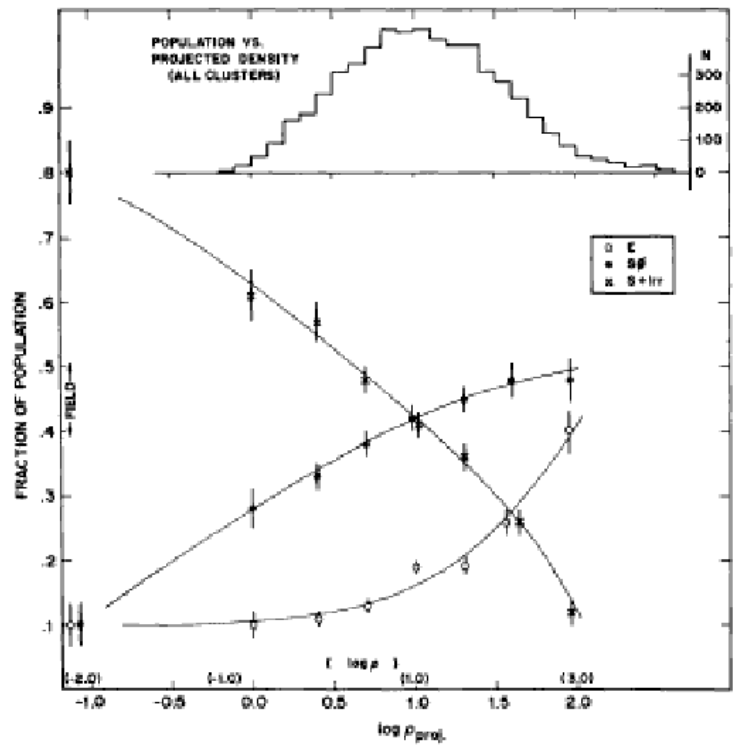,height=10cm,angle=0}
\end{center}
\caption{The fraction of E, S0, and S+I galaxies as a function
of the logarithm of the projected density. The upper histogram
shows the number distribution of the galaxies over the bins of 
projected density. From Dressler (1980a).
\label{fig-admorph}}
\end{figure}

In 1977 Oemler\cite{oem77} wrote that {\em ``density is the physical significant
parameter in determining the galaxy population of a cluster.''}
Figure~3 of his paper -- here reproduced in Fig.~\ref{fig-aomorph} --
is qualitatively very similar to Figure~4 in the
1980 paper of Dressler\cite{dre80a} -- here reproduced
in Fig.~\ref{fig-admorph}. Both figures show the fractional
variation of spirals, S0s and ellipticals as a function of the cluster
density. However, Oemler's density is the {\sl mean cluster density,}
and Dressler's density is the {\sl local density} around each
galaxy. Anyway, Oemler wrote (but did not show) that the same
morphology-density relation was also verified individually in clusters
dominated by early-type galaxies. The same year, even a spiral-rich
cluster (Abell~262) was found to display a {\em ``striking''}
morphological segregation (Moss \& Dickens\cite{mos77}).

Times were mature for Alan Dressler's milestone paper, {\sl Galaxy
morphology in rich clusters: implications for the formation and
evolution of galaxies}\cite{dre80a}, published in 1980, and based on
the evergreen {\sl Catalog of morphological types in 55 rich clusters
of galaxies}\cite{dre80b}. Dressler pointed out that: i) regular as
well as irregular clusters display the same morphology-density
relation; ii) it is not the radial distance, but the local density,
the basic parameter which determines the morphology mix. Dressler's
conclusions are still controversial nowadays (see, e.g. Sanrom\`a \&
Salvador-Sol\'e\cite{san90}), and it is possible that both global
cluster properties {\sl and} the local galaxy environment may play a
role in determining the galaxy morphology\cite{tho01}.

In the two following years, Bhavsar\cite{bha81} and
de~Souza\cite{des82} extended Dressler's morphology-density relation
into the low galaxy density regime, through the analysis of loose
groups.

\subsection{Luminosity segregation}
The idea that clusters form by gravitational clustering of field
galaxies led Zwicky\cite{zwi37} (and others) to suggest that cluster
galaxies are more massive than average, making their mutual
gravitational attraction stronger. The most massive
galaxies would cluster first, forming the cluster core, and other
galaxies would follow.  Assuming proportionality between a galaxy
luminosity and its mass, Zwicky then thought that luminosity
segregation must exist in clusters. Between 1942 and 1951 he found
some evidence for it in Virgo\cite{zwi42a}, and in
Coma\cite{zwi51}. At the same time he noted that also dwarf galaxies
are clustered\footnote{Reaves noted that the main problems for
the identification of dwarf galaxies were their low surface brightness,
and the fact that these galaxies {\em ``resemble water spots and
certain common emulsion defects''.}}, an evidence later confirmed by
Reaves\cite{rea56} and Hodge\cite{hod59}.

\begin{figure}
\begin{center}
\psfig{figure=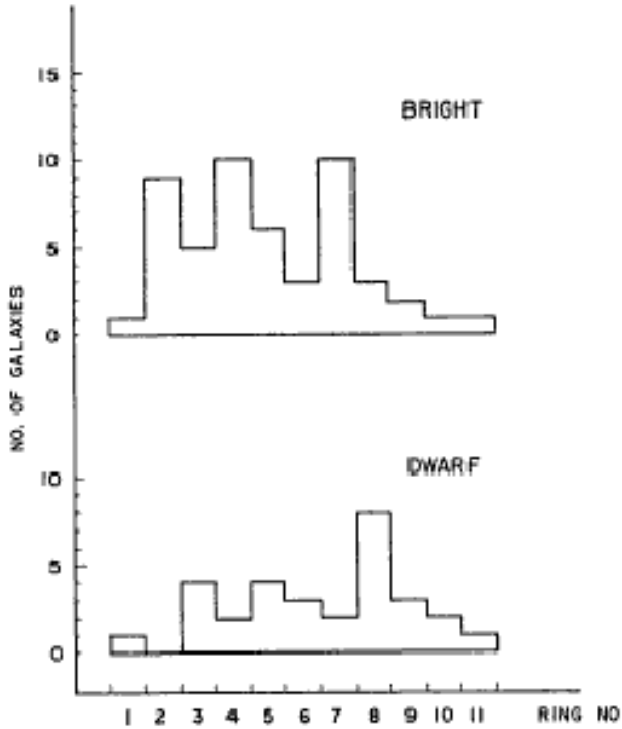,height=10cm,angle=0}
\end{center}
\caption{The radial distribution of bright and dwarf
galaxies in the Coma cluster. From Rood \& Turnrose (1968).
\label{fig-rtlumseg}}
\end{figure}

In the sixties, Reaves\cite{rea66} and Rood \& Turnrose\cite{roo68}
showed that dwarf galaxies are less clustered than giant galaxies
-- see Fig.~\ref{fig-rtlumseg}.
Not much later, Rood\cite{roo69} and Rood \& Abell\cite{roo73} noted
that the bright peak in the luminosity function of Coma galaxies
(first described by Shapley\cite{sha34} in 1934), is not present in the
outer regions of the cluster. This was interpreted as evidence for an
excess of bright galaxies in the cluster core, i.e. luminosity
segregation.

Oemler\cite{oem74} noted an increase of the mean radius of cluster
galaxies with galaxy magnitudes, another evidence for luminosity
segregation, which was not seen, however, in spiral-rich clusters.

Capelato et al.\cite{cap80a} examined in detail the luminosity
segregation in Coma, showing that it concerns the most luminous
galaxies in a range of about 2 magnitudes. They also enlightened the
role of the central cD in destroying the evidence of luminosity
segregation through cannibalism, as originally suggested by
Dressler\cite{dre78}.

Luminosity segregation also had opponents, like Noonan\cite{noo61},
Bahcall\cite{bah73}, and Sarazin\cite{sar80}, who suggested the
evidence for luminosity segregation to be spurious, and mostly due to
poor background subtraction. Recent analyses\cite{biv92,biv01}, based
on cluster members only, show that luminosity segregation is indeed
limited to the very bright galaxies only, $M_R < -22.6$.

\subsection{Kinematical segregation}
The issue of kinematical segregation also dates back to the 30's.
Smith\cite{smi36} pointed out that there was no evidence for
bright and faint galaxies in the Virgo cluster to have different
velocity distributions, and so did Zwicky\cite{zwi37} for galaxies in
the Coma cluster. The first evidence for kinematical segregation of
cluster galaxies came from Holmberg\cite{hol40} who, as early as in
1940, noticed that Virgo spirals had a larger velocity dispersion than
Virgo ellipticals, thus anticipating Tammann\cite{tam72}'s result.

Chandrasekhar\cite{cha43}'s paper on dynamical friction showed how
the more massive galaxies in a cluster could decelerate 
with respect to the less massive galaxies. However, a huge
observational effort was needed before a clear evidence for
kinematical segregation was established. In 1960, only 50 redshifts
were known for galaxies in the Coma cluster, each obtained through
$\simeq 2$ hours exposures\cite{may60}, leading Mayall to complain
that the {\em ``current rate of less than 10 velocities per year is
impracticably slow''.} 

In 1964, Zwicky \& Humason\cite{zwi64} had obtained 42 galaxy
redshfits in the cluster Abell~194. They claimed that the 21 brightest
galaxies had a higher velocity dispersion than the 21
faintest. Reanalyzing their data with a biweight estimator\cite{bee90}
proves their result was correct. In fact, there is a difference of
200 km/s between the velocity dispersions of the bright and faint
samples, and this is significant at the $\sim 95$~\% level. The
conclusions of Zwicky \& Humason were confirmed 13 years later by
Chincarini \& Rood\cite{chi77}, on a slightly larger sample of 57
redshifts for cluster members.  Meanwhile, in 1972 Rood et
al.\cite{roo72} had shown the velocity dispersion of bright
galaxies in the Coma cluster core to be as low as 231 km/s.

\begin{figure}
\begin{center}
\psfig{figure=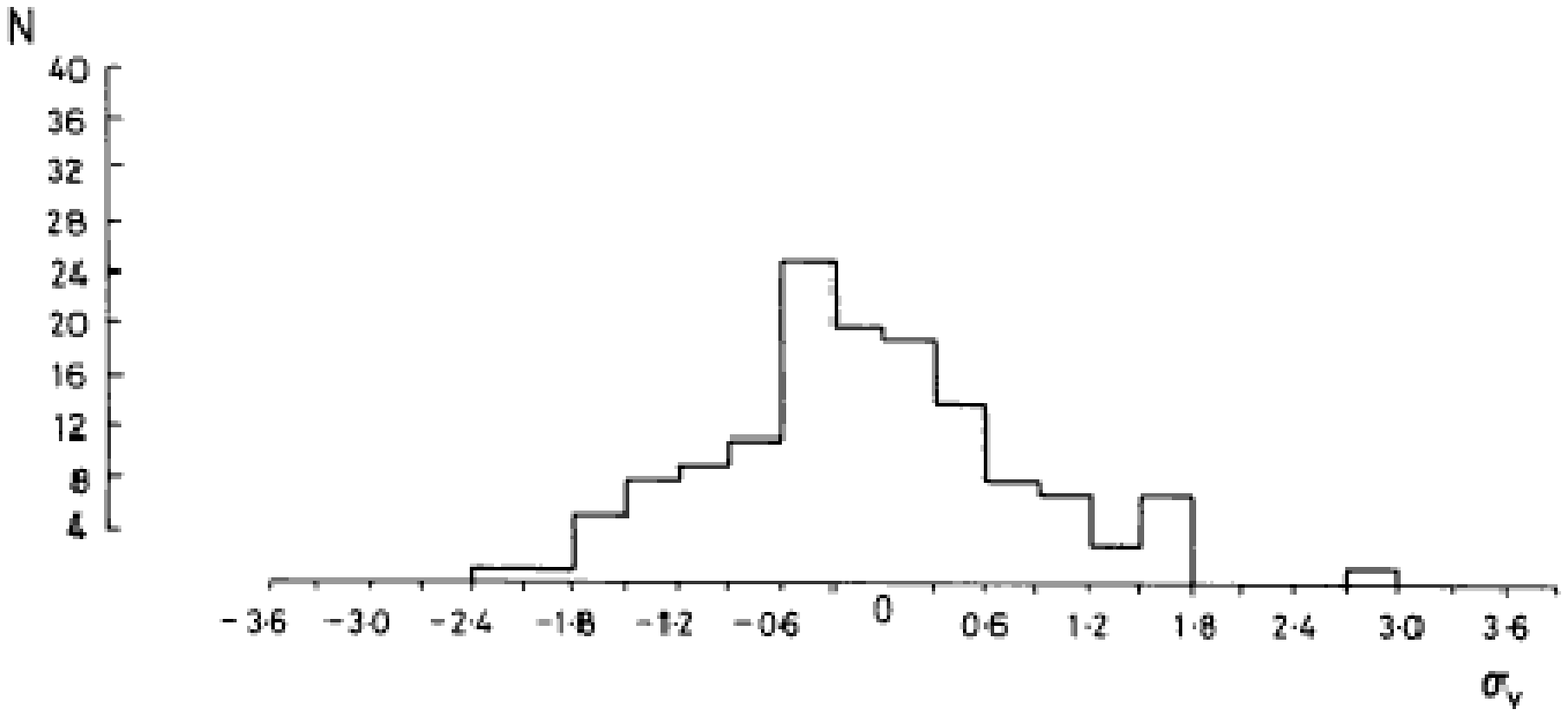,height=5cm,angle=0}
\psfig{figure=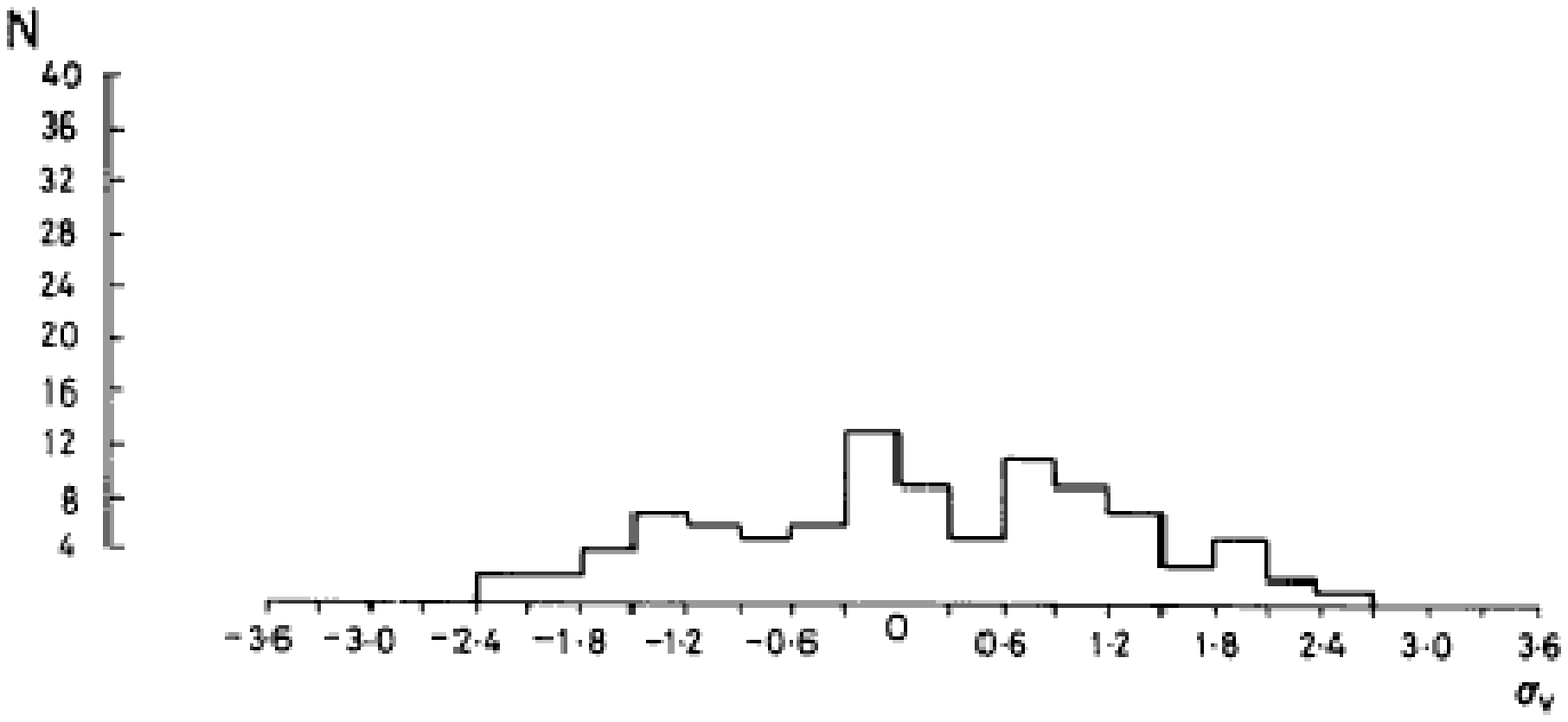,height=5cm,angle=0}
\end{center}
\caption{The combined velocity distribution of ellipticals and
S0s (top panel) and spirals (bottom panel) 
in five clusters. From Moss \& Dickens (1977).
\label{fig-mdsig}}
\end{figure}

In the same year, Tammann\cite{tam72} put Holmberg's early result on
solid bases, by analyzing a sample of 122 Virgo cluster members with
available velocities. Tammann showed that the velocity dispersion of
Virgo spirals was 40~\% higher than that of ellipticals and S0s.
Tammann's result was extended by Moss \& Dickens\cite{mos77} to
clusters in general. Moss \& Dickens showed that the velocity
distribution of ellipticals and S0s is broader than that of spirals
not only in Virgo, but also in Abell~194, 262, and 1367 --
see Fig.~\ref{fig-mdsig}.
Kent \&  Gunn\cite{ken82} later found the same effect in Coma.

Struble\cite{str79b} considered 13 galaxy clusters, each with at least
30 galaxy redshifts, up to a maximum of 325 in Coma. Using the
variance-ratio test he showed that there was no evidence for
kinematical segregation with luminosity, except in Coma. Since
Abell~194 was among the clusters he considered, his result was at odds
with those of Zwicky \& Humason\cite{zwi64} and Chincarini \&
Rood\cite{chi77}. Struble noticed that several clusters have a lower
velocity dispersion in their cores, and interpreted it as a product of
cannibalism and/or dynamical friction, a scenario that still
holds\cite{ada98}.

\begin{figure}
\begin{center}
\psfig{figure=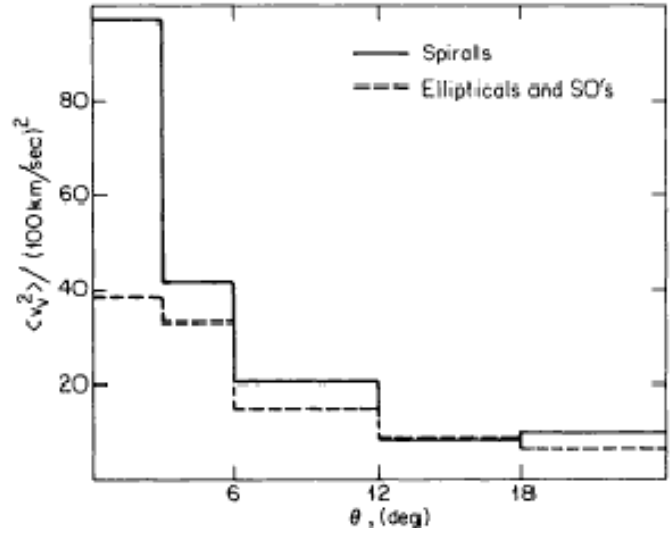,height=7cm,angle=0}
\end{center}
\caption{The velocity dispersion profiles for ellipticals and S0s
(dashed line) and spirals (solid line) in the Virgo cluster. From
Hoffman et al. (1980).
\label{fig-hofvdp}}
\end{figure}

Thanks to the huge observational effort of the 70's, in 1980 there
were more than 800 Virgo cluster galaxies with available
redshifts. Using this sample, Hoffman et al.\cite{hof80} constructed
the velocity dispersion profile of the Virgo cluster, for spirals and
early-type galaxies separately.  Not only the velocity dispersion of
spirals was confirmed to be higher than that of ellipticals and S0s,
but also the shapes of the velocity dispersion profiles were
different. By looking at Figure~9 in Hoffman et al.'s paper -- here
reproduced in Fig.~\ref{fig-hofvdp} --, we can notice that the
velocity dispersion profile of spirals is significantly steeper than
that of early-type galaxies. It almost took 20 years to extend the
validity of such a result to clusters in general (Adami et
al.\cite{ada98}).

\subsection{Star formation in cluster galaxies}
\begin{figure}
\begin{center}
\psfig{figure=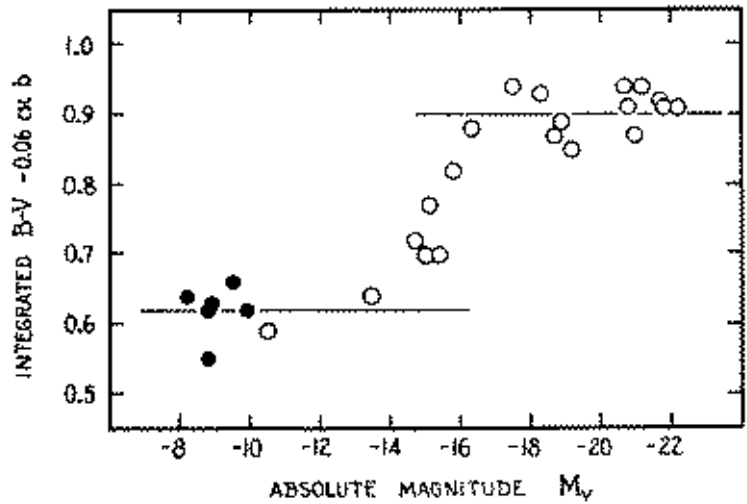,height=7cm,angle=0}
\end{center}
\caption{Intrinsic colour indices of old stellar systems
as a function of their absolute magnitudes. The circles represent
elliptical galaxies, and the dots globular clusters.
\label{fig-baumcm}}
\end{figure}

The first to notice the small spread of the colours of cluster
galaxies was Baade\cite{baa31} in the 30's. Such a small spread was
related to the predominance of ellipticals and S0s among cluster
galaxies, and the existence of a tight colour-magnitude relation,
discovered by Baum\cite{bau59} -- see Fig.~\ref{fig-baumcm} --
and \dV\cite{dev61c} around 1960, and
refined by Visvanathan \& Sandage\cite{vis77} in 1977.  Recently,
Stanford et al.\cite{sta98} confirmed the validity of the
colour-magnitude relation also for distant clusters ($z \simeq
0.9$). They also showed that the relation is one between the mass and
the metallicity of galaxies. The tightness of the colour-magnitude
relation and its mild evolution with redshift indicate that most
cluster ellipticals (and S0s) have formed at high redshifts, and they
evolve passively through the aging of their (old) stellar
populations (see, e.g., {\sc Dickinson}, these proceedings).

As far as cluster spirals are concerned, it was Erik
Holmberg\cite{hol58}, in 1958, the first to notice that Virgo spirals
are redder than field spirals. His result was confirmed by Chester \&
Roberts\cite{che69}, Davies \& Lewis\cite{dav73} and van den
Bergh\cite{vdb75} around 1970, and later
interpreted\cite{vdb76,ken83a} as a decreased star formation rate in
cluster spirals.

In 1973, Davies \& Lewis\cite{dav73} analyzed the HI-content of 25
Virgo galaxies and showed it to be 60~\% lower than in field galaxies,
on average. Three years later, van den Bergh\cite{vdb76} coined the
term {\em ``anemic spirals''} to indicate a class of galaxies with
intermediate characteristics between normal spirals and S0s.  He
attributed their anemic appearance to a reduced star formation rate,
probably a result of their HI-deficiency. A reduced star formation
rate could also naturally explain the redder colours of Virgo spirals,
an interpretation later supported by Kennicutt\cite{ken83a}. 

In following years, Davies \& Lewis' result was generalized to other
clusters by Sullivan and collaborators\cite{sul78,sul81}, Giovanelli
et al.\cite{gio81}, Chincarini et al.\cite{chi83}, and Giovanelli \&
Haynes\cite{gio83}. These authors also showed that HI-deficient
galaxies preferentially occur in high-density regions, i.e. the rich
cluster cores. A recent update on this topic can be found in {\sc Solanes}
(these proceedings).

\subsection{Density and velocity dispersion profiles}
\begin{figure}
\begin{center}
\psfig{figure=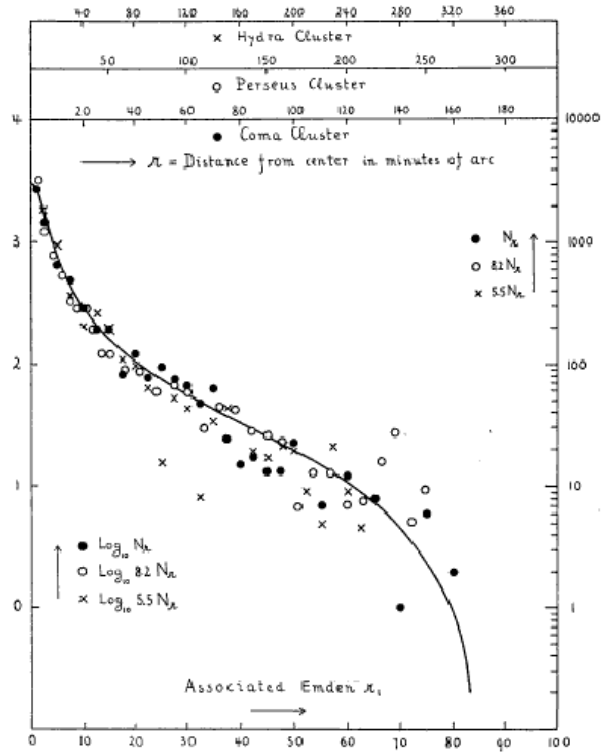,height=10cm,angle=0}
\end{center}
\caption{The number of nebul\ae~ per square degree vs. the distance
from the cluster centre and the best fit Emden model (solid line).
From Zwicky (1942b).
\label{fig-fzemden}}
\end{figure}

It was Zwicky\cite{zwi42b}, in 1942, the first to propose an
analytical form for the spatial distribution of galaxies in clusters,
i.e. Emden's model for a bounded isothermal gas sphere -- see
Fig.~\ref{fig-fzemden}. In 1954, Shane \& Wirtanen\cite{sha54} found
that the surface brightness profile of galaxy clusters could also be
fitted with the distribution function proposed by \dV\cite{dev48} as a
fit to the surface brightness profile of elliptical galaxies. As a
matter of fact, the similarity of the profiles of ellipticals and the
Coma cluster had already been noted by Zwicky\cite{zwi37} in 1937. In
1962 Abell\cite{abe62} pointed out that equally good fits could be
obtained using distribution formul\ae~ different from Emden's. The
fact that Emden's model fit the data well could not be taken as
evidence that clusters are isothermal spheres. One year later, as to
support Abell's conclusions, King\cite{kin62} published his empirical
density law for star clusters which proved very successful in
describing cluster density profiles as well.

One of the assumptions of all these models, spherical symmetry, was
called into question when Matthews et al.\cite{mat64} and
Sastry\cite{sas68} noted that the major axis of the central giant
galaxy was aligned with the galaxy distribution in cD clusters, thus
anticipating the results of Carter \& Metcalfe\cite{car80} and
Binggeli\cite{bin82}. Moreover, The NE--SW elongation of the Coma
cluster was remarked upon by Bahcall\cite{bah73}, Schipper \&
King\cite{sch78}, and Thompson \& Gregory\cite{tho78}.  Things
complicated even further when Sharov\cite{sha59}, Omer et
al.\cite{ome65}, and Clark\cite{cla68} found evidence for secondary
peaks in the density profiles of several clusters. Their findings were
later confirmed by Oemler\cite{oem74}. In 1978 Dressler\cite{dre78}
proposed subclustering as an explanation for irregularities in the
density profiles.

\begin{figure}
\begin{center}
\psfig{figure=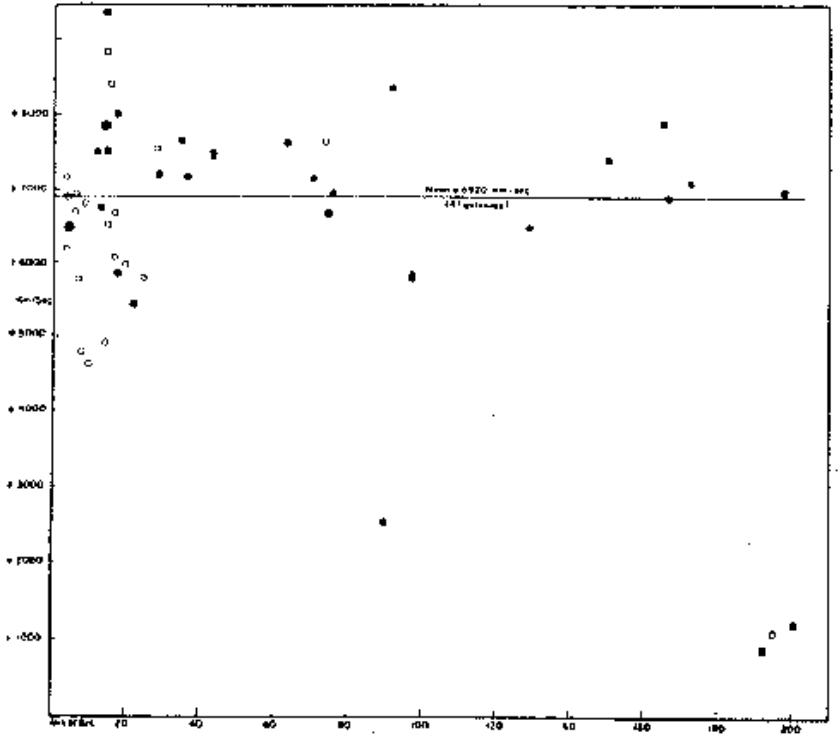,height=10cm,angle=0}
\end{center}
\caption{The velocities of galaxies in the Coma region vs.
clustercentric distances. The distances range from 0 to 200',
the velocities from 0 to 9000 km/s. From Mayall (1960).
\label{fig-mayvr}}
\end{figure}

Another assumption of Zwicky's model was isothermality, an hypothesis
supported by an early plot of the Virgo galaxy velocities
vs. clustercentric distances (Smith\cite{smi36}).  The validity of
Zwicky's assumption was shattered in 1960 by Mayall\cite{may60}'s
diagram of velocities vs. radii for 50 galaxies in the Coma
cluster. In this diagram -- here reproduced in Fig.~\ref{fig-mayvr} --
one could clearly see a decrease of the
velocity dispersion with radius. A similar trend was later found by
Karachentsev\cite{kar65} for the Virgo cluster.  On the other hand,
Zwicky \& Humason\cite{zwi64} found a flat velocity dispersion profile
in Abell~194.

In 1971 Chincarini \& Rood\cite{chi71} showed the Perseus cluster to
have a decreasing velocity dispersion profile, and one year later
Rood et al.\cite{roo72} confirmed Mayall\cite{may60}'s early
suggestion that the Coma cluster velocity dispersion profile is a
decreasing function of the clustercentric distance. These early
measurements of the Perseus and Coma velocity dispersion profile, were
later refined by Kent \& Sargent\cite{ken83b} and, respectively, Kent
\& Gunn\cite{ken82}, who confirmed deviation from isothermality.  The
velocity dispersion profiles of galaxy clusters were classified into
four different types by Struble\cite{str79b}.  He showed that
isothermal profiles are not a common feature of all clusters.

Density and velocity dispersion profiles have now been obtained for
the different galaxy populations\cite{ada98,biv01}.  Velocity
dispersion profiles are certainly not isothermal, and are different
for different galaxy populations\cite{biv01}, so that the global
velocity dispersion profile of a cluster changes according to its
galaxy morphological mix. So far, no analytical model has been
proposed for the cluster velocity dispersion profile. Recently Navarro
et al.\cite{nav96} have proposed a new analytical model for the
cluster density profiles, which is now extremely popular. Consistency
has been found between this new model and the data, but, once more,
other models provide equally good fits to the data\cite{car97}.

\subsection{The hot IC gas}
It was Limber\cite{lim59} in 1959 the first to suggest that diffuse
gas must be present among galaxies, and clusters be filled with a hot
IC diffuse gas component. He argued that galaxy formation from gas
cannot be 100~\% efficient, and some gas must be lost from galaxies
through collisions. The first detection of an X-ray source associated
with a cluster of galaxies came from Byram et al.\cite{byr66}, in
1966. They detected M~87, the central giant galaxy of the Virgo
cluster. In the same year, Boldt et al.\cite{bol66} claimed detection
of the Coma cluster in X-ray.  It took just one year to Friedman \&
Byram\cite{fri67} to show that Boldt et al.'s detection was spurious.
However, Boldt et al.'s spurious result inspired Felten et
al.\cite{fel66}'s correct theoretical estimate. Felten et
al. estimated that a thermalized diffuse gas in the Coma cluster
should have a temperature $\simeq 7 \times 10^7$ K, and would
therefore emit in the X-ray via thermal bremsstrahlung.

In 1971, Cavaliere et al.\cite{cav71} suggested that many
extragalactic X-ray sources are probably associated with clusters of
galaxies. The same year, the extended X-ray emission from the Coma IC
gas was detected, by Meekins et al.\cite{mee71}, with observations
from an {\sl Aerobee} 150 rocket, and, independently, by Gursky et
al.\cite{gur71}, with the {\sl Uhuru} satellite. Thanks to {\sl Uhuru}
many more clusters were X-ray detected, and as early as in 1972, Gursky
et al.\cite{gur72} suggested that
\begin{quote}
{\em ``most, if not all, rich clusters include an X-ray emission
region of large size and of net luminosity $10^{43}$--$10^{44}$ erg 
s$^{-1}$''}
\end{quote}
A first indication about the nature of the diffuse cluster X-ray
emission came from Solinger \& Tucker\cite{sol72} in 1972, with an
early indication of a correlation between the X-ray luminosities of
clusters and the velocity dispersions of their member galaxies.
Such a correlation is naturally expected if the gas is thermalized,
in equilibrium with the cluster gravitational potential, and the
emission mechanism is thermal bremsstrahlung. This correlation
was later improved by Cooke \& Maccagni\cite{coo76}.

Always in 1972, Syunyaev \& Zel'dovich\cite{sun72a} proposed {\sl The
observation of relic radiation as a test of the nature of X-ray
radiation from the clusters of galaxies.} Immediately after, an
over-enthusiast Parijsky\cite{par72} gave a start to a series of
spurious detections of the {\sl Syunyaev--Zel'dovich effect.} Other
early controversial detections were claimed by Gull \&
Northover\cite{gul76}, Lake \& Partridge\cite{lak77,lak80}, Birkinshaw
et al.\cite{bir78,bir81}, all regarded with much scepticism by
theorists (Gould \& Rephaeli\cite{gou78}, Tarter\cite{tar78b}). White
\& Silk\cite{whi80} noted that the combined X-ray and microwave
observations of Abell~576 would have implied an improbable value for
the Hubble constant of $\simeq$~1.5~km~s$^{-1}$~Mpc$^{-1}$!

There has been an impressive observational progress in this field over
the last decade. Nowadays, the rate of reliable Syunyaev--Zel'dovich
detections of clusters is very high, and techniques allow
Syunyaev--Zel'dovich imaging of galaxy clusters (see {\sc Carlstrom}, these
proceedings).

In 1973, Lea et al.\cite{lea73} analysed the distribution of the IC
gas and showed the gas to be less centrally concentrated than
galaxies. Their model of the IC gas distribution was the first of a
long series\cite{lea75,gul75,bah77a}, among which the $\beta$-model of
Cavaliere \& Fusco-Femiano\cite{cav76,cav78} proved the most
successful.  Lea et al.\cite{lea73}'s result was confirmed by
Bahcall\cite{bah74}, and by Gorenstein et al.\cite{gor79}, who
estimated the slope of the galaxy number density profile in Coma to be
twice the slope of the gas density profile.  Bahcall\cite{bah74} also
showed that the peak of the diffuse X-ray emission coincides with the
centre of the galaxy distribution, or with the position of the cD
galaxy.

Bahcall\cite{bah74} started a systematic comparison of optical and
X-ray cluster properties. She found richer clusters to be more likely
associated with X-ray sources, and cD-type clusters to have higher
X-ray luminosities. On the other hand, she confirmed Kellogg et
al.\cite{kel73}'s result that clusters of a given richness class span
a wide range of X-ray luminosities. Later, she found a relation
between the fraction of spirals in clusters and the X-ray
luminosity\cite{bah77b}.

\begin{figure}
\begin{center}
\psfig{figure=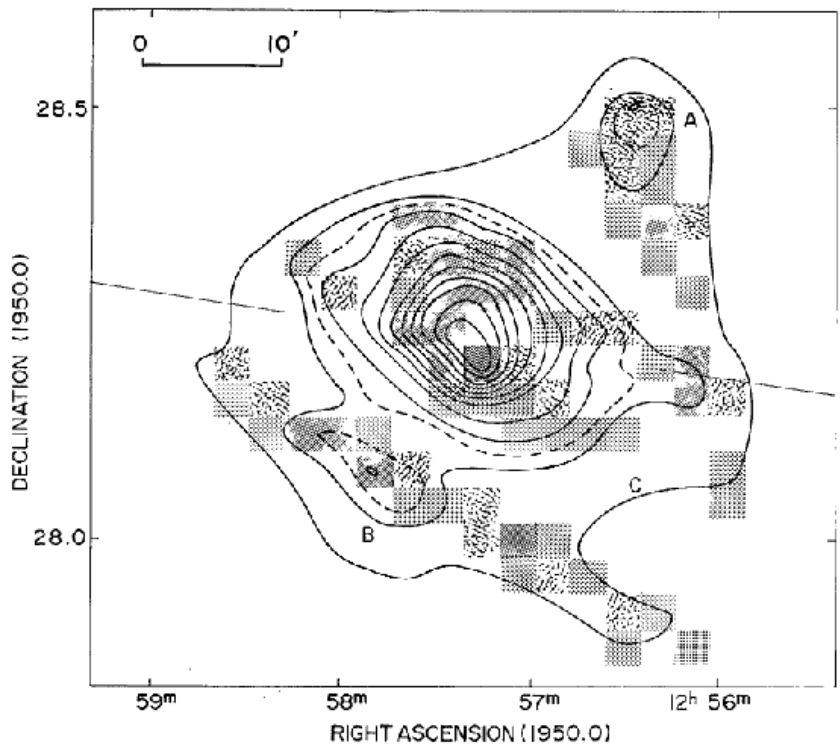,height=10cm,angle=0}
\end{center}
\caption{The Coma cluster X-ray brightness distribution, according to
two different reconstruction algorithms (contours and boxes). The straight
line is the major axis of the galaxy luminosity distribution.
From Johnson et al. (1979).
\label{fig-johcomax}}
\end{figure}

Wolff et al.\cite{wol74} were possibly the first to record a deviation
of the X-ray surface brightness distribution from spherical symmetry. They
showed the X-ray emission of Perseus to be elongated along the E--W
direction, like the galaxy distribution. Some years later, in 1979,
Gorenstein et al.\cite{gor79}, and Johnson et al.\cite{joh79} found a
good correspondence between the shape of the X-ray emission and the
galaxy distribution in Coma -- see Fig.~\ref{fig-johcomax}.
The {\em Einstein IPC} observations of
Jones et al.\cite{jon79} finally revealed all the complex cluster
X-ray morphologies.  The close correspondence between the X-ray
emission and the galaxy distribution was interpreted by Gioia et
al.\cite{gio82} as evidence for equilibrium of both the IC gas and the
cluster galaxies in the cluster gravitational potential.

\begin{figure}
\begin{center}
\psfig{figure=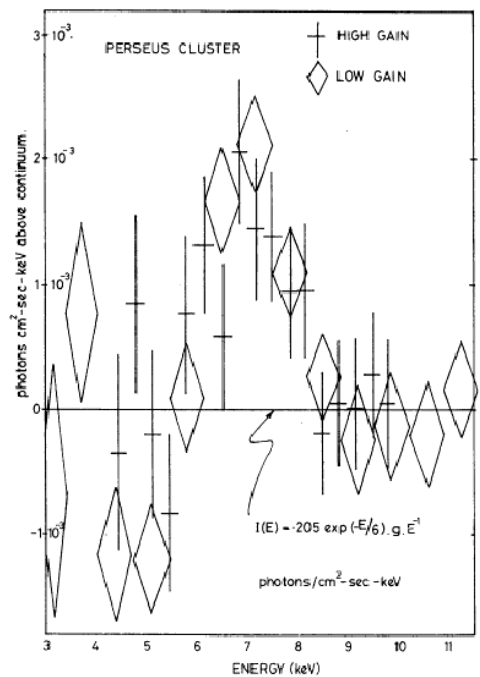,height=9cm,angle=0}
\end{center}
\caption{The deviation of the flux as a function of energy from the
flux predicted by the best fitting single temperature continuum
in the Perseus cluster. The Iron line feature is evident at
around 7 keV. From Mitchell et al. (1976).
\label{fig-mciron}}
\end{figure}

The thermal bremsstrahlung interpretation received further support by
the lack of detection of hard ($>$20~keV) X-ray emission from Coma and
Perseus by Scheepmaker et al.\cite{sch76a}'s balloon-borne X-ray
experiment. The thermal origin of the X-ray emission was finally
demonstrated in 1976 and 1977, with the {\sl Ariel~V} detection of the 7 keV
Iron line in Perseus and Centaurus by Mitchell et al.\cite{mit76} and
Mitchell \& Culhane\cite{mit77a} (see Fig.~\ref{fig-mciron}), and with
the analogous {\sl OSO~8} detections in Virgo, Perseus and Coma, by
Serlemitsos et al.\cite{ser77}.  In 1977, 30 clusters had been
identified as X-ray sources, 10 of them with extended
emission\cite{cul77}.  Mitchell et al.\cite{mit77b} and Mushotzky et
al.\cite{mus78} produced the first relations between the X-ray
temperatures and velocity dispersions of eight, and, respectively, 13
clusters. With much scatter, these relations looked however consistent
with $T_X \propto \sigma_v^2$ (where $T_X$ is the X-ray temperature
and $\sigma_v$ the galaxy velocity dispersion), as expected if the
X-ray emission was produced by an IC gas in equilibrium with the
gravitational potential traced by cluster galaxies.

In 1980, Schwartz et al.\cite{sch80a,sch80b} detected X-ray emission
from poor clusters and compact groups, at temperatures consistent with
the low velocity dispersions of their member galaxies.  The nature of
the X-ray emission from poor galaxy systems is still debated. Both the
contribution of individual galaxies to the total emission and
Supernova heating must be considered (see {\sc Ponman}, these proceedings).

\subsection{Radio components}
The idea that clusters could be associated with extragalactic
radio-sources dates back to 1960. At that time, it was generally
thought that galaxy-galaxy interactions and merging were a
pre-requisite for radio-source activity in galaxies. Spitzer \&
Baade\cite{spi51}'s work had shown that collisions must be frequent
among cluster galaxies.  It was then quite natural to suggest that
extragalactic radio-sources could be associated with galaxy clusters
(Minkowski\cite{min60}). Rogstad et al.\cite{rog65} however pointed
out that radio-galaxies in clusters are often associated with cDs.
Ko\cite{ko65} estimated an average of only one bright radio-galaxy per
cluster.

In their search for clusters of galaxies around radio-sources, Bahcall
et al.\cite{bah69} and Bahcall \& Bahcall\cite{bah70} found evidence
for significant galaxy clustering around quasars at $z \sim 0.1$--0.2.
In those years (the early 70's) the importance of this discovery was
that it provided evidence for a common origin of the galaxy and the
quasar redshifts. If the galaxy redshifts were cosmological, so were
the redshifts of quasars. R\'ozyczka\cite{roz72} extended the
quasar--cluster association up to redshifts $z \sim 0.5$. In 1980
Stockton et al.\cite{sto80} showed that while giant radio-galaxies are
often found in clusters, quasars live in intermediate density
environments, like galaxy groups.

A class of radio-sources that are exclusively found in clusters are
the head-tail radio-sources. Immediately after the IC gas discovery by
Meekins et al.\cite{mee71} and Gursky et al.\cite{gur71}, Miley et
al.\cite{mil72} were able to model this peculiar radio morphology in
terms of radio-trails of galaxies moving through the dense IC gas.

In 1959 Large et al.\cite{lar59} detected the extended radio-source
Coma~C at 408 MHz, in the direction of the Coma
cluster. Willson\cite{wil70} showed Coma~C to be a wide 40 arcmin
diffuse emission, not originating from the integrated emission of
individual galaxies. If located at the distance of the Coma cluster,
the size of Coma~C corresponds to 1.2 Mpc. For this reason, Willson
named it {\em ``the halo''}.

In those days, Coma was still considered as the typical cluster.
However, it was soon clear that clusters with radio-halos are rare.
Hanisch et al.\cite{han79,han82a} could list only four clusters with
detected radio-halos, and Jaffe \& Rudnick\cite{jaf79}'s extensive
search for radio-halos in 32 clusters did not detect any. Eventually,
two other cluster radio-halos were discovered in those years, by
Harris \& Miley\cite{har78} and Roland et al.\cite{rol81a}.

Cluster radio-halos were as difficult to model, as they were to find.
A first attempt was done by Jaffe\cite{jaf77}, who suggested that the
radio-halo could be created from the leakage of electrons out of
radio-galaxies, but the model could not really account for the wide
distribution of the radio-emission. Roland\cite{rol81b} proposed an
{\sl in situ} acceleration of relativistic electrons by magnetic field
fluctuations generated in the wakes of moving galaxies.  A hint to the
nature of radio-halos came from their rarity. In 1979 Smith et
al.\cite{smi79b} remarked that both Coma and Abell~2319 (two
radio-halo clusters) have too high an X-ray temperature for their
velocity dispersion. Three years later, Hanisch\cite{han82b} and
Vestrand\cite{ves82} noted that the rare clusters harbouring
a radio-halo have many other similar properties. These are: anomalous
high X-ray temperatures for their galaxy velocity dispersions, low
spiral contents, intermediate Bautz-Morgan types, large X-ray
core-radii, smooth X-ray distributions, without the central peak
typical of cD clusters. Hanisch and Vestrand suggested that the
presence of a radio-halo could be related to a short-lived dynamical
configuration, thus anticipating modern scenarios (see, e.g., {\sc Feretti},
these proceedings).

\section{Structure}
\subsection{Subclustering}
\begin{figure}
\begin{center}
\psfig{figure=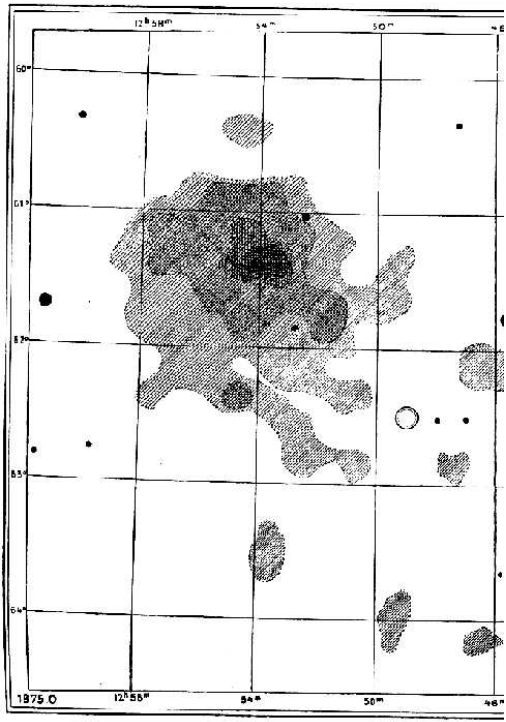,height=10cm,angle=0}
\end{center}
\caption{The density of nebulae in the region of Coma, according to
Wolf (1901). Note the south-western extension (north is up, east 
is to the left). Every grid element is 28'$\times$60'.
\label{fig-wolfsw}}
\end{figure}

\begin{figure}
\begin{center}
\psfig{figure=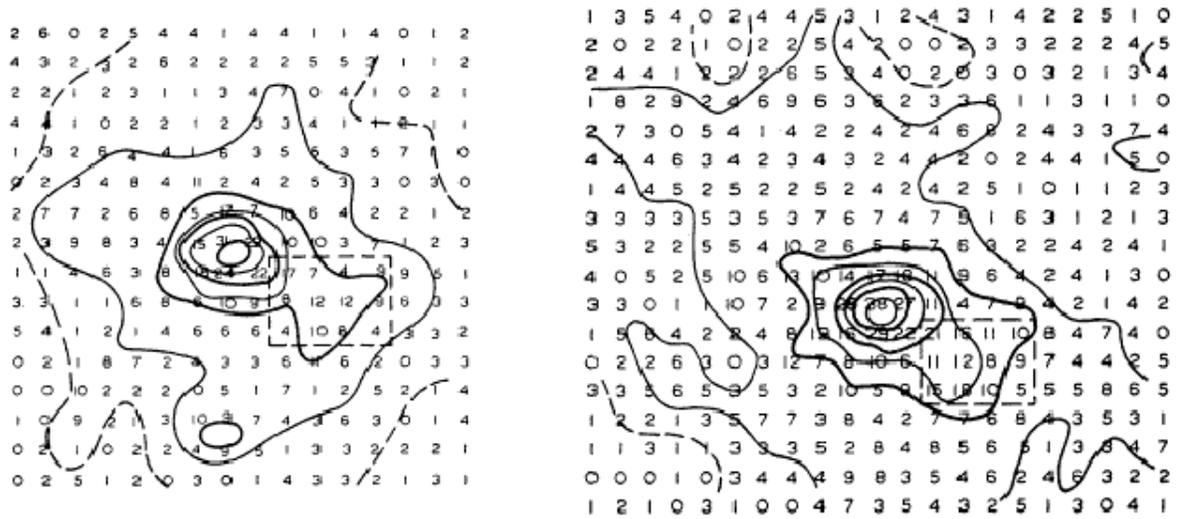,height=7cm,angle=0}
\end{center}
\caption{Contour maps of the Coma cluster of nebul\ae, based on
smoothed counts by 10' squares. Plate n.303 is on the left and no.1613
is on the right.  From Shane \& Wirtanen (1954).
\label{fig-swcoma}}
\end{figure}

The uneven internal structures of clusters were recognized quite early
on. By looking at Wolf\cite{wol01}'s plot of the galaxy distribution
in Coma it is easy to spot the south-western subcluster dominated by
NGC~4839 -- see Fig.~\ref{fig-wolfsw}.  This was re-discovered by
Shane \& Wirtanen\cite{sha54} in 1954, more than half a century
later. The subcluster is clearly visible in their Plates no.303 and
no.1613 -- here reproduced in Fig.~\ref{fig-swcoma} --, and the authors
suggested it could be a distant cluster seen in projection in the Coma
cluster region.  Shane \& Wirtanen\cite{sha54} classified clusters in
two broad classes: regular Coma-like and irregular Virgo-like
clusters. The uneven structure of the Virgo cluster had of course been
noticed very early (e.g. Zwicky\cite{zwi37}). However, it is
remarkable that subclustering in the prototype regular cluster was
also noticed very early, but apparently ignored until being
re-discovered in the X-ray\cite{whi93}.  A telling example is that of
Oemler\cite{oem76}. In 1976 he remarked that the giant galaxy NGC~4839
was quite an exception in his class, because there was not {\em ``any
evidence of clustering of galaxies around NGC~4839''}!

\begin{figure}
\begin{center}
\psfig{figure=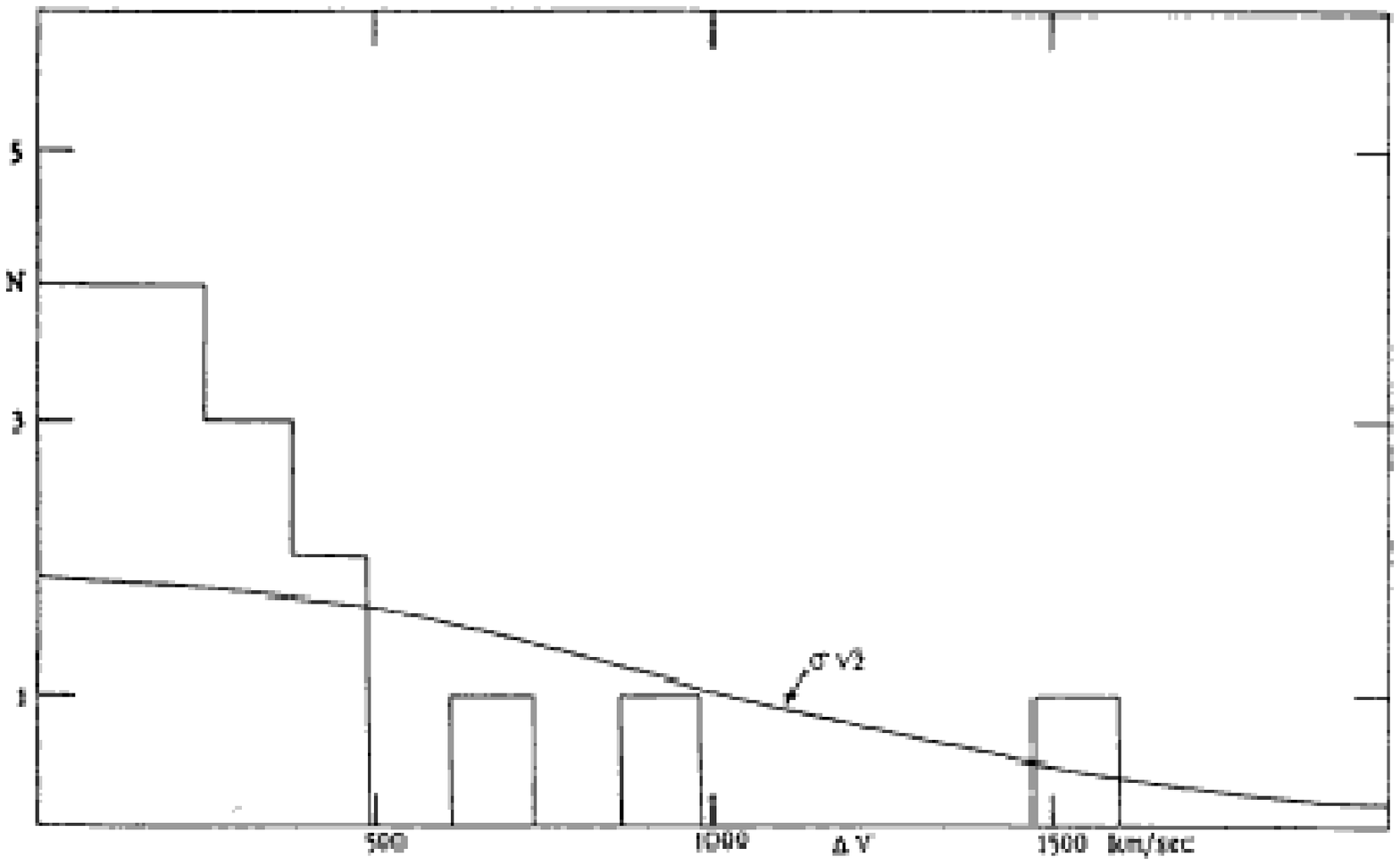,height=6cm,angle=0}
\end{center}
\caption{The observed distribution of velocity differences of pairs
of galaxies in Virgo with separation smaller than 10', compared to
the expected distribution for optical pairs. From van den Bergh (1960b).
\label{fig-vdbsubcl}}
\end{figure}

The first systematic analyses of subclustering in galaxy clusters date
back to the early 60's.  Sydney van den Bergh\cite{vdb60b,vdb61}
analyzed the distribution of velocity differences among pairs of
galaxies in the Virgo and Coma clusters. He compared the observed
distributions to those obtained from azimuthal scramblings of the
data-sets -- see Fig.~\ref{fig-vdbsubcl} -- and found evidence for
subclustering in both clusters, on $\sim 0.1$~Mpc scales: {\em ``Taken
at face value, this result implies that subclustering occurs in the
Coma cluster.''}  Abell et al.\cite{abe64} analyzed eight clusters and
found evidence for subclustering in six of them, but not in
Coma. However, Abell\cite{abe65} remarked that accounting for the
presence of subclusters could not remove the mass discrepancy problem
(see \S~4.2).

In 1973, Bahcall\cite{bah73} first noticed the existence of
substructures around the two central dominant galaxies of Coma,
NGC~4874 and NGC~4889. Her result was later confirmed by
Rood\cite{roo74a}, and refined, many years later, by Perea et
al.\cite{per86}, Fitchett \& Webster\cite{fit87}, and Mellier et
al.\cite{mel88}.  Bahcall also suggested that these subclusters should
be detectable as X-ray sources, independent from the cluster itself, a
suggestion confirmed by Vikhlinin et al.\cite{vik94} 21 years later.

According to Dressler\cite{dre78}, another evidence for subclustering
was given by the secondary peaks detected in the density
profiles of several clusters\cite{sha59,ome65,cla68}.

Subclusters became theoretically appealing after White\cite{whi76a}'s
n-body simulations showed that {\em ``clusters form by the progressive
amalgamation of an inhomogeneous system of subclusters''}.

Thanks to the increasing angular resolution of X-ray observations,
subclusters started to be found also in this band. In 1979 Gorenstein
et al.\cite{gor79} attributed the granularity in the Coma cluster
X-ray emission to subclustering, and a hint of the south-western
subcluster could already be seen in Johnson et al.\cite{joh79}'s X-ray
map of Coma. A major breakthrough came with the {\sl Einstein IPC}
images of Jones et al.\cite{jon79}. They showed that the X-ray
morphologies of clusters, far from being smooth and spherically
symmetric, were quite often irregular and clumpy. Subclustering was a
common feature of galaxy clusters!

In 1982 Geller \& Beers\cite{gel82} draw density-contour maps of the
galaxy distributions in 65 clusters and identified subclusters in
40~\% of them. The techniques for the detection of subclusters have
considerably improved in more recent years, but subsequent works have
roughly confirmed this fraction\cite{dre88a,esc94}. With gravitational
lensing techniques it is now possible to look for subcondensations
directly in the mass distribution, and the existence of dark
subcluster has been suggested (see {\sc Kneib}, these proceedings).

\subsection{Mass}
\begin{figure}
\begin{center}
\psfig{figure=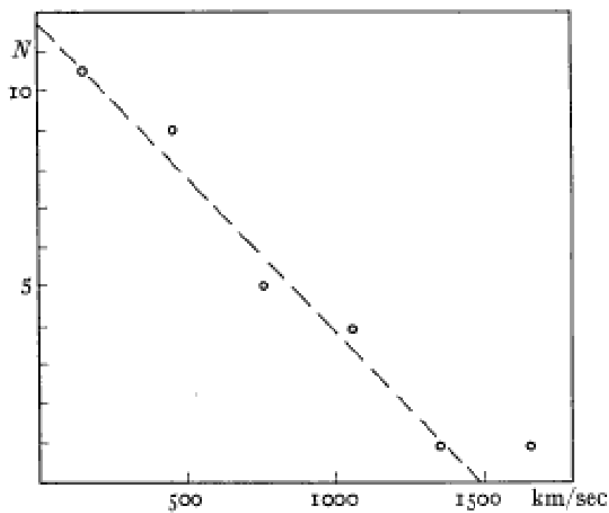,height=7cm,angle=0}
\end{center}
\caption{The distribution of velocities of Virgo cluster galaxies. 
From Smith (1936).
\label{fig-sssv}}
\end{figure}

In the 30's Hubble \& Humason, aiming at a high-redshift extension of
the velocity--distance relationship, measured several velocities for
galaxies in clusters. In 1931, they\cite{hub31} provided the first
estimates of the velocity dispersions in four clusters of
galaxies\footnote{Hubble \& Humason were interested in cluster
velocity dispersions because they wanted to estimate the uncertainties
in the cluster mean velocities, which were relevant to the
velocity--distance relationship.}. Hubble \& Humason noted that the
velocity range spanned by Coma galaxies was larger than in other
clusters (Virgo, Pegasus, Pisces). This was a first hint of the
relation between richness and velocity dispersion that
Bahcall\cite{bah81a} later established in 1981.  Hubble\cite{hub36}'s
early estimate of the cluster velocity dispersion was $\simeq
700$~km/s -- see Fig.~\ref{fig-sssv}, from Smith\cite{smi36} --,
a value remarkably close to modern estimates\cite{gir93}. 
Zwicky\cite{zwi33,zwi37} immediately saw the
great potentiality of Hubble \& Humason's data, and used them for
deriving the mass of the Coma cluster, via the application of the
virial theorem\footnote{The virial theorem had been first used in astronomy by
Poincar\'e in 1911.}. Smith\cite{smi36} followed Zwicky and derived
the virial mass of the Virgo cluster.

Zwicky\cite{zwi37}'s milestone paper: {\sl On the Masses of Nebulae
and of Clusters of Nebulae}, published in 1937, is an exceptional
work. In that paper, Zwicky correctly noticed that the masses of
nebul\ae, derived from rotation curves, are underestimated.
By assuming, {\em ``as a first approximation''}, that clusters
of nebul\ae~ are stationary systems, and using the virial theorem, he
derived a very conservative estimate of the Coma cluster mass.  This
implied a cluster mass-to-light ratio of 68~$M_{\odot}/L_{\odot}$
(after conversion to a modern value of the Hubble constant). Zwicky
had discovered the {\sl missing mass} problem.

His discovery relied very much on the hypothesis of cluster stability.
In support of his hypothesis, Zwicky noted that galaxies in the field
have a much lower velocity dispersion than galaxies in clusters.  This
indicated that field galaxies could not origin from cluster
disruption, or they would have much higher velocities than observed.
In this context, Zwicky implicitly criticized the work of
Smith\cite{smi36}, and emphasized the danger of applying the virial
theorem to irregular systems of galaxies, which are not likely to be
stable systems. Because of the possible biases inherent to the virial
mass estimates, Zwicky suggested to use gravitational lensing as the
{\em ``simplest and most accurate mass determination''}. He was half a
century in advance of observations\cite{lyn86,sou87}!

Smith\cite{smi36}'s paper essentially followed in the steps of
Zwicky\cite{zwi33}, but was published one year before the English
version of Zwicky\cite{zwi37}'s paper, and not surprisingly
Hubble\cite{hub36} quoted Smith and not Zwicky (although Zwicky was
quoted by Smith himself). Hubble remarked that galaxy mass estimates
were likely to be lower limits, while virial theorem estimates of
cluster masses were likely to be upper limits, so that eventually the
two might come into agreement. As a matter of fact, Zwicky's and
Smith's estimates of the Coma and, respectively, Virgo masses, were
quite correct, or, if anything, too low (Zwicky having tried to be
conservative). Anyway, a straightforward application of the virial
theorem was not without problems.  In 1959 Limber\cite{lim59} obtained
a more general expression for the virial theorem, in order to account
for the possible presence of diffuse IC matter. Much later Nezhinskii
\& Osipkov\cite{nez69} showed that the uncertainties in the virial
mass estimates are much larger than generally assumed if the diffuse
matter is not distributed like galaxies, and dominates the
potential. However, as it turned out, the cluster virial mass
estimates were essentially correct, and it was the galaxy mass
estimates which had to be revised upwards.

\begin{figure}
\begin{center}
\psfig{figure=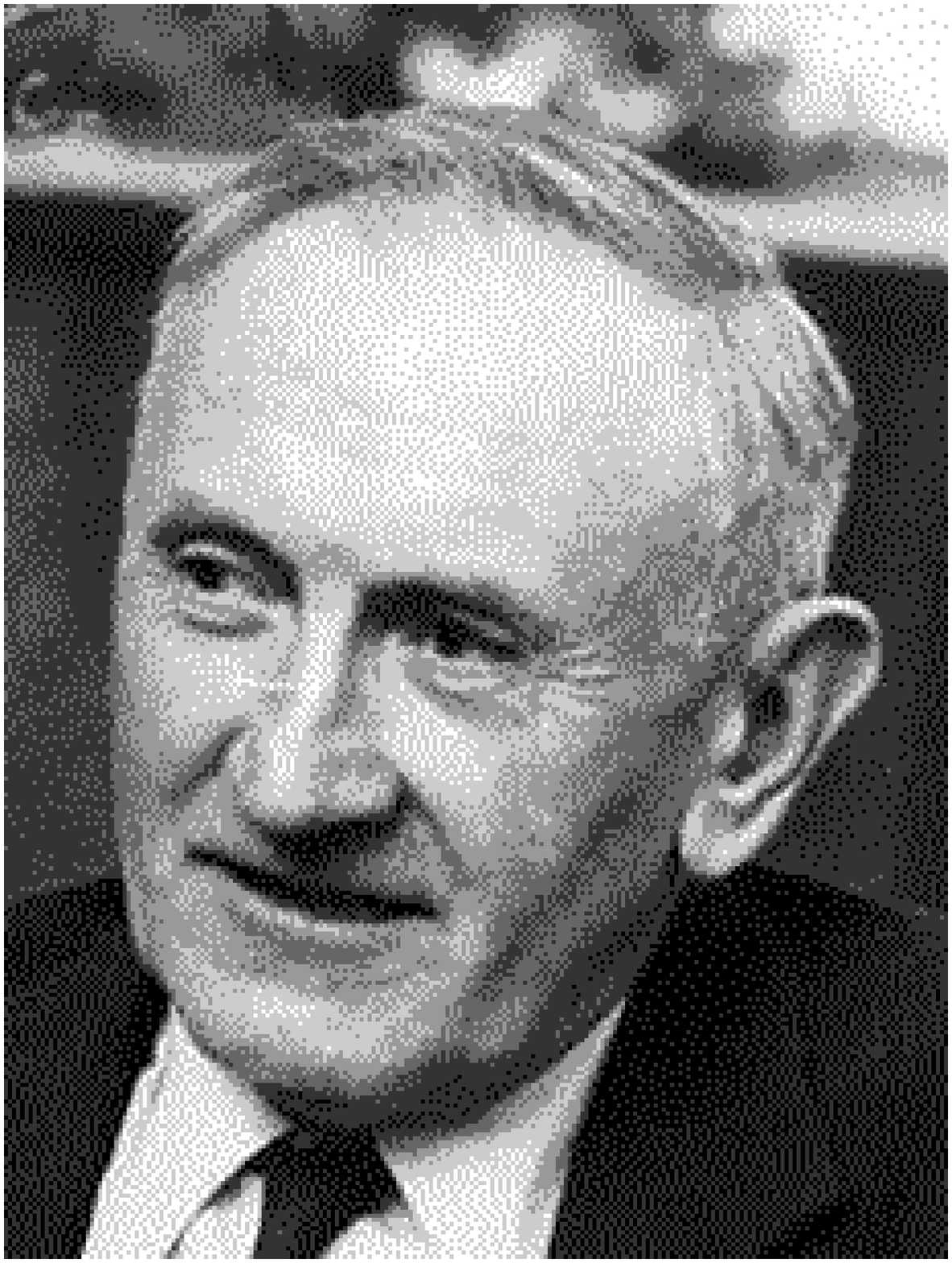,height=7cm,angle=0}
\psfig{figure=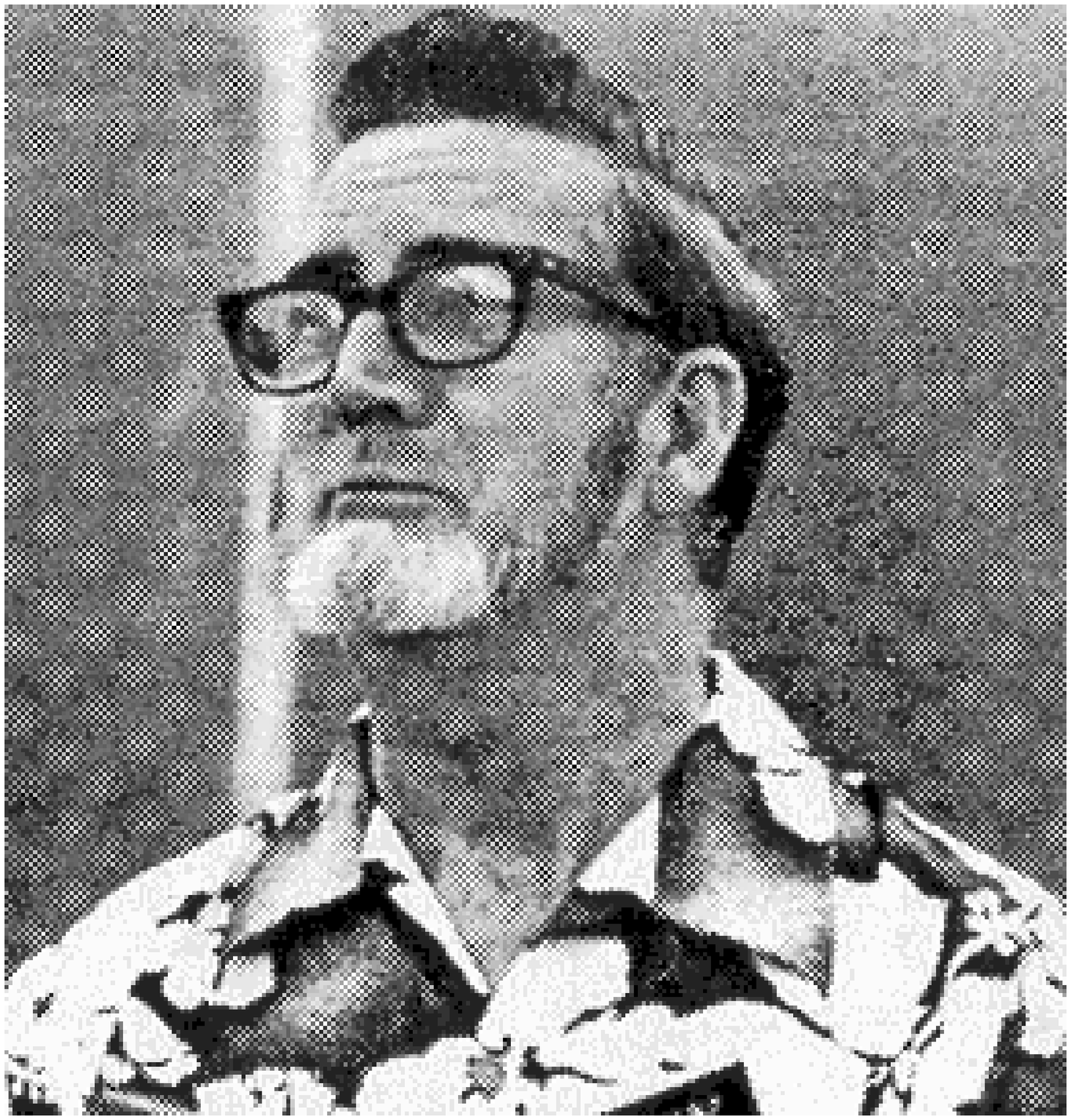,height=7cm,angle=0}
\end{center}
\caption{Portraits of Fritz Zwicky (left) and George O. Abell.}
\end{figure}

Holmberg\cite{hol40} was possibly the first to criticize Zwicky's dark
mass hypothesis, that he considered an {\em ``unlikely assumption''}.
He attributed the high velocity dispersion of cluster galaxies to the
presence of a large number of galaxies on hyperbolic orbits, i.e.
interlopers. In 1954 Schwarzschild\cite{sch54} tried to get rid of
{\em ``interlopers''} to improve the estimate of the Coma cluster
velocity dispersion.  After eliminating many supposed interlopers from
the Coma cluster sample (far too many, in fact) he came to the wrong
estimate of 630 km/s for the velocity dispersion of the Coma cluster.
Some years later Abell\cite{abe61} pointed out that the existence of
superclusters enhances the probability of projection effects, leading
to overestimate the cluster velocity dispersions. In 1977 Yahil \&
Vidal\cite{yah77} devised a method for getting rid of interlopers in
galaxy clusters that remained in use until recently\cite{gir93}.

Schwarzschild's estimate was too low, yet not enough to solve the
discrepancy between the mass-to-light ratios of clusters and those of
individual galaxies, or galaxy pairs. Page\cite{pag52} had just found
that galaxy pairs have a much lower mass-to-light ratios than
clusters.  Of course, estimating the masses of galaxy pairs was not
simpler than estimating the masses of clusters\footnote{The work of
Page required 165 hours of observations!}, as Limber\cite{lim62}
pointed out. Despite the intrinsic uncertainties due to poorly
controlled selection biases, Page's work strongly influenced the
astronomical community, leading to a diffuse scepticism towards the
cluster mass estimates. Interestingly, however, the nearest galaxy
pair (M~31 and the Milky Way) was shown in those years to display the
same missing mass problem of clusters (Kahn \& Woltjer\cite{kah59}).
The mass estimate of Kahn \& Woltjer relied on the simple assumption
that M~31 and the Milky Way are on a bound orbit. Apparently, Kahn \&
Woltjer were unaware of Zwicky's and Smith's results on the mass of
galaxy clusters.

Around 1960, Ambartsumian\cite{amb58,amb61} reversed Zwicky's
hypothesis on the stability of clusters. According to Ambartsumian,
the large velocity dispersions of clusters indicate they have positive
total energy, i.e. they are disintegrating, and missing mass is not
needed.  In those years astronomers were discovering the wild world of
radio-galaxies, with their jets, suggestive of a mechanism to emit
matter out of galaxies. Similarly, interacting galaxies looked to many
as the result of a fragmentation process rather than the result of
encounters. Somewhat later, Noerdlinger\cite{noe70} invoked quasars as
the source of the energy leading to the cluster disruption.
Ambartsumian's hypothesis became quite popular in the astronomical
community because 
\begin{quote}
{\em ``unless one is prepared to make wild hypotheses outside the
realm of verification by direct observation} [\ldots] {\em the
'hidden-mass' hypothesis must be ruled out''} (\dV\cite{dev60})
\end{quote}

The stability of groups and irregular clusters started to be
questioned. Zwicky\cite{zwi50a,zwi50b} insisted on the stability of
clusters, even the Cancer cluster, which Bothun et al.\cite{bot83}
much later proved to be just {\em ``an unbound collection of
groups''.} On the other hand, the Burbidge's\cite{bur59} suggested
that the Hercules cluster was just an unbound collection of groups,
but in fact it is not, it is only rich in substructures\cite{bir93}.
\dV\cite{dev60,dev61a,dev61b} suggested that groups might result from
random encounters of unbound field galaxies. He also provided marginal
evidence that Virgo was not a single dynamical unit, but two different
clusters seen in projection.  His hypothesis was turned down first by
Kowal\cite{kow69} who used Supernov\ae~ to estimate the distances of
Virgo galaxies, and then by Sandage \& Tammann\cite{san76} who used a
much larger sample of Virgo galaxy velocities. Finally Helou et
al.\cite{hel79} closed this issue by determining the relative
distances of galaxies in Virgo with the Tully-Fisher
relation\cite{tul77}.

At variance with irregular clusters and small groups, the stability of
Coma was never in question, given the high degree of symmetry and
regularity of this cluster. This implied that the Coma cluster
contains a large quantity of unseen mass, and so {\em ``why should not
the others?''}  (Burbidge \& Sargent\cite{bur69}). Abell\cite{abe65}
used the cluster virial mass estimates to provide an estimate of the
mean density of the Universe, $\Omega_0 \simeq 0.1$.

A possible solution to the missing mass problem was to revise the
estimates of cluster velocity dispersions. Internal subclustering was
known to be a potential source of error in the velocity dispersion
estimates\footnote{A detailed account of the topic of subclustering is
given in \S~4.1.}.  However, subclustering in Coma took long to be
recognized, and Abell\cite{abe65} pointed out that the correction for
subclustering, while important, was nevertheless too small to get rid
of the missing mass (Ozernoy \& Reinhardt\cite{oze78} later came to
the same conclusion). Godfredsen\cite{god61} and Holmberg\cite{hol61}
suggested that the cluster velocity dispersion estimates were boosted
up by large errors in the galaxy velocities. Their hypothesis was
rejected by \dV~ \& \dV\cite{dev63} and, later, by
Kirshner\cite{kir77}, who found a similar mass discrepancy in groups,
despite a considerable improved determination of galaxy velocities.
Finally, Rood\cite{roo69} pointed out that an {\em a-priori}
assumption of isotropic galaxy orbits could lead to overestimate a
cluster velocity dispersion, if these orbits were instead mainly
radial.

In the early 60's Burbidge \& Burbidge\cite{bur59,bur61} and
Limber\cite{lim62} advanced the major argument in favour of the
stability of galaxy clusters. If clusters have positive energy, the
time-scale for their disruption is very short. Clusters must therefore
be young systems. However, clusters are populated by ellipticals,
which are old galaxies, as inferred from their stellar populations.
This argument seemed ironclad, yet many astronomers still preferred to
question the old age of ellipticals (and the models of stellar
evolution), rather than accepting the existence of dark matter
(see, e.g., Neyman et al.\cite{ney61})!

After 1965 the growing evidence for dark matter in single galaxies
started to change the situation. As early as in 1939
Babcock\cite{bab39} had shown that the rotation curve of M~31, as
measured in the optical, was still raising at the last measured
point. But the observational evidence for non-Keplerian galaxies
rotation curves really came from radio-observations.  In 1965
Seielstad \& Whiteoak\cite{sei65} noted that the turn-over radii of
the galaxy rotation curves were larger when measured in the radio than
when measured in the optical. More 21cm measurements accumulated, in
particular through the work of Roberts\cite{rob69} and Roberts \&
Rots\cite{rob73}.  In 1969 Vorontsov-Velyaminov\cite{vor69} argued
that the 21cm measurements indicated flat rotation curves for galaxies
and Freeman\cite{fre70} and Lewis\cite{lew72} suggested that this
implied an increasing mass-to-light ratio with radius. Arp \&
Bertola\cite{arp69} and \dV\cite{dev69} argued for a high mass of the
giant elliptical M~87, a suggestion later confirmed by Fabricant et
al.\cite{fab80}.  Hunt \& Sciama\cite{hun72} suggested that the
brighter galaxies may have X-ray coron\ae, a prediction later
confirmed by Mathews\cite{mat78a}. In 1973, Ostriker \&
Peebles\cite{ost73} argued for the need of a massive halo to stabilize
the spiral disks.

Progress was also being made in the dynamical modeling of galaxy
systems. In 1970 Allen\cite{all70} derived a velocity-independent
distance for NGC~7320, based on the hydrogen-mass to
optical-luminosity ratio.  He found that this galaxy lies at a
different distance from other galaxies of the Stephan's quintet, thus
reducing the mass discrepancy in this system. On the other hand, the
n-body simulations of Aarseth \& Saslaw\cite{aar72b} indicated that
the group masses were underestimated by the use of the virial theorem,
thus anticipating the conclusions of Tully\cite{tul80}, and Giuricin
et al.\cite{giu88}.  A few years later, Geller \& Peebles\cite{gel73}
obtained a robust statistical estimate of the masses of groups, and
showed that interlopers cannot cause the whole of the mass discrepancy
problem. Gott et al.\cite{got73} and Turner \& Sargent\cite{tur74}
however argued that only a fraction of all groups are bound, and of
these, very few are virialized.

\begin{figure}
\begin{center}
\psfig{figure=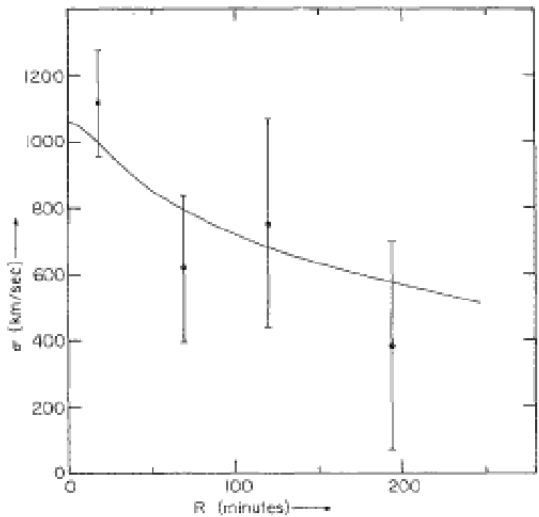,height=7cm,angle=0}
\end{center}
\caption{The Coma cluster velocity dispersion profile. A model with
isotropic galaxy orbits is also plotted.  From Rood et al. (1972).
\label{fig-roodvdp}}
\end{figure}

In 1966 Aarseth\cite{aar66}'s simulations had established that a
cluster in equilibrium should be characterized by a Gaussian
distribution of galaxy velocities. Six years later Rood et
al.\cite{roo72} proved the velocity distribution of galaxies in the
Coma cluster to be Gaussian, lending support to the idea that the Coma
cluster was a stable dynamical system.  Using a larger data-set, they
confirmed Mayall\cite{may60}'s earlier suggestion that the velocity
dispersion of Coma decreases with increasing radius. Previously, a
similar trend in the Virgo cluster had been explained by
Karachentsev\cite{kar65} as an indication of the expansion of the
cluster. Rood et al. instead correctly pointed out that the decreasing
velocity dispersion profile was due to the finiteness of the
cluster. They fitted the profile with a model where galaxies on
isotropic orbits trace the mass distribution -- see Fig.~\ref{fig-roodvdp}.

Despite the observational and theoretical progress, still in the early
70's the general feeling of the astronomical community about the dark
matter issue was quite negative. As an example, here are Chincarini \&
Rood\cite{chi71}'s conclusions from their 1971 paper on the dynamics
of the Perseus cluster:
\begin{quote}
{\em ``We are not inclined to admit this possibility of adequate
intergalactic mass in the cluster} [\ldots] {\em The large 'mass' of 
the Perseus cluster therefore is explained with difficulty if
the cluster is bound, and may suggest instability''}
\end{quote}
Another telling example is the obituary of Fritz Zwicky, written by
Cecilia Payne-Goposchkin\cite{pay74} in 1974. Many of Zwicky's major
contributions to astrophysics were mentioned, but not the discovery of
dark matter.

I do not know how Zwicky managed to change astronomers' minds from
Heaven.  It is a fact, however, that only a few months after his
death, Einasto et al.\cite{ein74} and Ostriker et al.\cite{ost74}
published two papers that catalyzed a paradigm change in favour of the
existence of dark matter in the Universe. Einasto et al. and,
independently, Ostriker et al.  summarized the evidence supporting the
existence of galaxy dark halos, and argued that the mass-to-light
ratio increases with scale, from galaxies to galaxy clusters.  Despite
some residual criticism from Burbidge\cite{bur75}, the existence of
dark matter became rapidly accepted, to such a point that in 1980
Jim~Gunn\cite{gun80} claimed that {\em ``observations now leaves
little doubt of its presence.''}

The paradigm had changed, and dark matter rapidly became a very
popular subject in astronomy. Many different determinations of the
galaxy system masses reached very similar conclusions.
Peebles\cite{pee76a} developed the {\em ``cosmic virial theorem''} and
performed the first analysis of the peculiar velocity field in the
Local Supercluster\cite{pee76b}. Davis et al.\cite{dav80} followed in
his steps a few years later.  Capelato et al.\cite{cap79,cap80b,cap81}
developed their {``Multi-Mass Model''} which accounted for a
distribution of the masses of cluster galaxies. Ozernoy \&
Reinhardt\cite{oze79} and, independently, Valtonen \& Byrd\cite{val79}
developed a binary model for Coma, later shown to be inconsistent with
the X-ray and optical data by Tanaka et al.\cite{tan82} and The \&
White\cite{the88}, respectively. Bahcall \& Tremaine\cite{bah81b}
invented the {\em ``projected mass estimator'',} as an alternative to
the virial theorem. In 1982 Kent \& Gunn\cite{ken82} analyzed the
phase-space distribution of galaxies in Coma, and found that an
isotropic mass-follows-light model was the best fit to the data, thus
confirming Rood et al.\cite{roo72}'s result. On the other hand,
Bailey\cite{bai82}, using the same data, showed that many other
dynamical models were equally acceptable, and the cluster mass was
poorly constrained. One year later, Kent \& Sargent\cite{ken83b} found
that radial orbits were needed to model the dynamics of another
cluster, Perseus. Beers et al.\cite{bee82}, following in the steps of
Kahn \& Woltjer\cite{kah59}, applied a two-body dynamical analysis to
the double cluster Abell~98. In 1980 Lucey et al.\cite{luc80} showed
Centaurus to be another example of a double cluster.

The virial mass estimates of galaxy clusters received a definitive
confirmation through the gravitational lensing analyses (see, e.g.,
Fort \& Mellier\cite{for94}), just as predicted by a visionary Fritz
Zwicky some 60 years earlier. New methods of cluster mass determinations
are reviewed by {\sc Geller} (these proceedings).

\subsection{Luminosity}
The first studies on the luminosity function (LF, hereafter) of
cluster galaxies aimed at determining the population of cluster
galaxies, and, in particular, if dwarf galaxies were clustered like
bright galaxies. When Zwicky\cite{zwi33} discovered the missing mass
problem, it became very important to evaluate the total cluster
luminosity, in order to understand how much of the missing mass could
be accounted for by galaxies fainter than the highest observed
magnitude, or by diffuse IC light. 

\begin{figure}
\begin{center}
\psfig{figure=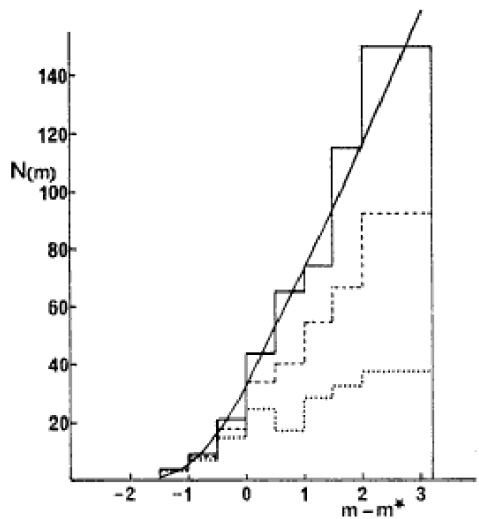,height=7cm,angle=0}
\end{center}
\caption{Luminosity function for a cluster numerical model.  
The solid histogram is the overall luminosity function, and the smooth 
curve is the Schechter function from which it is derived. The other
two histograms correspond to luminosity functions constructed using
only particles within 3.9 Mpc (dashed line) and 1 Mpc (dotted line).
From White (1976a).
\label{fig-swlfev}}
\end{figure}

In 1931, Carpenter\cite{car31} analyzed the LF of the newly discovered
Cancer cluster, and noted that it was a steeply rising function at
faint magnitudes, with no maximum. Hubble \& Humason\cite{hub31} and
Hubble\cite{hub36}, on the other hand, advocated for a LF with a
maximum around the magnitude $\simeq 17$. Such a maximum was also
noted by Baade\cite{baa28} in Ursa Major, and by Shapley\cite{sha34}
in Coma, but only in the inner region, while the LF seemed to increase
to fainter magnitudes in the surrounding regions. Such a phenomenology
was later confirmed by Rood \& Abell\cite{roo73}, and reproduced by
White\cite{whi76a}'s numerical simulations. White explained the
difference between the inner and outer LFs as an effect of dynamical
friction and merging, leading to an excess of bright galaxies in the
core -- see Fig.~\ref{fig-swlfev}.
Recently, the non-monotonous behaviour of the Coma LF has been
reconsidered\cite{tho93,biv95}. 

\begin{figure}
\begin{center}
\psfig{figure=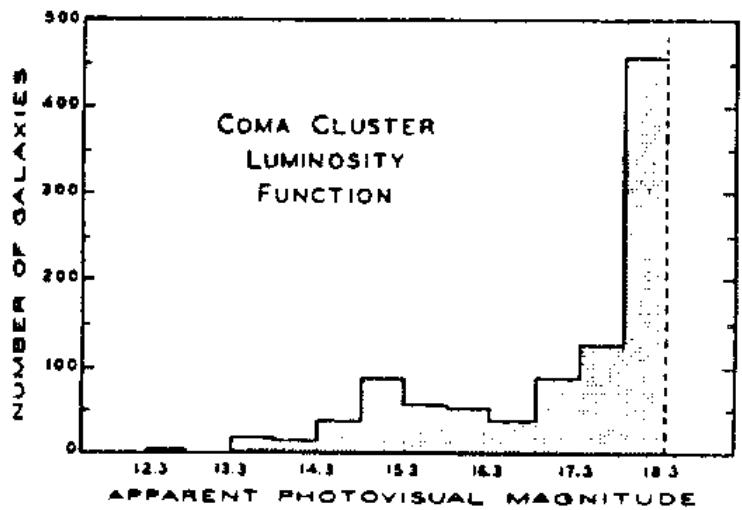,height=7cm,angle=0}
\end{center}
\caption{Abell's estimate of the differential LF of Coma galaxies.
From Sky \& Telescope (1959).
\label{fig-galf}}
\end{figure}

In 1951 Zwicky\cite{zwi51} denied the existence of a maximum in the
Coma cluster LF. He advocated for a LF rising all the way down the
faintest magnitudes reached by observations. This was in agreement
with Holmberg\cite{hol50}'s recent analysis of the LF of the M~81 and
M~101 groups, which indicated a considerable fraction of dwarf
galaxies. As a matter of fact, the large fraction of dwarf galaxies in
the Local Group was already known in the 30's, and clearly at odds
with Hubble's Gaussian LF. In the late 50's dwarf galaxies were also
found in Virgo (Reaves\cite{rea56,rea62}) and Fornax
(Hodge\cite{hod59,hod60}).

In 1959 Abell\cite{abe59a,abe59b,abe60} showed that the cluster LF
increased down to a photovisual magnitude of 19.2, despite a secondary
maximum around magnitude 15 -- see Fig.~\ref{fig-galf}.  Two years
later Abell\cite{abe62,abe64} analyzed several cluster LFs, and
confirmed Zwicky's view of a LF steeply raising down to very faint
magnitudes. However, Abell noted the existence of a particular
magnitude where the LF changes slope, in disagreement with
Zwicky\cite{zwi64}, who did not consider the LF secondary maximum to
be statistically significant. Abell also explained the apparent
Gaussian shape of Hubble's LF as a result of a selection effect.

In 1952 Zwicky\cite{zwi52b} first claimed the detection of IC light in
Coma. Twenty years later, his finding was confirmed by Welch \&
Sastry\cite{wel71}. \dV \& \dV\cite{dev70} showed that most IC light
was due to the extended halos of the two central dominant
galaxies. They estimated that the IC light accounts for less than
40~\% of the total cluster luminosity.  Mattila\cite{mat77} and,
independently, Thuan \& Kormendy\cite{thu77} remarked that the blue
colour of this IC light suggested it could be originated in dwarf
galaxies. Rood et al.\cite{roo72} had previously estimated that dwarf
galaxies could contribute at most 15~\% of the total cluster light.

\begin{figure}
\begin{center}
\psfig{figure=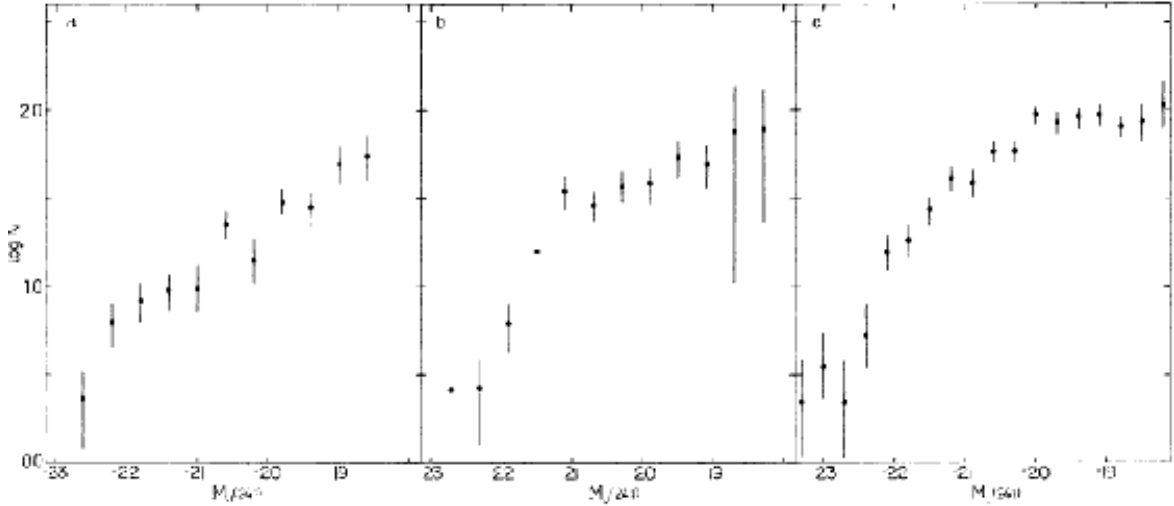,height=7cm,angle=0}
\end{center}
\caption{Composite differential luminosity functions for spiral-rich
(panel a), spiral-poor (panel b), cD-clusters (panel c).
From Oemler (1974).
\label{fig-ao3lf}}
\end{figure}

\begin{figure}
\begin{center}
\psfig{figure=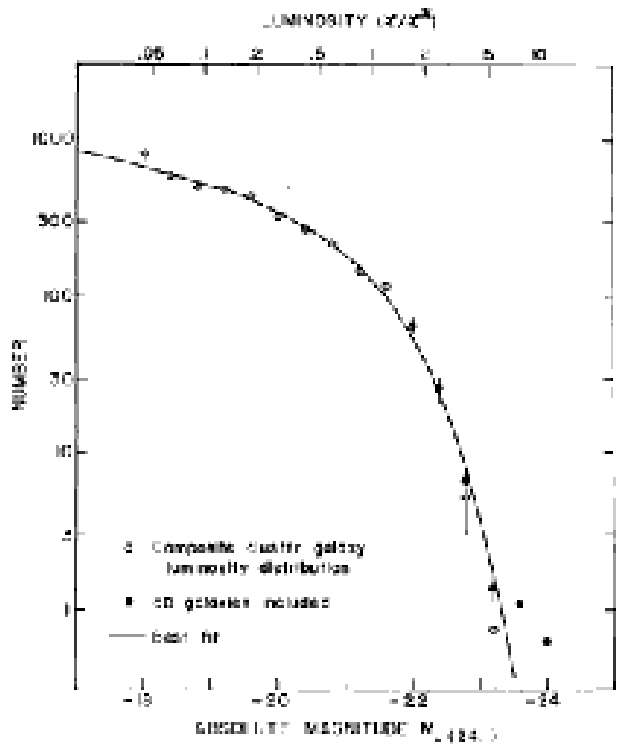,height=10cm,angle=0}
\end{center}
\caption{Best fit of Schechter's analytic expression to Oemler's observed
composite cluster luminosity distribution.
From Schechter (1976).
\label{fig-schelf}}
\end{figure}

In 1974, Austin \& Peach\cite{aus74} found a secondary maximum in the
LF of Abell~1413. This was the second cluster, after Coma, to show a
non-monotonous behaviour of its LF. However, three major works put the
LF irregularities into oblivion. First, Oemler\cite{oem74} insisted
upon the similarity of the LFs of clusters of different type. However,
this is not apparent from Figure~11 in his paper -- here reproduced
in Fig.~\ref{fig-ao3lf}. Possibly Oemler
overlooked differences among the observed LFs, in order to emphasize
the overall remarkable similarity with the theoretical mass function
recently worked out by Press \& Schechter\cite{pre74}. In their paper,
Press \& Schechter compared their model to Oemler's LF for Coma, and
explained Abell's exponential cut-off magnitude $M^{\star}$ as a
characteristic feature of the {\em ``self-similar gravitational
condensation''} model. Finally, in 1976, Schechter\cite{sch76b}
condensed the results of Oemler and Press \& Schechter. He built a
composite LF from Oemler's data for 13 clusters, and show it to be
consistent with a soon-to-be famous {\em ``analytic expression for the
luminosity function for galaxies''} -- see Fig.~\ref{fig-schelf}.

Schechter's universal LF was readily accepted, probably because it was
not purely phenomenological, like the previous ones of
Zwicky\cite{zwi57b} and Abell\cite{abe62}, but based on Press \&
Schechter's physical model.  Several authors\cite{kru74,tur76b}
stressed the similarities of the LFs of different clusters and groups.
Nonetheless, the numerical simulations of Simon~White\cite{whi76a}
indicated that an evolution of the LF in clusters was expected,
because of dynamical friction and merging -- see Fig.~\ref{fig-swlfev}.
In the discussion following
a talk of Ostriker\cite{ost77a} White remarked upon the similarity of
his results and the recent determination of the Coma LF by Godwin \&
Peach\cite{god77}. In 1980, Thompson \& Gregory\cite{tho80} remarked
that Schechter's analytic form can fit the LF of all cluster galaxies,
but not the LFs of separate morphological classes. In particular, they
noted that Hubble's Gaussian LF could provide a good fit to the LF of
bright cluster ellipticals.  Since different clusters have different
fractions of ellipticals, their LFs should be different. The idea of
an universal LF was being shattered. Later
works\cite{san85,tho93,and98} confirmed Thompson \& Gregory's
result. The universality of the LF may still hold within each
morphological class separately (e.g. Krupp\cite{kru74},
Andreon\cite{and98}).

In 1977 Abell\cite{abe77} claimed evidence for a steepening of the
Coma LF beyond magnitude 17.5. A few years later, Heiligman \&
Turner\cite{hei80} noted on the contrary a lack of faint galaxies in
compact groups. These two papers anticipated the current discussion on
the faint end of the cluster LF\cite{ber95,dep95,lob97,ada00}, which
seems to be quite steeper than the field LF\cite{zuc97,lov00} (see
{\sc Andreon}, {\sc Ulmer}, these proceedings). If true, this difference can be
explained in the context of Cavaliere et al.\cite{cav97}'s model
for the evolution of galaxies in clusters, a model supported by the
observations of Wilson et al.\cite{wil97}.

\subsection{On the nature of the dark matter in clusters}
The early papers by Zwicky\cite{zwi33,zwi37} and Smith\cite{smi36} did
not convince the astronomical community that dark matter existed in
clusters. Most astronomers favoured the alternative hypothesis,
cluster instability, until the 21cm measurements proved the galaxy
rotation curves to be non-Keplerian (see \S~4.2 and the excellent
reviews of Sidney van den Bergh\cite{vdb99b,vdb00}). However, many
astronomers took the dark matter hypothesis very seriously and tried
to elucidate its nature.

When Zwicky\cite{zwi37} discovered the missing matter problem, his
first reaction was to question the validity of Newton's gravitational
law. Later he turned his attention to possible forms of dark matter,
that could also provide IC obscuration and thus explain the
non-uniform sky distribution of clusters\cite{zwi52a,zwi53,zwi57a}.
During the 50's he could not find much observational evidence for a
significant quantity of IC matter\cite{zwi52b}, so he\cite{zwi57a}
again suggested abandoning the general theory of relativity.

In 1956 Heeschen\cite{hee56}, motivated by Stone\cite{sto55}'s
theoretical argument, searched for and detected HI emission from Coma.
The detected emission implied a mass of $\simeq 5 \times 10^{13}$
solar masses. Heeschen's detection was however shown to be spurious by
Muller\cite{mul59}, three years later. From a theoretical point of
view, Limber\cite{lim59} noted that clusters are likely to contain IC
gas, because the galaxy formation process is unlikely to be 100~\%
efficient, and because galaxy-galaxy collisions sweep gas out of
galaxies. He pointed out that if this IC gas remained undetected at
21~cm, it could be hot and ionized. Extensive searches for
intergalactic material by Zwicky \& Humason\cite{zwi60} did not prove
very successful. In 1961 the total amount of IC cold gas was
constrained by Penzias\cite{pen61} to be less than a tenth of the
virial mass in the Pegasus~I cluster. Penzias remarked that the
integrated 21~cm emission from individual cluster galaxies could well
account for the total cluster emission.

Dark matter was also searched for in the form of diffuse optical
luminosity\cite{zwi52b,wel71,mat77,thu77} and dwarf
galaxies\cite{hol50,rea56,hod59,hod60,roo72}, but these components
were found to account for less than half the total cluster luminosity
(see \S~4.3).

In 1960 \dV\cite{dev60} summarized the observational situation by
noting that the missing mass could neither consist of cold HI, nor of
dust, nor of diffuse (optical) luminosity. The total mass of all these
components only makes a small fraction of the total cluster mass, and
therefore {\em ``a large number of essentially dark bodies must also
be assumed.''} If the existing observations could not establish the
nature of dark matter, they were anyway not in conflict with the
hypothesis that only a small fraction of the matter in the Universe is
in bright galaxies (Layzer\cite{lay62}).

The idea that galaxies have dark halos gained a hold upon the
astronomical community in a very short time, from 1969 to 1974,
through the work of Arp \& Bertola\cite{arp69}, \dV\cite{dev69},
Vorontsov-Velyaminov\cite{vor69}, Freeman\cite{fre70},
Lewis\cite{lew72}, Ostriker \& Peebles\cite{ost73}, Einasto et
al.\cite{ein74} and Ostriker et al.\cite{ost74} (see \S~4.2).
Rood\cite{roo65} had already demonstrated in 1965 that not all cluster
dark matter can be attached to galaxies, or relaxation processes could
produce much more energy equipartition (and hence, luminosity
segregation) than observed (Rood's early finding was later confirmed
by White\cite{whi77}). Moreover, Lecar\cite{lec75} pointed out that
galaxies in clusters should loose their halos via tidal stripping.

\begin{figure}
\begin{center}
\psfig{figure=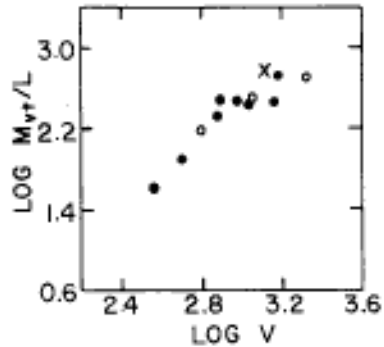,height=5cm,angle=0}
\end{center}
\caption{The logarithm of the mass-to-light ratios for rich clusters
vs. the logarithm of their velocity dispersions.
From Rood (1974b).
\label{fig-roodml}}
\end{figure}

The idea of a scale-dependent mass-to-light ratio took a step forward
through the works of Zwicky \& Humason\cite{zwi64},
Karachentsev\cite{kar66}, Rood et al.\cite{roo70}, Rood\cite{roo74b}
-- see Fig.~\ref{fig-roodml} --,
Ostriker et al.\cite{ost74}, Einasto et al.\cite{ein74},
Bahcall\cite{bah81a} and Davis et al.\cite{dav80}. The mass-to-light
ratio seemed to increase from galaxies to groups and to rich clusters.
This evidence seemed to indicate that the dark matter does not follow
the distribution of bright galaxies. Dressler\cite{dre78} however
noted that including the IC gas mass would reduce the mass discrepancy
in galaxy clusters, and destroy the evidence for a scale dependence of
the mass-to-light ratio.  {\sc Bahcall} (these proceedings) has recently
shown that the mass-to-blue light ratio increases with scale up to the
size of galaxy clusters, but not beyond. The trend can be explained as
an age effect (galaxies in high density environments form earlier,
so that their blue luminosity fades earlier).

The apparently different distribution of dark matter and bright
galaxies strengthened the idea that the dark matter consists of
diffuse gas. A significant amount of IC HI had been ruled out by
observations. Astronomers then started looking for ionized gas. In
1967 Woolf\cite{woo67} put the first constraints on diffuse ionized
gas, by looking at H$\alpha$ and H$\beta$ emission from clusters. He
concluded that if the cluster dark matter was in the form of ionized
gas, the temperature of this gas had to be below $10^6$~K. Three years
later, Turnrose \& Rood\cite{tur70} confirmed Woolf's estimate, using
X-ray data. When diffuse X-ray cluster emission was detected (Meekins
et al.\cite{mee71}, Gursky et al.\cite{gur71}) it was immediately
clear that the IC gas could not account for all the cluster missing
mass.

It was at this point that astronomers really started to grope in the
dark. In 1969, van den Bergh\cite{vdb69} considered massive collapsed
objects (of $10^8$--$10^{12} M_{\odot}$ each) as dark matter
candidates, but ruled them out on the basis of the limited tidal
distortion of galaxies in Virgo. Peebles\cite{pee71} suggested frozen
HI snowballs as dark matter candidates, a possibility never really
ruled out (see, e.g., Wright et al.\cite{wri74}). Another form of
baryonic dark matter was proposed by Tarter \& Silk\cite{tar74} (M8
dwarf stars), but they also frankly remarked that {\em ``nothing
better''} could be said on this topic than had already said thirty
years earlier by Zwicky.  A scaled-down version of Tarter \& Silk's
dark-matter candidates were Napier \& Guthrie\cite{nap75}'s 10$^{-2}
M_{\odot}$ {\em ``black dwarfs''}.  Tarter \& Silk's dark matter
candidates were later suggested by Sarazin \& O'Connell\cite{sar83} to
be the end-products of the cooling flows onto cD galaxies (see
\S~5.4).  In 1981 Gott\cite{got81} proposed a gravitational lensing
experiment to detect an hypothetical huge population of low-mass stars
in galaxy halos, thus anticipating the recent {\sl AGAPE}\cite{ans97},
{\sl EROS}\cite{alc98} and {\sl MACHO}\cite{alc98} projects. 

Baryons as candidate for dark matter are however ruled out by the
theory of primordial nucleosynthesis (see, e.g., Cavaliere et
al.\cite{cav98}), and therefore more exotic dark-matter candidates
were proposed. Here is a short list of them: a variable $G$
(Lewis\cite{lew76}); MOdified Newtonian Dynamics
(Milgrom\cite{mil83a}); vacuum strings (Vilenkin\cite{vil81});
magnetic monopoles (Hoyle\cite{hoy83}); heavy neutrinos (Cowsik \&
McClelland\cite{cow73}, Szalay \& Marx\cite{sza76}, Doroshkevich et
al.\cite{dor80}) -- eventually unstable (Sciama\cite{sci82});
gravitinos (Pagels \& Primack\cite{pag82}), axions (Stecker \&
Shafi\cite{ste83}), and cold dark matter in general (Bond et
al.\cite{bon82}).

Recent observations of the cosmic microwave background (de~Bernardis
et al.\cite{deb00}) have added considerable constraints on the nature
of the missing mass, which is now thought to consist of a mixture of
cold dark matter and dark energy (in the form of a cosmological
constant or quintessence, see, e.g., Bahcall et al.\cite{bah00}). It
is nevertheless wise to close this section with 
a statement of Alan~Dressler\cite{dre78}:
\begin{quote}
{\em ``The answer to the mass discrepancy problem awaits more
data and more inspiration, not necessarily in that order.''}
\end{quote}

\section{Evolution}
\subsection{The formation and evolution of galaxy clustering}
The question of the origin of clusters of galaxies was addressed as
soon as the extragalactic nature of nebul\ae~ was established.  In
1927 Lundmark\cite{lun27} suggested that clusters could form through
many subsequent galaxy--galaxy encounters. These encounters would lead
to a loss of the orbital energy of the galaxies, which would then form
a bound system. Nine years later, the theory had not progressed much,
and Hubble\cite{hub36} was unable to be very specific on this topic:
\begin{quote}
{\em ``condensations in the general field may have produced the clusters,
or the evaporation of clusters may have populated the general field.''}
\end{quote}
The different velocity dispersions of cluster and field nebul\ae~ led
however Zwicky\cite{zwi37} to reject the latter of Hubble's
possibilities.  He also considered extremely unlikely that the rich,
regular, centrally concentrated clusters could be just an effect of
geometrical chance alignments of galaxies along the line-of-sight.
His favourite scenario was that of Lundmark\cite{lun27}. By requiring
the mass of cluster galaxies to be higher than the mass of field
galaxies, gravitational clustering could be made more efficient. The
large cluster virial masses obviously seem to support this
view. Despite the large masses implied for cluster galaxies, Zwicky
however realized that the formation of great clusters by subsequent
capture of field galaxies would take a very long time, much larger
than the age of the Universe. In 1943, by using
Chandrasekhar\cite{cha43}'s theory of dynamical friction,
Tuberg\cite{tub43} indeed estimated the cluster relaxation timescale
to be $10^{11}$--$10^{12}$ years, i.e. orders of magnitude larger than
the estimated age of the Universe at that time.

In 1941, Erik~Holmberg\cite{hol41}, a supporter of the capture theory
for cluster formation, published his remarkable paper {\sl A study of
encounters between laboratory models of stellar systems by a new
integration procedure.} Two decades before the first n-body numerical
simulation of von Hoerner\cite{vho60}, Holmberg ideated an ingenious
device to simulate galaxy--galaxy encounters. His idea was bright and
simple: gravitation was replaced by light in his model. The mass
elements (37 per stellar system, set in circular annuli on a plane)
were represented by light-bulbs, the candle power being proportional
to mass. By modulating the bulb candle power with the distance from
the centre of the system of bulbs, Holmberg was able to simulate a
given density profile.  The two stellar systems were given a certain
approach velocity, and were also set in rotation.  All measurements
were performed on a plane surface, so that the simulation was
2-dimensional.  The acceleration on a given element was measured by
integrating the light at the position of that element with a
photocell. The light bulb was then moved accordingly.

\begin{figure}
\begin{center}
\psfig{figure=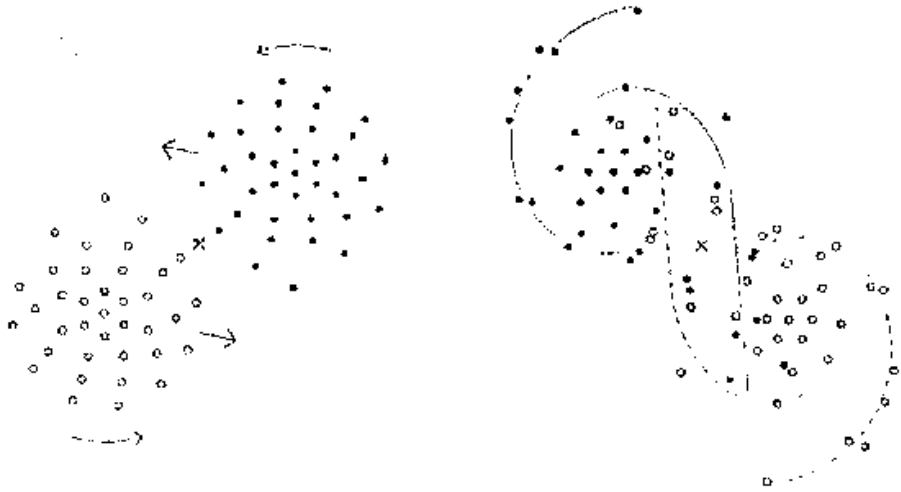,height=7cm,angle=0}
\end{center}
\caption{Results of the simulation of a collision between two
nebul\ae. Left panel: two disky galaxies approaching. Right
panel: after the collision.
From Holmberg (1941).
\label{fig-ehsim}}
\end{figure}

Holmberg's results were very interesting. By looking at Figure~4 in
his paper -- here reproduced in Fig.~\ref{fig-ehsim} --, we can see
clear evidence for the tidal features that Toomre \&
Toomre's\cite{too72} n-body simulations reproduced only 30 years
later. Holmberg however mistook tidal features for spiral arms in the
process of formation. Moreover, even if {\em ``in favorable cases,
captures may occur''}, the experiment essentially ruled out the
capture theory for cluster formation (which was Holmberg's favourite
scenario).

In 1952 and 1956 Zwicky\cite{zwi52a,zwi56} remarked upon the
similarity of distant and nearby clusters. This lack of evolution
seemed difficult to reconcile with an expanding Universe. Zwicky was
trying to rule out Hubble's expanding Universe, because its short age
was clearly inconsistent with the long dynamical timescales he thought
necessary to build the rich regular galaxy clusters. Detecting the
evolution of the cluster number density was to prove very
difficult. Some observational evidence in this sense was claimed by
Just\cite{jus59} in 1959, by Paal\cite{paa64} in 1964, and by
Rowan-Robinson\cite{row72} eight years later. Rowan-Robinson however
warned against possible selection effects that could have biased his
result. The preferential selection of the richest clusters as
spectroscopic targets was shown by Reaves\cite{rea74} to account for
the evidence for evolution. Anyway, these first attempts opened
the way to modern investigations of the cluster number density
evolution (see, e.g., Borgani et al.\cite{bor99}).

In 1961 van Albada\cite{val61} performed a numerical integration of a
model for the cluster evolution, and draw the first modern
scenario of cluster formation:
\begin{quote}
{\em ``Clusters can be formed by gravitational amplification of
statistical density fluctuations in an initial homogeneous field
of galaxies''}
\end{quote}

In 1963 Aarseth\cite{aar63} performed the first of a long series of
n-body simulations of galaxy (or stellar) clusters. His first
simulations contained at most 100 point-masses. Twenty years later,
thanks to the advances in computer technology, Miller\cite{mil83b}
could run a $10^{5}$-body simulation. The increase rate of {\sl n} in
{\sl n-}body simulations over the last thirty years is described in
{\sc Moore} (these proceedings).

\begin{figure}
\begin{center}
\psfig{figure=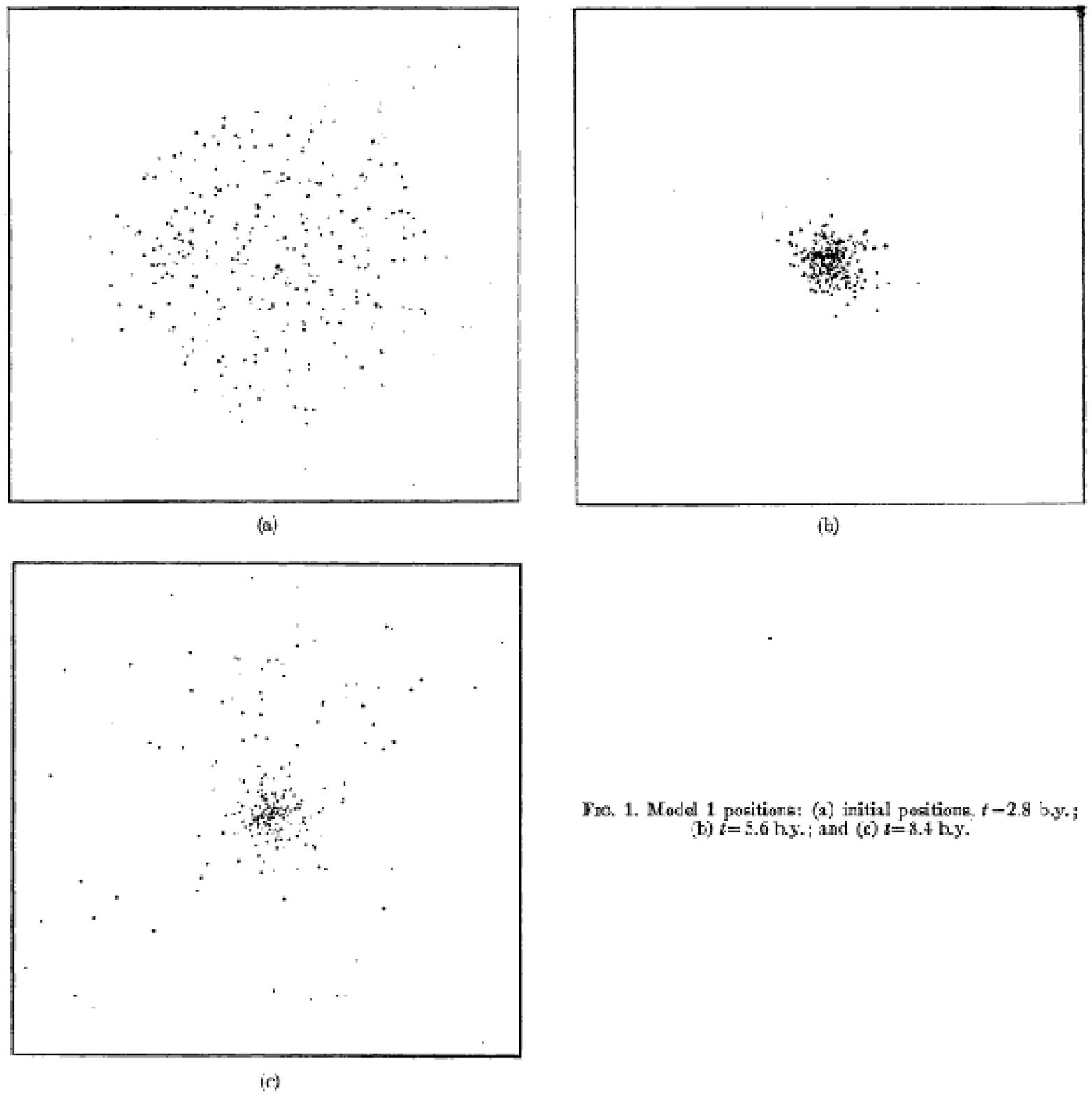,height=10cm,angle=0}
\end{center}
\caption{300-body numerical simulation of the Coma cluster, at three
different times (increasing from panel a to c).
From Peebles (1970).
\label{fig-peesim}}
\end{figure}

One year later, H\'enon\cite{hen64} performed numerical computations
of the dynamical mixing in spherically symmetric clusters. He noted
that phase-mixing rapidly leads to a steady-state configuration after
the initial system contraction. He prepared the field to
Lynden-Bell\cite{lyn67}'s milestone paper {\sl Statistical mechanics
of violent relaxation in stellar systems,} published in 1967.
Lynden-Bell showed that:
\begin{quote}
{\em ``the violently changing gravitational field of a newly formed galaxy
is effective in changing the statistics of stellar orbits}
[which] {\em in the relevant limit} [\ldots] {\em becomes
Maxwell's distribution but with temperature proportional to mass''.}
\end{quote}
Lynden-Bell showed that the predicted density distribution approached
the modified isothermal sphere, or King\cite{kin66}'s recently
published distribution. Violent relaxation removes the need of very
long timescales for a cluster to reach stable, relaxed dynamical
configurations. Lynden-Bell's results was confirmed by
Peebles\cite{pee70}'s 300-body simulation of a Coma-like cluster.
Peebles showed that 10 Gyr suffice to form a rich symmetric cluster
-- see Fig.~\ref{fig-peesim}.

The cluster collapse and subsequent infall of material into clusters
were theoretically examined by Gunn \& Gott\cite{gun72} (see also
\S\S~5.2, 5.3, 5.4). They were probably the first to remark that {\em
``the present is very much the epoch of cluster formation''.} Their
statement was based on the idea that the many existing irregular
clusters were still in a pre-collapse phase.  This idea was later
developed by Richstone et al.\cite{ric92}, who saw the possibility of
constraining the density of the Universe by estimating the fraction of
substructured (i.e. irregular) clusters.  Oemler\cite{oem74} also
elaborated Gunn \& Gott's idea by identifying the irregular clusters
with the spiral-rich, and the regular, evolved ones with the cD-type,
which {\em ``must have begun as the densest fluctuations in the early
Universe''}.

Between 1965 and 1975, two opposite scenarios for the formation of
structures were developed, mainly by Peebles\cite{pee65}, Peebles \&
Dicke\cite{pee68}, Silk\cite{sil68} and Gott \& Rees\cite{got75}, on
one side, and Zel'dovich \& Syunyaev\cite{zel70,sun70,sun72b} on the
other side. Peebles and collaborators advocated for a hierarchical
bottom-up formation of the galaxy structures, while Zel'dovich and
collaborators developed a theory for the evolution of large density
perturbations leading to a top-down scenario, with the formation of
galaxies from fragmentation of {\sl pancakes}. In their original
purely baryonic versions, the hierarchical scenario predicted an
evolution of structures from isothermal primordial density
fluctuations, while in the top-down scenario the primordial
fluctuations were adiabatic. The bottom-up scenario was soon proven by
Aarseth \& Hills\cite{aar72a}'s simulations to be a viable scenario
for the formation of a cluster via the merging of separate
subclusters.  It then received a formidable support from Press \&
Schechter\cite{pre74}'s 1974 paper {\sl Formation of galaxies and
clusters of galaxies by self-similar gravitational condensation.}
Press \& Schechter obtained their famous mass function, and compared
it with observations, finding {\em ``rather striking agreements''}
(see \S~4.3). Also the Russian {\sl pancake} theory (with the added
ingredient of massive neutrinos -- see Klinkhamer \&
Norman\cite{kli81}) had many supporters.  As an example, Thompson \&
Gregory\cite{tho78} argued that Coma is {\em ``a Zel'dovich
disk''}. The popularity of the model started to decline in 1983, when
Frenk et al.\cite{fre83} showed it implied a very late formation of
galaxies, much too late to reconcile with observations. In the end, a
hierarchical structure formation from primordial adiabatic density
fluctuations has emerged, a sort of compromise between the two
original scenarios, where the Zel'dovich approximation is still valid
for describing the initial evolution of structure on large scales, and
cold dark matter plays a leading role in shaping the structure of the
Universe (Bond et al.\cite{bon82}).

\begin{figure}
\begin{center}
\psfig{figure=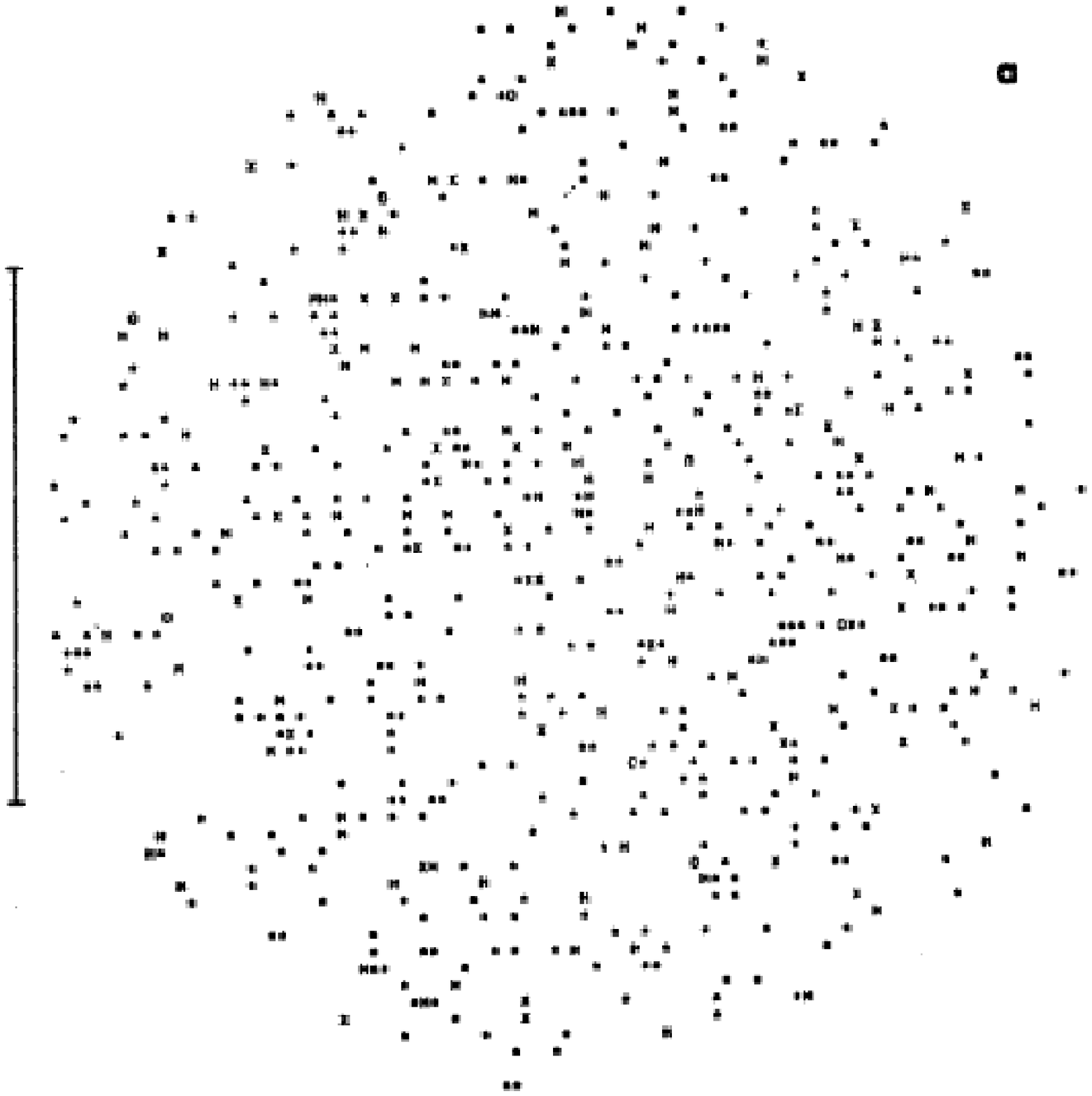,height=3.5cm,angle=0}
\psfig{figure=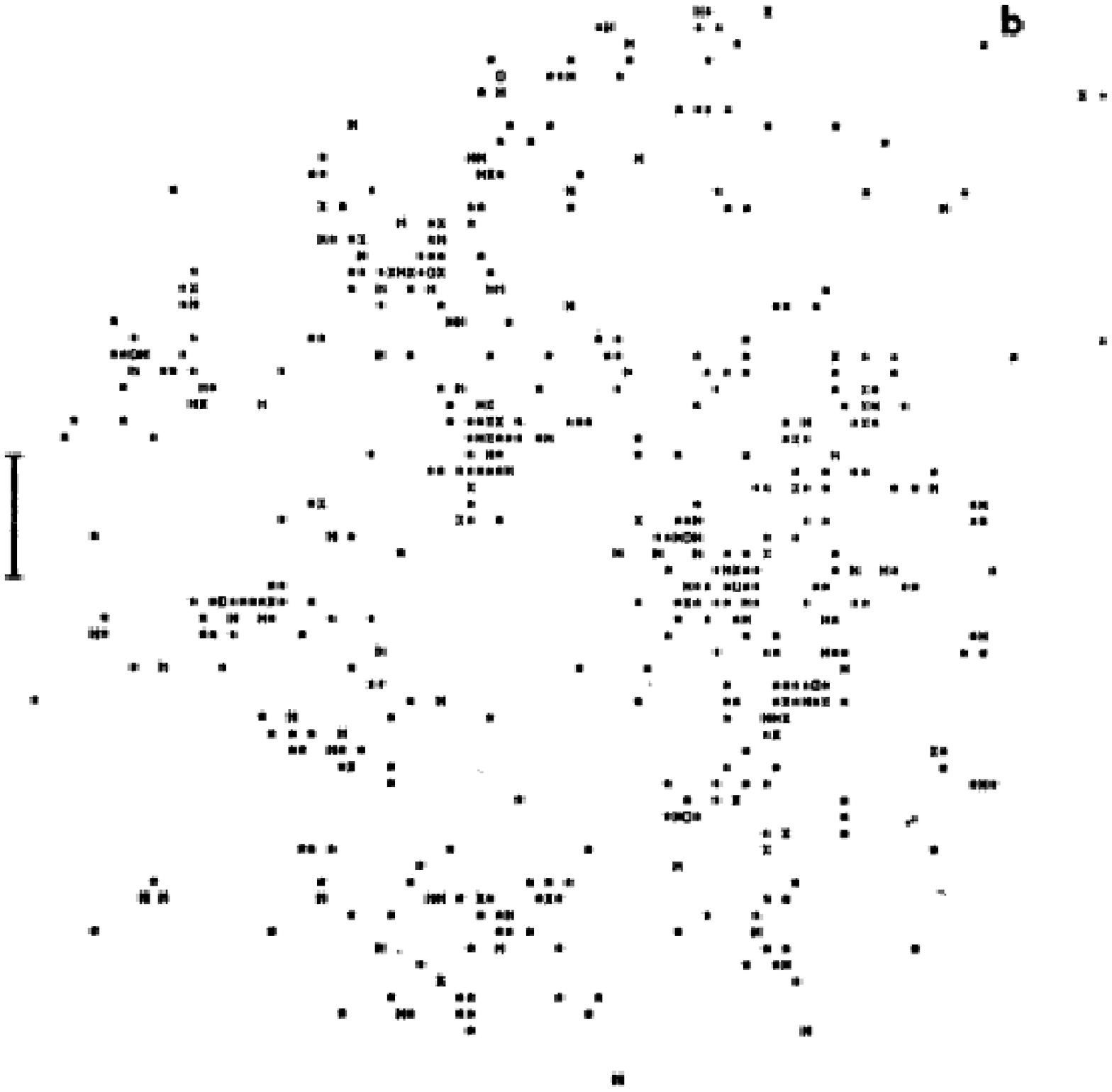,height=3.5cm,angle=0}
\psfig{figure=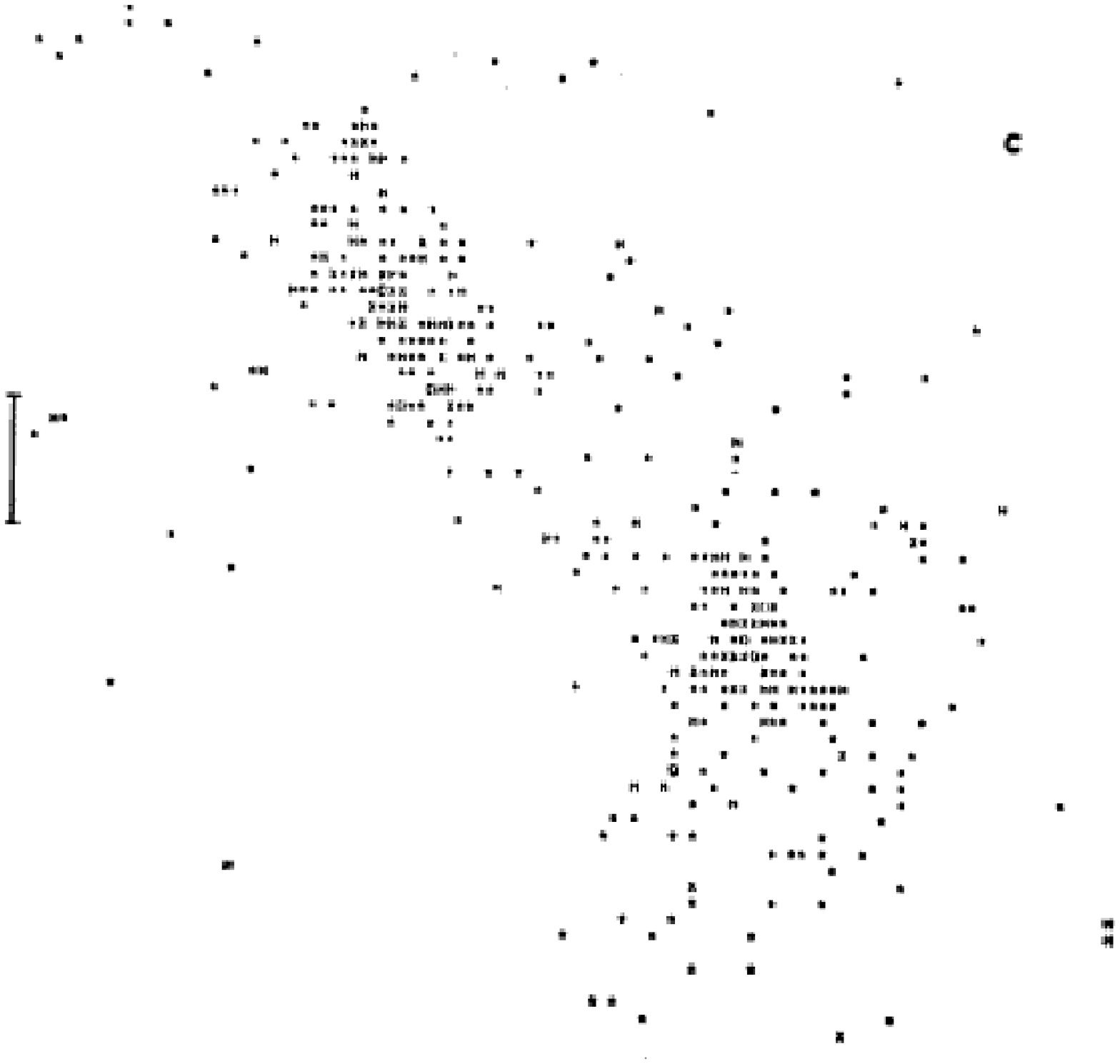,height=3.5cm,angle=0}
\psfig{figure=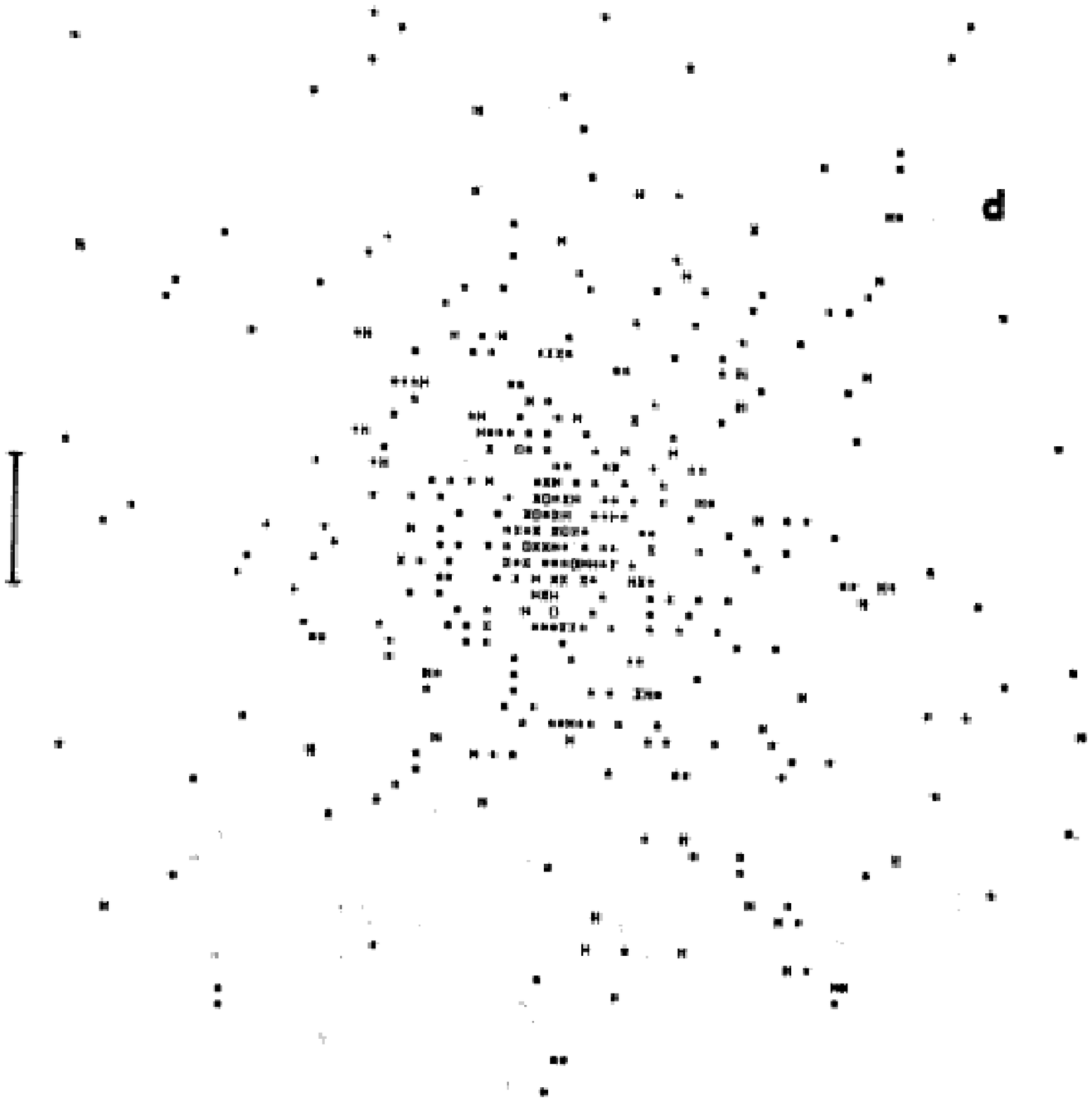,height=3.5cm,angle=0}
\end{center}
\caption{700-body numerical simulation of a cluster, at four
different times. From White (1976a).
\label{fig-swsim}}
\end{figure}

In 1976, further support to the hierarchical clustering scenario came
from White\cite{whi76a}'s 700-body simulations. He showed that {\em
``clusters form by the progressive amalgamation of an inhomogeneous
system of subclusters''}. The direction of the final major merger
defines the direction of the cluster elongation, and there is no need
to invoke cluster rotation or tidal distortions to explain the cluster
elongations -- see Fig.~\ref{fig-swsim}.
Following White's result, Forman et al.\cite{for81}
interpreted the double structure of some X-ray emitting clusters as an
evidence for an intermediate evolutionary stage.

In 1978 Fall\cite{fal78} reproduced the shape of Peebles' covariance
function in his 1000-body simulations. The following year, the
4000-body simulations by Aarseth et al.\cite{aar79} not only confirmed
Fall's results, but also reproduced the recently discovered huge
voids\cite{tif76,gre78,tho78,tar78a,chi79,tar80,gre81} in the galaxy
distribution. Aarseth et al. noted that if the Universe has
$\Omega_0=1$ {\em ``the clustering is proceeding at the present
time''}, while this is not the case if $\Omega_0$ is low.  An
$\Omega_0$-dependence of the covariance function was noted by Gott et
al.\cite{got79} and Efstathiou\cite{efs79} in their n-body
simulations. The cellular, filamentary appearance of the structure of
the Universe was reproduced in the $10^5$-body simulations of
Miller\cite{mil83b}.

The cluster number density evolution has now become a strong constrain
for cosmological theories. Most observational evidence of this kind
points to a low-$\Omega_0$ Universe (see {\sc Bahcall}, {\sc Borgani},
these proceedings), and the extensive ongoing surveys will soon
improve the statistics and probe deeper in space (see, in these
proceedings, {\sc Bartlett}, {\sc B\"ohringer}, {\sc Carlstrom}, {\sc
Dickinson}, {\sc Gal}, {\sc Gioia}, {\sc Jones}, {\sc Lobo}, {\sc
Schuecker}, and {\sc Zaritsky}).

\subsection{The evolution of galaxies in clusters}
The importance of collisions for the evolution of cluster galaxies was
understood quite early. Since a cluster of galaxies is a dense
environment, {\em ``collisions must necessarily enter as a factor in
the evolution of the system''} (Shapley\cite{sha3?}, 1935). In 1937
Zwicky\cite{zwi37} imagined that collisions might lead to the
disruption of certain types of nebul\ae, which could explain why the
morphological mix of cluster galaxies is different from the field.
The first observational evidence for this effect came only thirty
years later, when Reaves\cite{rea66} found that dwarf galaxies avoid
the cluster centres.

In 1943 Chandrasekhar\cite{cha43} developed his theory of {\em
``dynamical friction''}, {\em ``the systematic decelerating
effect of the fluctuating field of force acting on a star in
motion''.}  Chandrasekhar derived his formula on the basis of the
two-body approximation for stellar collisions. More than thirty years
later, with the discovery of massive halos around galaxies,
Lecar\cite{lec75} suggested that galaxies gradually settle to the
cluster centres by dynamical friction through a sea of
tidally-stripped galaxy halos.  The validity of Chandrasekhar's
formula was confirmed through numerical simulations by
White\cite{whi76b,whi77}.

In 1940 Holmberg\cite{hol40} had remarked that spirals must transform
into ellipticals, if clusters form by the capture of field
galaxies. Spitzer \& Baade\cite{spi51}, in 1951, were the first to
suggest collisions as a mechanism to transform a galaxy type into
another. They thought that collisions would affect primarily the gas
content of a galaxy, and not so much its stellar structure, leading to
the formation of irregular galaxies. A year later Zwicky\cite{zwi52b}
found evidence for intergalactic matter in small galaxy groups, and
attributed it to material stripped from galaxies during close
encounters. This was confirmed 20 years later by the simulations of
Toomre \& Toomre\cite{too72}.  Spitzer \& Baade's analysis was revised
twice between 1963 and 1965.  First Aarseth\cite{aar63} revised
downward Spitzer \& Baade's estimate of the number of galaxy-galaxy
collisions, as a consequence of the revised distance scale. Then,
Alladin\cite{all65} revised upwards Spitzer \& Baade's estimate of the
internal energy change of a galaxy during a collision. 

In 1970 Tinsley\cite{tin70} developed her theory for the evolution of
the spectral energy distribution of galaxies and showed that strong
evolutionary corrections were to be expected for the colours of
ellipticals, because of the aging of the stellar
population\footnote{As Spinrad\cite{spi77} noted in 1977, Tinsley's
work led to an {\em ``amusing''} conceptual inversion of the classical
cosmological quest: instead of comparing the properties of nearby and
distant galaxies to constrain the cosmological model, one must adopt a
cosmological model in order to constrain the evolution of
galaxies.}. The following year, Oke\cite{oke71} devised to compare the
colours of nearby and distant cluster ellipticals with evolutionary
models, and thus infer their (photometric) redshifts.

\begin{figure}
\begin{center}
\psfig{figure=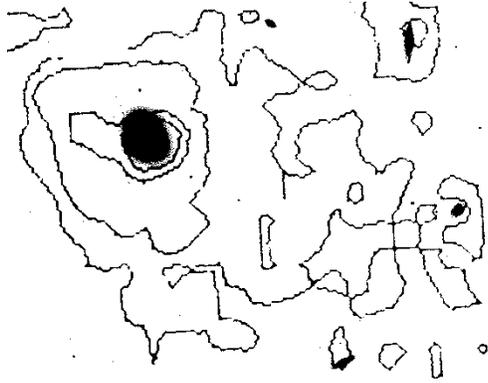,height=5cm,angle=0}
\end{center}
\caption{Contours of X-ray emission around the galaxy M~86 in Virgo
The extended emission was interpreted  as
evidence for ram pressure stripping of hot gas from the galaxy.
From Forman et al. (1979).
\label{fig-form86}}
\end{figure}

In 1972 Rood et al.\cite{roo72} noticed that the Coma cluster S0s were
not confined to the cluster core, where collisions were expected to be
most effective, and questioned the validity of the collision model for
the formation of lenticular galaxies. In the same year, Gunn \&
Gott\cite{gun72} and Larson\cite{lar72} presented two alternative
models for the evolution of galaxy morphologies. Gunn \& Gott proposed
ram pressure stripping of the interstellar gas by the hot IC medium as
a mean of transforming spirals into S0s. The first direct
observational evidence of such an effect came seven years later, with
Forman et al.\cite{for79}'s X-ray observations of the Virgo galaxy
M~86 -- see Fig.~\ref{fig-form86}.
Larson, on the other hand, suggested a relation between the
morphological type of a galaxy, and the collapse time of the gas
during galaxy formation. Galaxies with a short collapse time would
have their material used up early, leading to old stellar populations
and little gas left (like in ellipticals and S0s). The
morphology--density relation could then follow by relating the collapse
time to the ambient density. According to Oemler\cite{oem74}, the {\em
``birthrate of elliptical galaxies} [\ldots] {\em increases with density
relative to the other galaxy types'',} and collisions may be
sufficient to transform spirals into S0s but not into ellipticals.
Larson's ideas were later developed by Gott \& Thuan\cite{got76}.

In 1975 Biermann \& Tinsley\cite{bie75} remarked upon the similarity
of the colours of ellipticals and S0s. This implies that ellipticals
and S0s have similar stellar populations, and therefore similar old
ages, so that a recent transformation of spirals into S0s is out of
question.  The issue is certainly not closed, with independent
evidences in favour\cite{ell97} and against\cite{dre97} an ancient
origin of S0s.

In 1976 White\cite{whi76a}'s n-body simulations showed that the
formation process of a cluster leads to an increasing ellipticity of
galaxy orbits with clustercentric radius, i.e. radial motions are
predominant in the outer cluster regions. The observations of Moss \&
Dickens\cite{mos77} seemed to confirm White's findings.  Moss \&
Dickens observed that late-type galaxies have a higher velocity
dispersion than early-types, and interpreted it as an evidence for an
infalling population of field galaxies into the clusters. Recently
Biviano et al.\cite{biv97} have shown that emission-line galaxies in
clusters are characterized by predominantly radial orbits. A thorough
determination of the orbits of different types of cluster galaxies,
through the solution of the Jeans equation, is in
preparation\cite{biv99}.

White\cite{whi76a,whi76b}'s simulations also showed that a marginal
mass segregation can establish in clusters through dynamical
friction. Merging of the slowed-down galaxies would then follow in the
cluster core, eventually with the formation of a cD galaxy (see
\S~5.4). Struble\cite{str79b}'s observation of a low velocity
dispersion in the core of some galaxy clusters was taken as supporting
evidence for these effects. A few years later Roos \&
Aarseth\cite{roo82} re-examined the issue of mass segregation by
running n-body simulations of a galaxy system with a Schechter-like
distribution of galaxy masses. They noted that segregation establishes
in subclusters before these merge to form the final
cluster. Segregation is then conserved while the cluster evolves,
because tidal stripping predominantly affects the outer regions of
subclusters. Such an evolutionary scenario was found to be consistent
with Capelato et al.\cite{cap80a}'s observations of luminosity
segregation in Coma, and with recent analyses of the Coma cluster
structure\cite{mel88,biv96}.

In 1980 Dressler\cite{dre80a} noted that ram-pressure stripping could
not account for the different bulge-to-disk ratios of spirals and
S0s. Richstone\cite{ric76} and Marchant \& Shapiro\cite{mar77} had
already shown that collisions of spirals can fatten the galaxy disks,
so that Dressler's observation was not a problem in the collision
scenario. Farouki \& Shapiro\cite{far80}'s simulations showed however
that also the ram-pressure mechanism would lead to a thickening of the
galaxy disks. Finally, in 1982 Nulsen\cite{nul82} noted that other
interaction mechanisms between cluster galaxies and the hot IC gas
medium (viscosity, thermal conduction, turbulence) could be even more
effective than ram-pressure in stripping gas from galaxies.

In 1980, Larson et al.\cite{lar80} noted that if star formation
continued in galaxy disks at the rate determined in the local
Universe, spirals would run out of gas in a relatively short time.
Disk replenishment of gas is therefore needed.  An early generation of
spirals, formed in high density regions, would be characterized by
small disks, and such spirals could evolve into nowadays S0s by the
loss of their gaseous halos through collisions. According to Roos \&
Norman\cite{roo79}'s n-body simulations, ellipticals could instead form
via mergers during the early stage of cluster collapse, before the
dispersion of galaxy velocities becomes too high.

\begin{figure}
\begin{center}
\psfig{figure=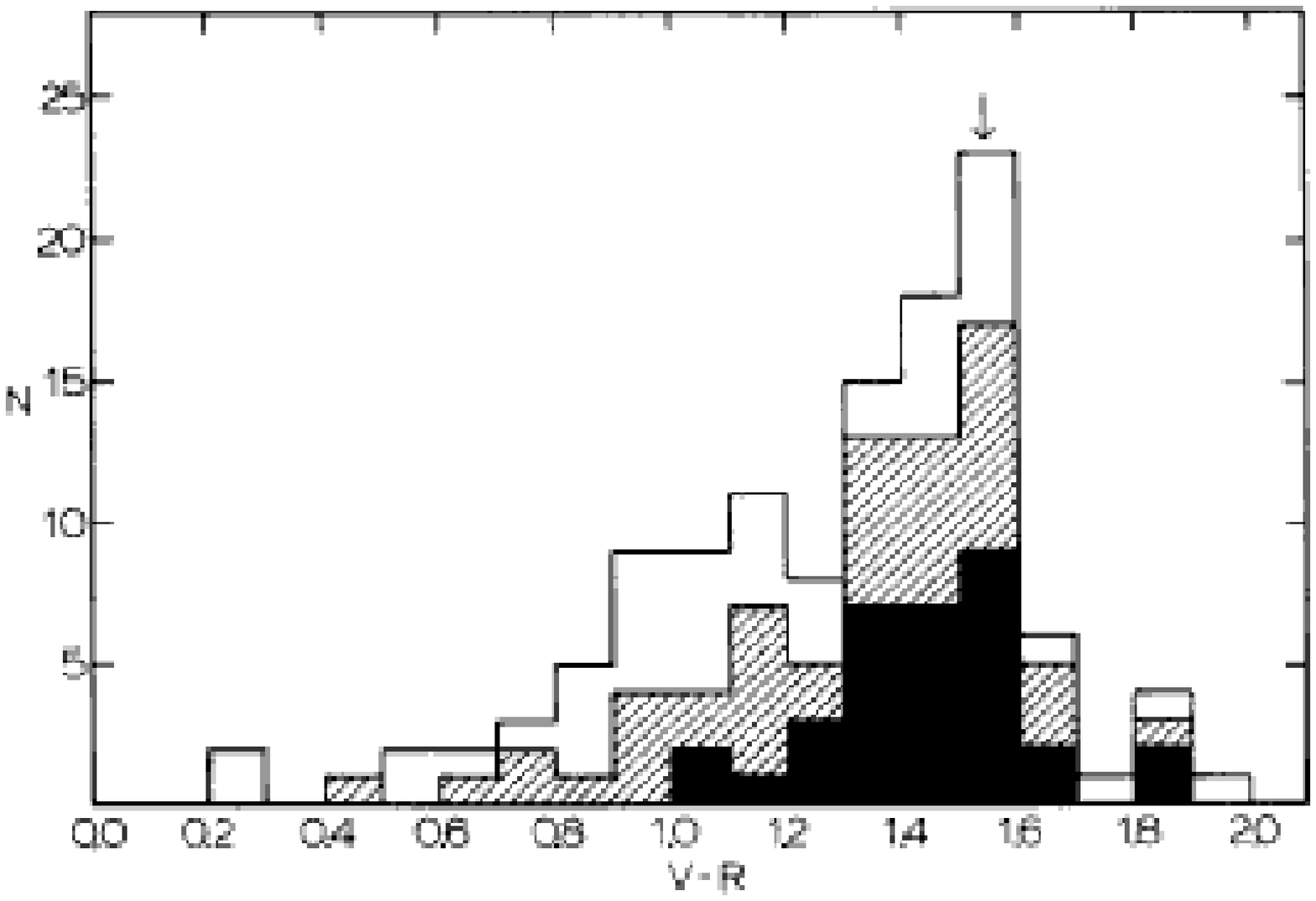,height=4cm,angle=0}
\psfig{figure=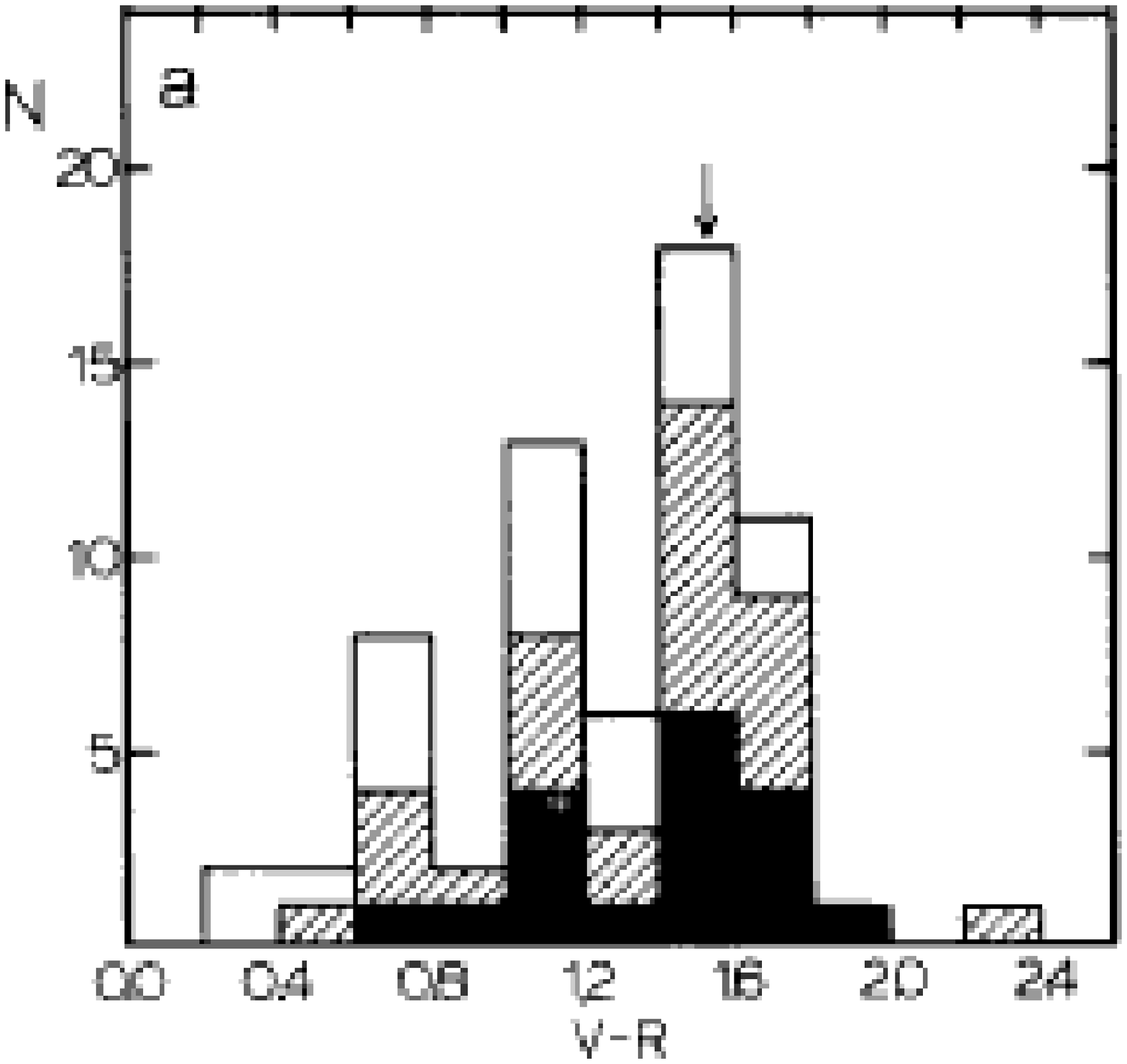,height=4cm,angle=0}
\psfig{figure=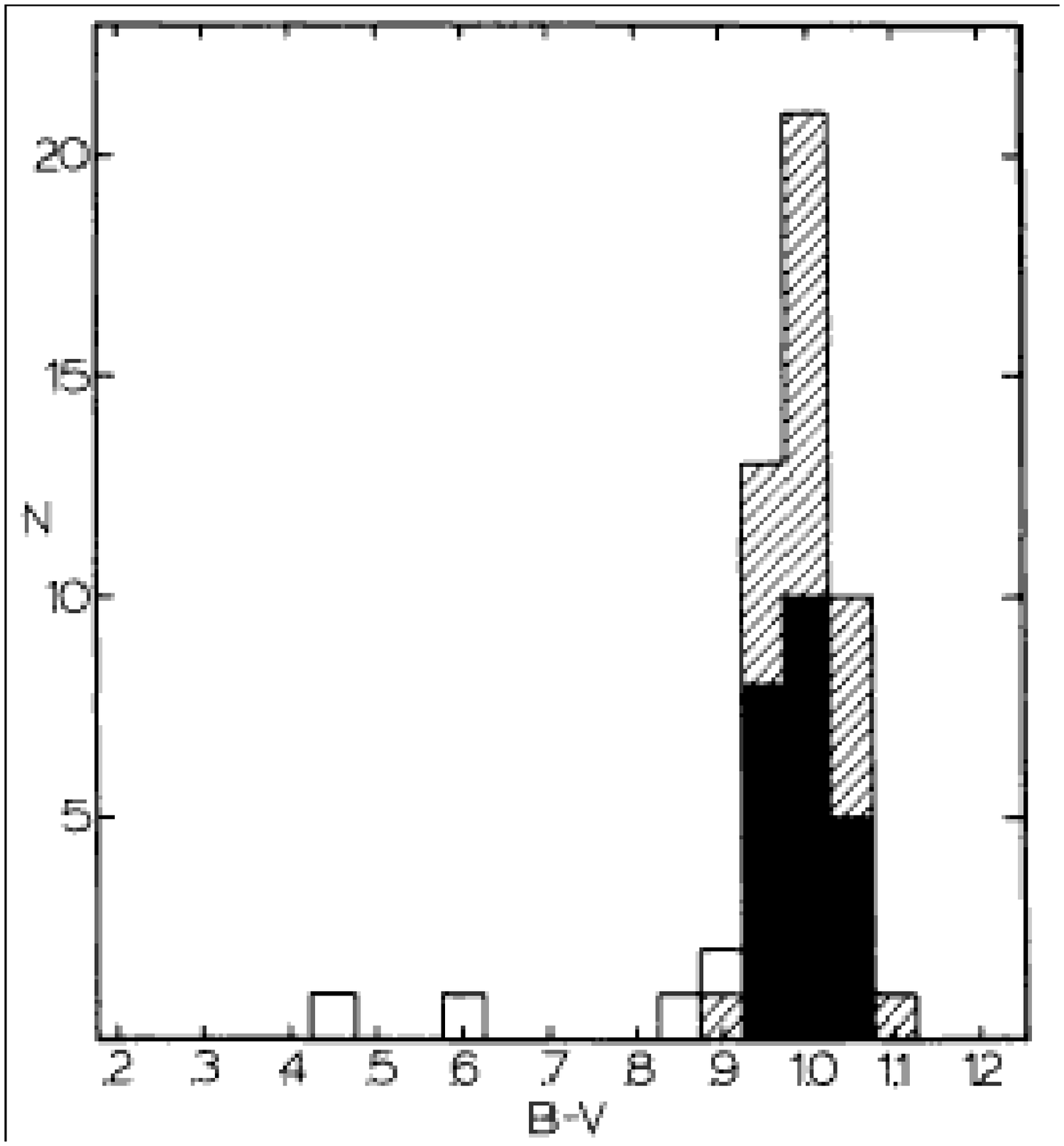,height=4cm,angle=0}
\end{center}
\caption{The V-R colour distribution of galaxies in the
cluster Cl0024+1654 (left), and in the cluster 3C295
(middle). Different shadings correspond to subsamples
of galaxies at different distances from the cluster centres.
The B-V distribution of galaxies in the Coma cluster (right).
Solid area: ellipticals; hatched area: S0s; remainder: spirals.
From  Butcher \& Oemler (1978a).
\label{fig-bo}}
\end{figure}

All these theoretical efforts to determine the evolution of galaxies
received a formidable acceleration with the first direct observational
evidence for the evolution of the cluster galaxy population.  In 1978,
Butcher \& Oemler\cite{but78a} published the first of a series of
papers on {\sl The evolution of galaxies in clusters.} Their
photometric observations of two regular, centrally concentrated, $z
\simeq 0.4$ clusters, showed an excess of blue galaxies, as compared
to nearby rich clusters -- see Fig.~\ref{fig-bo}.
Butcher \& Oemler\cite{but78a,but78b} noted
that such a high fraction of blue galaxies was more typical of nearby
poor irregular clusters like Hercules. They later confirmed their
finding through photometric observations of seven more clusters at
redshifts beyond 0.2 (Butcher et al.\cite{but80}).

Butcher \& Oemler's result was greeted with much scepticism.  Even
before Butcher \& Oemler's paper was published, Baum (in the
discussion following a talk of Spinrad\cite{spi77}) suggested that
their result could be due to contamination by field
galaxies. Koo\cite{koo81} imaged another distant cluster, where he did
not find evidence for the Butcher-Oemler effect. Mathieu \&
Spinrad\cite{mat81} re-examined the fraction of blue galaxies in one
of Butcher-Oemler clusters, and showed it to be much lower than
originally estimated. Lucey's critical {\em ``assessment of the completeness
and correctness of the Abell catalogue''} led him to conclude that the
Butcher-Oemler effect was due to an erroneous assignment of cluster
membership.

Theorists were however not discouraged by potential observational
biases.  In the models of Norman \& Silk\cite{nor79} and Himmes \&
Biermann\cite{him80} the IC gas gradually build-up from the gas
stripped through collisions of cluster galaxies. Norman \&
Silk\cite{nor79} noted that such a gradual built-up of the IC gas can
delay the effectiveness of ram-pressure stripping until $z \sim
0.5$. If ram-pressure transforms spirals into S0s, this would explain
the excess of spirals in high-redshift clusters. However, Henry et
al.\cite{hen79}'s X-ray observations showed the existence of a dense
IC medium in one of the clusters showing the Butcher-Oemler effect.

In 1982, 1983, Dressler \& Gunn\cite{dre82,dre83} finally performed
spectroscopic observations of galaxies in Butcher-Oemler clusters.
The fraction of blue galaxies which are cluster members was found to
be lower than predicted by Butcher \& Oemler, but still higher than in
nearby rich clusters. The Butcher-Oemler effect was confirmed.

More than twenty years after the original discovery, the
Butcher-Oemler effect is well established (see {\sc Ellingson},
{\sc Margoniner}, these proceedings), and a considerable progress has been
made in determining the nature of the excess blue galaxies (see, e.g.,
Poggianti et al.\cite{pog99}). The physical mechanisms responsible for
the evolution of cluster galaxies are not yet determined with
certainty, but it is likely that collisions, as initially suggested by
Shapley\cite{sha3?}, are of fundamental importance (see {\sc Moore},
{\sc Kauffman}, {\sc Lanzoni}, these proceedings).

\subsection{The evolution of the IC gas}
Many years before its detection, Limber\cite{lim59} had argued that IC
gas must exist because galaxy formation cannot be 100~\% efficient,
and that it must evolve through the loss of gas from colliding
galaxies. The IC gas was eventually detected\cite{mee71,gur71} in
1971. In those years, Gott \& Gunn\cite{got71,gun72} developed their
theory of intergalactic gas infall into clusters. They argued that
this infall could generate a hot IC gas through shock heating. They
suggested that irregular clusters are seen in a pre-collapse phase, so
that their IC gas had not yet reached high temperatures. In this way
they hinted at the existence of a class of X-ray faint clusters (which
are now being discovered, see Holden et al.\cite{hol97}). Gunn \&
Gott\cite{gun72} also suggested ram pressure as a mean to strip gas
from cluster galaxies and enrich the IC medium.

An early gas infall became a common feature of models in which the IC
gas is in hydrostatic equilibrium in the cluster gravitational field
(Lea\cite{lea75}, Gull \& Northover\cite{gul75}, Cavaliere \&
Fusco-Femiano\cite{cav76,cav78}). On the other hand, Yahil \&
Ostriker\cite{yah73} developed a theory with an IC gas outflow.  They
argued that the gas shed from the galaxies would feed an outflow wind
from the cluster. Such a radial outflow of the IC gas was soon found
to be at odds with the random direction of the cluster galaxy
radio-tails (Lea\cite{lea77}).

In 1973 Lea et al.\cite{lea73} remarked that since
the mass of IC gas is comparable to the total mass in cluster
galaxies, not all of the IC gas can originate from cluster galaxies,
and most of it must be primordial. On the other hand, Larson \&
Dinerstein\cite{lar75} advocated for a galaxy origin of a significant
fraction of the IC gas, through supernova explosions and stellar
winds. Their model predicted a significant abundance of heavy elements
in the IC gas. The hydrodynamic numerical simulations by Lea \&
De~Young\cite{lea76} indicated that as much as 90~\% of the gas can be
removed from galaxies moving through the IC gas at transonic speed.

In 1977, Iron was found in the IC gas\cite{mit76,mit77a,ser77},
proving that at least some of the IC gas had been processed in
stars. A purely primordial origin of the IC gas was thus ruled out. As
a matter of fact, observations seemed to indicate that the IC Iron
mass was larger than could be produced in cluster galaxies.  This led
Vigroux\cite{vig77} to suggest an early heavy-element enrichment of
the IC gas by a pre-galactic population of massive stars. Fabian \&
Pringle\cite{fab77a} noted however that the estimates of the total
cluster Iron mass were very uncertain, being based on extrapolations
from the inner regions. Recently, Gibson \& Matteucci\cite{gib97} have
shown that even a large population of dwarf cluster galaxies, as
implied by the steep cluster LF, could account for the bulk of the IC
gas and metals.

Norman \& Silk\cite{nor79} and Himmes \& Biermann\cite{him80}
developed models for the temporal evolution of the IC gas.  An initial
amount of IC gas would first originate from galaxies through
supernov\ae~ emission. Only then, ram pressure stripping could start.
This model was proposed as an explanation of the Butcher-Oemler effect
(see \S~5.2).

In 1980, White \& Silk\cite{whi80} noted, in disagreement with Gingold
\& Perrenod\cite{gin79}, that mergers of subclusters can lead to
strong heating of the IC gas in the compression region. This was later
observed\cite{bri94}.

Cowie \& Perrenod\cite{cow78}'s models indicated a mild evolution of
the X-ray cluster luminosity with redshift. Perrenod\cite{per78}'s
more refined model, now including a cluster gravitational potential
varying in time, predicted instead a very strong evolution of the
X-ray cluster luminosity, a factor ten from $z \sim 1$ to the
present. Perrenod\cite{per80} later showed that the evolution rate of
the cluster X-ray luminosities was related to the density of the
Universe, so that X-ray observations of distant clusters could be used
to put useful cosmological constraints.

Perrenod's prediction of a strong evolution in the cluster X-ray
properties was first tested observationally by Henry et
al.\cite{hen79}. Unfortunately, the wide range of X-ray luminosities
for distant clusters made it impossible to test the model. Two years
later, in 1981, Perrenod \& Henry\cite{per81} argued for an X-ray
temperature negative evolution with redshift, based on a limited
sample of seven clusters observed at $z > 0.1$. Such an evolution was
however not confirmed in other investigations. First, White et
al.\cite{whi81} detected an extremely bright and hot X-ray cluster at
$z=0.54$, then, Henry et al.\cite{hen82} did not detect any change in
the slope of the cluster X-ray luminosity function with redshift.

The first observational evidence for a cosmological evolution of the
X-ray cluster properties dates back to 1982.  Anticipating the results
that were to be published in their entirety by Gioia et
al.\cite{gio90} many years later, Stocke et al.\cite{sto82} noted that
the clusters detected in the flux-limited {\sl Einstein Medium Survey
Sample} have a low average X-ray luminosity and a low average
redshift, and their total number is half that expected for a uniform
distribution of sources. This was interpreted as evidence for a
negative evolution of the cluster X-ray luminosity function.

This evolution is now confirmed for the high-luminosity tail of
the X-ray clusters only (see {\sc Mullis}, these proceedings).
The high fraction of hot X-ray clusters at high redshift is now
considered to be a strong evidence for a low-$\Omega_0$ Universe
(see {\sc Gioia}, these proceedings).

\subsection{Cooling flows and the evolution of cD galaxies}
\begin{figure}
\begin{center}
\psfig{figure=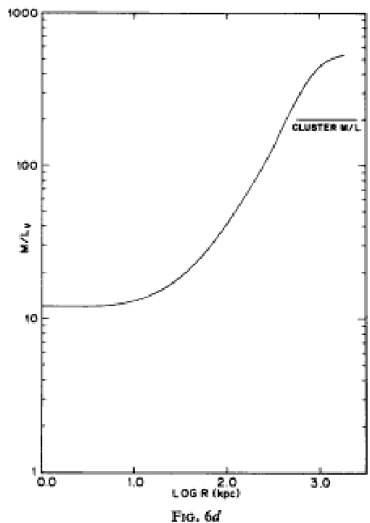,height=7cm,angle=0}
\end{center}
\caption{The rising mass-to-light ratio with radius in the cD galaxy
of the cluster Abell~2029.
From  Dressler (1979).
\label{fig-adcd}}
\end{figure}

The phenomenology of cD galaxies was first described in 1964 by
Matthews et al.\cite{mat64}. Eight years later Gunn \& Gott\cite{gun72} and
Gallagher \& Ostriker\cite{gal72} proposed two alternative mechanisms
for the formation and evolution of these cDs. Gunn \&
Gott\cite{gun72} were possibly the first to suggest the existence of a
physical link between the IC gas and cD galaxies. They showed that the
cooling of IC gas, by thermal bremsstrahlung, would produce a flow of
material in the central cluster region, that might accrete onto the cD
galaxy. An alternative mechanism for the formation of cD galaxies was
proposed by Gallagher \& Ostriker\cite{gal72} who suggested that the
cD might form out of stars stripped from other galaxies. In this case
one expects the outer parts of the cD to be in equilibrium with the
cluster (rather than the galaxy) gravitational
potential. Consistently, Dressler\cite{dre79}'s observations of the cD
in Abell~2029 showed a rapidly growing galaxy velocity dispersion with
radius, implying that the mass-to-light ratio of the cD was also rising
with distance from the galaxy centre -- see Fig.~\ref{fig-adcd}.
A year later, in 1980, Gallagher et al.\cite{gal80} showed the
envelopes of cDs to be bluer than the mean galaxy colour, again
consistent with the tidal debris scenario.

Another popular scenario for the formation of cD galaxies was proposed
in the 70's by Ostriker \& Tremaine\cite{ost76}, and developed by
Ostriker \& Hausman\cite{ost77b}. The cD galaxy would grow by
cannibalism of its neighbours.  This scenario was supported by the
n-body simulations of White\cite{whi76b}.  White showed that as the
cluster evolves, the dynamical friction mechanism can drive galaxies
to the centre, and thus favour merging phenomena.  Carnevali et
al.\cite{car81} modelled the evolution of small groups, and showed
that the {\em ``merging instability''} leads to the formation of a
large central object (they anticipated the discovery of fossil compact
groups, see {\sc Ponman}, these proceedings).  In the 80's the merging
scenario for the formation of cD galaxies was re-examined by Roos \&
Aarseth\cite{roo82} who concluded for an early creation of cDs via
merging in small groups of galaxies, before the cluster formation.

The merging scenario was supported by several observational evidences.
Oemler\cite{oem76} determined the luminosities of cD envelopes and
showed them to be correlated with the total luminosities of their parent
clusters. Dressler\cite{dre78} pointed out that the lack of
significant luminosity segregation in cD-type clusters was another
indication that cD galaxies had cannibalized neighbouring galaxies.
Carter \& Metcalfe\cite{car80} showed the cD major axis to be aligned
with the distribution of surrounding galaxies.

The merging scenario for the formation of cDs was shattered in 1978,
when White\cite{whi78}'s simulations showed that merging can produce
giant elliptical galaxies, but not the the cD extended halos. In those
years, Lea et al.\cite{lea73}, Silk\cite{sil76}, Cowie \&
Binney\cite{cow77} and Fabian \& Nulsen\cite{fab77b} estimated the
cooling time of the IC gas in the dense X-ray emitting clusters to be
lower than a Hubble time.  Fabian \& Nulsen noted that {\em
``slow-moving galaxies in core of X-ray emitting clusters can accrete
large quantities of cooling gas''}, and Quintana \& Lawrie\cite{qui82}
showed cD galaxies to be characterized by small velocities relative
to the cluster mean. This gave new strength to the hypothesis of cD
growth via accretion of the cooling IC gas.

A first observational evidence for the existence of cool gas in the
cluster centres came in 1979 with the detection of soft X-ray
components in the spectrum of the Perseus galaxy NGC~1275 (Mushotzky
\& Smith\cite{mus79}). Another observational evidence came with the
detection of optical emission-line filaments near the centre of
clusters, that were interpreted by Cowie et al.\cite{cow80} and Fabian
et al.\cite{fab81} as arising from the IC gas cooling down to $\sim
10000$~K.

Gorenstein et al.\cite{gor79} remarked upon the different X-ray
emissions of the central galaxies in Virgo and Perseus, on one side,
and the two dominant galaxies in Coma, on the other side. They
correctly pointed out that the difference was related to the lack of
cooling flows in the Coma cluster. If NGC~4874 and NGC~4889 were
moving through the IC gas, their motion could prevent the formation of
a cooling flow (Mathews \& Bregman\cite{mat78b}). A significant motion
of the two dominant Coma galaxies with respect to the cluster was
later discovered\cite{biv96}.

In the 80's Lea et al.\cite{lea82} and Sarazin \&
O'Connell\cite{sar83} noted that the inferred mass deposition rates in
cooling flows were much higher than the inferred star formation rates
as derived from UV observations (e.g. Bertola et al.\cite{ber80} for
M87). The hypothesis was made\cite{sar83} that only low-mass stars,
characterized by small UV emission, can form in the high-density
cooling flow regions.

Cooling flows have since become a major research topic in cluster
astrophysics. Two thirds of all clusters contain a cooling flow at
their centre. The deposited mass is still unaccounted for, but there
exist evidence for X-ray absorption in the centres of cooling-flow
clusters, which might be related to the deposited material (see {\sc
Fabian}, these proceedings). Maybe the active nucleus which is often
present in galaxies with cooling flows plays a significant role in
re-distributing the accreted material (see {\sc McNamara}, these
proceedings).

\section{Conclusions}
In the course of time, the concept of cluster of galaxy has
significantly evolved. A concentration of nebul\ae, maybe galactic
objects, like a star cluster, in the early days of the XX century. A
remarkable (but relatively rare) concentration of external galaxies,
which nevertheless were much smaller than our own, in Hubble's
times. Or rather the extreme of a continuous clustered distribution of
galaxies, according to Carpenter.  A stable, bound dynamical system,
with an incredible mass, according to Zwicky. Or instead, a light,
rapidly disrupting system, whose explosion was powered by unknown
mechanisms operating in the centres of its galaxies, according to
Ambartsumian. A galaxy incubator, according to Zel'dovich' top-down
scenario, or rather an association of free galaxies, according to
Peebles' bottom-up scenario. A dangerous place to live, for spiral
galaxies, according to {\sl nurture} scenarios for galaxy evolution.
Or maybe a very quiet place, where old ellipticals can passively
evolve for billions of years, according to {\sl nature} scenarios for
galaxy formation. A knot in the filamentary structure of the Universe,
when the Large Scale Structure was finally unveiled by observations in
the 80's.  A cluster of gas, rather than a cluster of galaxies, in the
90's, when X-ray surveys became more effective in finding high
redshift clusters than the traditional optical methods. And now,
finally, a cluster of dark matter, a dark cluster, which will be found
through the weak lensing surveys (see {\sc Mellier}, these
proceedings).

If the evolution of clusters is slow (see, e.g., {\sc Dickinson},
these proceedings), not so slow is the evolution of science. Moreover,
this evolution is often discontinuous and non-monotonic. Zwicky's
missing mass was re-discovered in galaxy halos after 40 years; the
existence of significant subclustering in clusters was demonstrated in
the 60's by van den Bergh, but the irregular X-ray morphologies of
clusters came as a surprise to many astronomers. The Local
Supercluster was hinted at by J.~Herschel in the XIX century, and
rediscovered several times before \dV~ re-affirmed its existence, in
the 50's.  And many other examples can be found by reading this
review.  We certainly need to keep track of the evolution of science, or
we risk to forget about fundamental results that might take years to
be re-discovered. I hope this modest review can be helpful in this
respect.

\begin{quote}
{\em ``Quello che lei non sa \`e il vero scopo del nostro
lavoro} [\ldots] {\em \`E perch\`e tutto non sia stato
inutile, per trasmettere tutto quello che sappiamo ad altri
che non sappiamo chi sono n\`e cosa sanno.''} 

\hspace{5cm} Italo Calvino, {\sl La memoria del mondo}
\end{quote}

\section*{Acknowledgments}
This paper is dedicated to my wife Patrizia, who has tolerated me
sharing my free time among Abell, Herschel, Zwicky, and herself.

I wish to thank Florence Durret and Daniel Gerbal, for organizing such
an interesting, exciting, and memorable (ah, la guinguette sous
l'orage!)  meeting. I also thank the Scientific Organizing Committee,
for giving me the opportunity of preparing this review.

Special thanks to Sandro Bardelli and Renata Longo for sending me
copies of W.~Herschel's works, and Hubble's {\sl Realm of the
Nebul\ae}, respectively. Piotr Flin's remarks have been very useful
for the writing of \S~2.1.  Stefano Borgani's careful reading of
\S~5.1 is gratefully acknowledged.  Many thanks also to the librarians
of the Trieste Observatory, Laura Abrami and Chiara Doz, who assisted
me in my bibliographic research.

Et un grand merci pour tout \`a Daniel.

This research has made use of NASA's Astrophysics Data System Abstract
Service.


\section*{References}

\begin{thebibliography}{99}
\bibitem{aar63} S.J. Aarseth, \Journal{\MN}{126}{223}{1963}.
\bibitem{aar66} S.J. Aarseth, \Journal{\MN}{132}{35}{1966}.
\bibitem{aar78} S.J. Aarseth, \Journal{{\em IAU Symp.}}{79}{189}{1978}.
\bibitem{aar79} S.J. Aarseth, J.R. Gott III, \& E.L. Turner,
\Journal{\APJ}{228}{664}{1979}.
\bibitem{aar72a} S.J. Aarseth \& J.G. Hills \Journal{\AA}{21}{255}{1972}.
\bibitem{aar72b} S.J. Aarseth \& W.C. Saslaw \Journal{\APJ}{172}{17}{1972}.
\bibitem{abe58} G.O. Abell, \Journal{\APJS}{3}{211}{1958}.
\bibitem{abe59a} G.O. Abell, \Journal{\AJ}{64}{125}{1959a}.
\bibitem{abe59b} G.O. Abell, cited in \Journal{\SKY}{18}{495}{1959b}.
\bibitem{abe60} G.O. Abell, \Journal{\AJ}{65}{481}{1960}.
\bibitem{abe61} G.O. Abell, \Journal{\AJ}{66}{607}{1961}.
\bibitem{abe62} G.O. Abell, \Journal{{\em IAU Symp.}}{15}{213}{1962}.
\bibitem{abe64} G.O. Abell, \Journal{\AJ}{69}{529}{1964}.
\bibitem{abe65} G.O. Abell, \Journal{\ARAA}{3}{1}{1965}.
\bibitem{abe75} G.O. Abell, in {\em Galaxies and the Universe,}
A. Sandage, M. Sandage, \& J. Kristian eds. (Chicago: The Univ. of
Chicago Press, 1975), p.601.
\bibitem{abe77} G.O. Abell, \Journal{\APJ}{213}{327}{1977}.
\bibitem{abe83a} G.O. Abell, \Journal{{\em Highl. Astron.}}{6}{753}{1983}.
\bibitem{abe83b} G.O. Abell \& H.G. Corwin Jr., 
\Journal{{\em IAU Symp.}}{104}{179}{1983}.
\bibitem{abe89} G.O. Abell, H.G. Corwin Jr., \& R.P. Olowin, 
\Journal{\APJS}{70}{1}{1989}.
\bibitem{abe67} G.O. Abell \& C.E. Seligman, \Journal{\AJ}{72}{288}{1967}.
\bibitem{abr83} F. Abramopoulos \& W. H.-M. Ku, \Journal{\APJ}{271}{446}{1983}.
\bibitem{ada98} C. Adami, A. Biviano, \& A. Mazure, \Journal{\AA}{331}{439}{1998}.
\bibitem{ada00} C. Adami, M.P. Ulmer, F. Durret, R.C. Nichol, A. Mazure,
B.P. Holden, A.K. Romer, \& C. Savine, \Journal{\AA}{353}{930}{2000}.
\bibitem{alb77} C.E. Albert, R.A. White, \& W.W. Morgan, 
\Journal{\APJ}{211}{309}{1977}.
\bibitem{alc98} C. Alcock et al., \Journal{\APJ}{499}{L9}{1998}.
\bibitem{all65} S.M. Alladin, \Journal{\APJ}{141}{768}{1965}.
\bibitem{all70} R.J. Allen, \Journal{\AA}{7}{330}{1970}.
\bibitem{amb58} V.A. Ambartsumian, in {\em La structure et l'evolution de
l'univers,} R. Stoops ed. (Brussels: Coudenberg, 1958), p.241.
\bibitem{amb61} V.A. Ambartsumian, \Journal{\AJ}{66}{536}{1961}.
\bibitem{and98} S. Andreon, \Journal{\AA}{336}{98}{1998}.
\bibitem{ans97} R. Ansari et al., \Journal{\AA}{324}{843}{1997}.
\bibitem{arp69} H. Arp \& F. Bertola, 
\Journal{{\em Ap. Letters}}{4}{23}{1969}.
\bibitem{aus74} T.B. Austin \& J.V. Peach, \Journal{\MN}{168}{591}{1974}.
\bibitem{baa28} W. Baade, \Journal{\AN}{233}{67}{1928}.
\bibitem{baa31} W. Baade, \Journal{\AN}{243}{303}{1931}.
\bibitem{bab39} H.W. Babcock, \Journal{{\em Lick Obs. Bull.}}{498}{41}{1939}.
\bibitem{bah70} J.N. Bahcall \& N.A. Bahcall, \Journal{\PA}{82}{721}{1970}.
\bibitem{bah69} J.N. Bahcall, M. Schmidt, \& J.E. Gunn, 
\Journal{\APJ}{157}{77}{1969}.
\bibitem{bah81b} J.N. Bahcall \& S. Tremaine, \Journal{\APJ}{244}{805}{1981}.
\bibitem{bah73} N.A. Bahcall, \Journal{\APJ}{183}{783}{1973}.
\bibitem{bah74} N.A. Bahcall, \Journal{\APJ}{193}{529}{1974}.
\bibitem{bah77b} N.A. Bahcall, \Journal{\ARAA}{15}{505}{1977}.
\bibitem{bah79a} N.A. Bahcall, \Journal{\APJ}{232}{689}{1979a}.
\bibitem{bah79b} N.A. Bahcall, \Journal{\APJ}{232}{L83}{1979b}.
\bibitem{bah81a} N.A. Bahcall, \Journal{\APJ}{247}{787}{1981}.
\bibitem{bah93} N.A. Bahcall \& R. Cen, \Journal{\APJ}{407}{L49}{1993}
\bibitem{bah00} N.A. Bahcall, J.P. Ostriker, S. Perlmutter, \& P. Steinhardt,
\Journal{{\em Science}}{284}{1481}{1999}.
\bibitem{bah77a} N.A. Bahcall \& C.L. Sarazin, \Journal{\APJ}{213}{L99}{1977}.
\bibitem{bah82} N.A. Bahcall \& R.M. Soneira, \Journal{\APJ}{262}{419}{1982}.
\bibitem{bah83} N.A. Bahcall \& R.M. Soneira, \Journal{\APJ}{270}{20}{1983}.
\bibitem{bai82} M.E. Bailey, \Journal{\MN}{201}{271}{1982}.
\bibitem{bar94} S. Bardelli, E. Zucca, G. Vettolani, G. Zamorani, R. Scaramella, 
C.A. Collins, \& H.T. MacGillivray, \Journal{\MN}{267}{665}{1994}.
\bibitem{bau58} W.A. Baum, \Journal{\PA}{70}{450}{1958}.
\bibitem{bau59} W.A. Baum, \Journal{\PA}{71}{106}{1959}.
\bibitem{bau70} L.P. Bautz \& W.W. Morgan, \Journal{\APJ}{162}{L149}{1970}.
\bibitem{bee90} T.C. Beers, K. Flynn, \& K. Gebhardt, 
\Journal{\AJ}{100}{32}{1990}.
\bibitem{bee82} T.C. Beers, M.J. Geller, \& J.P. Huchra, 
\Journal{\APJ}{257}{23}{1982}.
\bibitem{ber95} G.M. Bernstein, R. Nichol, J.A. Tyson, M.P. Ulmer, \& 
D. Wittman, \Journal{\AJ}{110}{1507}{1995}.
\bibitem{ber80} F. Bertola, M. Capaccioli, A.V. Holm, \& J.B. Oke, 
\Journal{\APJ}{237}{L65}{1980}.
\bibitem{bha81} S.P. Bhavsar, \Journal{\APJ}{246}{L5}{1981}.
\bibitem{bie75} P. Biermann \& B.M. Tinsley \Journal{\AA}{41}{441}{1975}.
\bibitem{bin82} B. Binggeli, \Journal{\AA}{107}{338}{1982}.
\bibitem{bin85} B. Binggeli, A. Sandage, \& G.A. Tammann, 
\Journal{\AJ}{90}{1681}{1985}.
\bibitem{bir81} M. Birkinshaw, S.F. Gull, \& Moffet, \Journal{\APJ}{251}{L69}{1981}
\bibitem{bir78} M. Birkinshaw, S.F. Gull, \& K.J.E. Northover, 
\Journal{\MN}{185}{245}{1978}.
\bibitem{bir93} C.M. Bird, J.M. Dickey, \& E.E. Salpeter, 
\Journal{\APJ}{404}{81}{1993}
\bibitem{biv95} A. Biviano, F. Durret, D. Gerbal, O. Le F\`evre, C. Lobo, 
A. Mazure, \& E. Slezak, \Journal{\AA}{297}{610}{1995}
\bibitem{biv96} A. Biviano, F. Durret, D. Gerbal, O. Le F\`evre, C. Lobo, 
A. Mazure, \& E. Slezak, \Journal{\AA}{311}{95}{1996}
\bibitem{biv92} A. Biviano, M. Girardi, G. Giuricin, F. Mardirossian, 
\& M. Mezzetti, \Journal{\APJ}{396}{35}{1992}
\bibitem{biv93} A. Biviano, M. Girardi, G. Giuricin, F. Mardirossian, 
\& M. Mezzetti, \Journal{\APJ}{411}{L13}{1993}
\bibitem{biv97} A. Biviano, P. Katgert, A. Mazure, M. Moles, R. den Hartog,
J. Perea, P. Focardi, \Journal{\AA}{321}{84}{1997}
\bibitem{biv99} A. Biviano, P. Katgert, T. Thomas, A. Mazure, \& C. Adami,
in {\em Formazione ed evoluzione delle galassie,} C. Chiosi, L. Portinari,
\& R. Tanatalo eds., p.45 (Padova University, 1999).
\bibitem{biv01} A. Biviano, P. Katgert, T. Thomas, A. Mazure, \& C. Adami,
{\em in preparation} (2001).
\bibitem{bog73} R.S. Bogart \& R.V. Wagoner, \Journal{\APJ}{181}{609}{1973}.
\bibitem{bol66} E. Boldt, F.B. McDonald, G. Riegler, \& P. Serlemitsos, 
\Journal{\PRL}{17}{447}{1966}.
\bibitem{bon82} J.R. Bond, A.S. Szalay, \& M.S. Turner, 
\Journal{\PRL}{48}{1636}{1982}.
\bibitem{bor99} S. Borgani, M. Girardi, R.G. Carlberg, H.K.C. Yee, \&
E. Ellingson, \Journal{\APJ}{527}{561}{1999}.
\bibitem{bot83} G.D. Bothun, M.J. Geller, T.C. Beers, \& J.P. Huchra, 
\Journal{\APJ}{268}{47}{1983}.
\bibitem{bri93} U.G. Briel \& P.J. Henry, \Journal{\AA}{278}{379}{1993}.
\bibitem{bri94} U.G. Briel \& P.J. Henry, \Journal{\NAT}{372}{439}{1994}.
\bibitem{bur61} E.M. Burbidge \& G.R. Burbidge \Journal{\AJ}{66}{541}{1961}.
\bibitem{bur75} G.R. Burbidge, \Journal{\APJ}{196}{L7}{1975}.
\bibitem{bur59} G.R. Burbidge \& M. Burbidge \Journal{\APJ}{130}{629}{1959}.
\bibitem{bur69} G.R. Burbidge \& W.L.W. Sargent 
\Journal{{\em Comm. Ap. Space Phys.}}{1}{220}{1969}.
\bibitem{but78a} H. Butcher \& A. Oemler Jr., \Journal{\APJ}{219}{18}{1978a}.
\bibitem{but78b} H. Butcher \& A. Oemler Jr., \Journal{\APJ}{226}{559}{1978b}.
\bibitem{but80} H. Butcher, D. Wells, \& A. Oemler Jr., 
\Journal{{\em IAU Symp.}}{92}{49}{1980}.
\bibitem{byr66} E.T. Byram, T.A. Chubb, \& H. Friedman, 
\Journal{\AJ}{71}{379}{1966}.
\bibitem{cap80b} H.V. Capelato, D. Gerbal, G. Mathez,
A. Mazure, J. Roland, \& E. Salvador-Sol\'e, \Journal{\AA}{87}{132}{1980}.
\bibitem{cap81} H.V. Capelato, D. Gerbal, G. Mathez,
A. Mazure, J. Roland, \& E. Salvador-Sol\'e, \Journal{\AA}{96}{235}{1981}.
\bibitem{cap80a} H.V. Capelato, D. Gerbal, G. Mathez,
A. Mazure, E. Salvador-Sol\'e, \& H. Sol, \Journal{\APJ}{241}{521}{1980}.
\bibitem{cap79} H.V. Capelato, D. Gerbal, E. Salvador-Sol\'e, G. Mathez,
A. Mazure, \& J. Roland, \Journal{\AAS}{38}{295}{1979}.
\bibitem{car97} R.G. Carlberg, H.K.C. Yee, \& E. Ellingson, 
\Journal{\APJ}{478}{462}{1997}.
\bibitem{car81} P. Carnevali, A. Cavaliere, \& P. Santangelo, 
\Journal{\APJ}{249}{449}{1981}.
\bibitem{car31} E.F. Carpenter, \Journal{\PA}{43}{247}{1931}.
\bibitem{car38} E.F. Carpenter, \Journal{\APJ}{88}{344}{1938}.
\bibitem{car80} D. Carter \& N. Metcalfe, \Journal{\MN}{191}{325}{1980}.
\bibitem{cav76} A. Cavaliere \& R. Fusco-Femiano, \Journal{\AA}{49}{137}{1976}.
\bibitem{cav78} A. Cavaliere \& R. Fusco-Femiano, \Journal{\AA}{70}{677}{1978}.
\bibitem{cav71} A. Cavaliere, H. Gursky, \& W.H. Tucker, 
\Journal{\NAT}{231}{437}{1971}. 
\bibitem{cav97} A. Cavaliere \& N. Menci, \Journal{\APJ}{480}{132}{1997}.
\bibitem{cav98} A. Cavaliere, N. Menci, \& P. Tozzi, 
\Journal{\APJ}{501}{493}{1998}.
\bibitem{cha43} S. Chandrasekhar, \Journal{\APJ}{97}{255}{1943}.
\bibitem{che69} C. Chester \& M.S. Roberts, \Journal{\AJ}{69}{635}{1969}.
\bibitem{chi78} G. Chincarini, \Journal{\NAT}{274}{452}{1978}.
\bibitem{chi83} G. Chincarini, R. Giovanelli, \& M.P. Haynes, 
\Journal{\APJ}{269}{13}{1983}.
\bibitem{chi71} G. Chincarini \& H.J. Rood, \Journal{\APJ}{168}{321}{1971}.
\bibitem{chi77} G. Chincarini \& H.J. Rood, \Journal{\APJ}{214}{351}{1977}.
\bibitem{chi79} G. Chincarini \& H.J. Rood, \Journal{\APJ}{230}{648}{1979}.
\bibitem{cla68} E.E. Clark, \Journal{\AJ}{73}{1011}{1968}.
\bibitem{coo76} B.A. Cooke \& D. Maccagni, \Journal{\MN}{175}{65P}{1976}.
\bibitem{cow77} L.L. Cowie \& J. Binney, \Journal{\APJ}{215}{723}{1977}.
\bibitem{cow80} L.L. Cowie, A.C. Fabian, \& P.E.J. Nulsen,
\Journal{\MN}{191}{399}{1980}.
\bibitem{cow78} L.L. Cowie \& S.C. Perrenod, \Journal{\APJ}{219}{354}{1978}.
\bibitem{cow73} R. Cowsik \& J. McClelland, \Journal{\APJ}{180}{7}{1973}.
\bibitem{cul77} J.L. Culhane, \Journal{{\em Highl. Astron.}}{4}{293}{1977}.
\bibitem{cur18a} H.D. Curtis, \Journal{{\em Pub. Lick Obs.}}{13}{11}{1918a}.
\bibitem{cur18b} H.D. Curtis, \Journal{{\em Pub. Lick Obs.}}{13}{15}{1918b}.
\bibitem{dar65} H. d'Arrest, \Journal{\AN}{65}{1}{1865}.
\bibitem{dav73} R.D. Davies \& B.M. Lewis, \Journal{\MN}{165}{231}{1973}.
\bibitem{dav76} M. Davis \& M.J. Geller, \Journal{\APJ}{208}{13}{1976}.
\bibitem{dav82} M. Davis, J. Huchra, D.W. Latham, \& J. Tonry,
\Journal{\APJ}{253}{423}{1982}
\bibitem{dav80} M. Davis, J. Tonry, J. Huchra, \& D.W. Latham, 
\Journal{\APJ}{238}{L113}{1980}
\bibitem{deb00} P. de Bernardis et al., \Journal{\NAT}{404}{955}{2000}.
\bibitem{dep95} R. De Propris, C.J. Pritchet, W.E. Harris, \& R.D. McClure,
\Journal{\APJ}{450}{534}{1995}.
\bibitem{des82} R.E. de Souza, H.V. Capelato, L. Arakaki, \& C. Logullo,
\Journal{\APJ}{263}{557}{1982}.
\bibitem{dev48} G. \dV, \Journal{{\em Ann. Astroph.}}{11}{247}{1948}.
\bibitem{dev53} G. \dV, \Journal{\AJ}{58}{30}{1953}
\bibitem{dev58} G. \dV, \Journal{\AJ}{63}{253}{1958}
\bibitem{dev60} G. \dV, \Journal{\APJ}{131}{585}{1960}
\bibitem{dev61a} G. \dV, \Journal{\APJS}{6}{213}{1961a}
\bibitem{dev61b} G. \dV, \Journal{\AJ}{66}{629}{1961b}
\bibitem{dev61c} G. \dV, \Journal{\APJS}{5}{233}{1961c}
\bibitem{dev63} G. \dV~ \& A. \dV, \Journal{\AJ}{68}{278}{1963}
\bibitem{dev64} G. \dV~ \& A. \dV, {\em Reference catalogue of bright galaxies,}
(Austin: Univ. Texas Press, 1964)
\bibitem{dev69} G. \dV, \Journal{{\em Ap. Letters}}{4}{17}{1969}.
\bibitem{dev75} G. \dV, in {\em Galaxies and the Universe,}
A. Sandage, M. Sandage, \& J. Kristian eds. (Chicago: The Univ. of
Chicago Press, 1975), p.557.
\bibitem{dev70} G. \dV~ \& A. \dV \Journal{{\em Astroph. Lett.}}{5}{219}{1970}
\bibitem{dor80} A.G. Doroshkevich, Ya.B. Zel'dovich, R.A. Syunyaev, \& 
M.Yu. Khlopov, \Journal{\SAL}{6}{257}{1980}.
\bibitem{dre78} A. Dressler, \Journal{\APJ}{226}{55}{1978}.
\bibitem{dre79} A. Dressler, \Journal{\APJ}{231}{659}{1979}.
\bibitem{dre80a} A. Dressler, \Journal{\APJ}{236}{351}{1980a}.
\bibitem{dre80b} A. Dressler, \Journal{\APJS}{42}{565}{1980b}.
\bibitem{dre82} A. Dressler \& J.E. Gunn, \Journal{\APJ}{263}{533}{1982}.
\bibitem{dre83} A. Dressler \& J.E. Gunn, \Journal{\APJ}{270}{7}{1983}.
\bibitem{dre97} A. Dressler, A. Oemler Jr., J.W. Couch, I. Smail, R.S. Ellis,
A. Barger, H. Butcher, B.M. Poggianti, \& R.M. Sharples, 
\Journal{\APJ}{490}{577}{1997}.
\bibitem{dre88a} A. Dressler \& S.A. Shectman, \Journal{\AJ}{95}{284}{1988}.
\bibitem{dre88b} J.L.E. Dreyer, 
\Journal{{\em Mem. R. Astron. Soc.}}{49}{1}{1888}.
\bibitem{duu77} A. Duus \& B. Newell, \Journal{\APJS}{35}{209}{1977}.
\bibitem{eas04} C. Easton, \Journal{\AN}{166}{131}{1904}.
\bibitem{efs79} G. Efstathiou, \Journal{\MN}{187}{117}{1979}.
\bibitem{ein80} J. Einasto, M. J\^oeveer, \& E. Saar, 
\Journal{\MN}{193}{353}{1980}.
\bibitem{ein74} J. Einasto, A. Kaasik, \& E. Saar, 
\Journal{\NAT}{250}{309}{1974}.
\bibitem{ell97} R.S. Ellis, I. Smail, A. Dressler, W.J. Couch, A. Oemler Jr.,
H. Butcher, R.M. Sharples, \Journal{\APJ}{483}{582}{1997}.
\bibitem{esc94} E. Escalera, A. Biviano, M. Girardi, G. Giuricin,
F. Mardirossian, A. Mazure, \& M. Mezzetti, \Journal{\APJ}{423}{539}{1994}.
\bibitem{fab77b} A.C. Fabian \& P.E.J. Nulsen, \Journal{\MN}{180}{479}{1977}.
\bibitem{fab81} A.C. Fabian, P.E.J. Nulsen, G.C. Stewart, W.H.-M. Ku, 
D.F. Malin, R.F. Mushotzky, \Journal{\MN}{196}{P35}{1980}
\bibitem{fab77a} A.C. Fabian \& J.E. Pringle, \Journal{\MN}{181}{5P}{1977}.
\bibitem{fab80} D. Fabricant, M. Lecar, \& P. Gorenstein, 
\Journal{\APJ}{241}{552}{1980}.
\bibitem{fal78} S.M. Fall, \Journal{\MN}{185}{165}{1978}.
\bibitem{far80} R. Farouki \& S.L. Shapiro, \Journal{\APJ}{241}{928}{1980}.
\bibitem{fel66} J.E. Felten, R.J. Gould, W.A. Stein, \& N.J. Woolf, 
\Journal{\APJ}{146}{955}{1966}.
\bibitem{fit87} M. Fitchett \& R. Webster, \Journal{\APJ}{317}{653}{1987}.
\bibitem{fli88} P. Flin, \Journal{{\em Acta Cosmologica}}{15}{25}{1988}.
\bibitem{for81} W. Forman, J. Bechtold, W. Blair, R. Giacconi, L. van 
Speybroeck, \& C. Jones, \Journal{\APJ}{243}{L133}{1981}.
\bibitem{for79} W. Forman, J. Schwarz, C. Jones, W. Liller, \& A.C. Fabian, 
\Journal{\APJ}{234}{L27}{1979}.
\bibitem{for94} B. Fort \& Y. Mellier, \Journal{\AAR}{5}{239}{1994}.
\bibitem{fre70} K.C. Freeman, \Journal{\APJ}{160}{811}{1970}.
\bibitem{fre83} C.S. Frenk, S.D.M. White, \& M. Davis, 
\Journal{\APJ}{271}{417}{1983}.
\bibitem{fri67} H. Friedman \& E.T. Byram, \Journal{\APJ}{147}{399}{1967}.
\bibitem{gal80} J.S. Gallagher III, D. Burstein, \& S.M. Faber, 
\Journal{\APJ}{235}{743}{1980}.
\bibitem{gal72} J.S. Gallagher III \& J.P. Ostriker, 
\Journal{\AJ}{77}{288}{1972}.
\bibitem{gel82} M.J. Geller \& T.C. Beers, \Journal{\PA}{94}{421}{1982}.
\bibitem{gel83} M.J. Geller \& J.P. Huchra, \Journal{\APJS}{52}{61}{1983}.
\bibitem{gel73} M.J. Geller \& P.J.E. Peebles, \Journal{\APJ}{184}{329}{1973}.
\bibitem{gib97} B.K. Gibson \& F. Matteucci, \Journal{\APJ}{475}{47}{1997}.
\bibitem{gin79} R.A. Gingold, \& S.C. Perrenod, \Journal{\MN}{187}{371}{1979}.
\bibitem{gio90} I.M. Gioia, J.P. Henry, T. Maccacaro, S.L. Morris, J. Stocke,
\& A. Wolter, \Journal{\APJ}{356}{L35}{1990}.
\bibitem{gio82} I.M. Gioia, T. Maccacaro, M.J. Geller, J.P. Huchra, J. Stocke,
\& J.E. Steiner, \Journal{\APJ}{255}{L17}{1982}.
\bibitem{gio83} R. Giovanelli \& M.P. Haynes, \Journal{\AJ}{88}{881}{1983}.
\bibitem{gio81} R. Giovanelli, M.P. Haynes, \& G. Chincarini, 
\Journal{\APJ}{247}{383}{1981}.
\bibitem{gir93} M. Girardi, A. Biviano, G. Giuricin, F. Mardirossian, 
\& M. Mezzetti, \Journal{\APJ}{404}{38}{1993}.
\bibitem{giu88} G. Giuricin, P. Gondolo, F. Mardirossian, M. Mezzetti, \&
M. Ramella, \Journal{\AA}{199}{85}{1988}.
\bibitem{god61} E.A. Godfredsen, \Journal{\AJ}{66}{285}{1961}.
\bibitem{god77} J.G. Godwin \& J.V. Peach, \Journal{\MN}{181}{323}{1977}.
\bibitem{gor79} P. Gorenstein, D. Fabricant, K. Topka, \& F.R. Harnden Jr., 
\Journal{\APJ}{230}{26}{1979}.
\bibitem{got81} J.R. Gott III, \Journal{\APJ}{243}{140}{1981}.
\bibitem{got71} J.R. Gott III \& J.E. Gunn, \Journal{\APJ}{169}{L13}{1971}.
\bibitem{got75} J.R. Gott III \& M.J. Rees, \Journal{\AA}{45}{365}{1975}.
\bibitem{got76} J.R. Gott III, T.X. Thuan, \Journal{\APJ}{204}{649}{1976}.
\bibitem{got79} J.R. Gott III, E.L. Turner, \& S.J. Aarseth, 
\Journal{\APJ}{234}{13}{1979}.
\bibitem{got73} J.R. Gott III, G.T. Wrixon, P. Wannier, 
\Journal{\APJ}{186}{777}{1973}.
\bibitem{gou78} R.J. Gould \& Y. Rephaeli, \Journal{\APJ}{219}{12}{1978}.
\bibitem{gre75} S.A. Gregory, \Journal{\APJ}{199}{1}{1975}.
\bibitem{gre78} S.A. Gregory \& L.A. Thompson, \Journal{\APJ}{222}{784}{1978}.
\bibitem{gre81} S.A. Gregory, L.A. Thompson, \& W.G. Tifft, 
\Journal{\APJ}{243}{411}{1981}.
\bibitem{gul75} S.F. Gull \& K.J.E. Northover, \Journal{\MN}{173}{585}{1975}.
\bibitem{gul76} S.F. Gull \& K.J.E. Northover, \Journal{\NAT}{263}{572}{1976}.
\bibitem{gun80} J.E. Gunn, 
\Journal{{\em Phil. Trans. R. Soc. London, Ser.A}}{296}{313}{1980}.
\bibitem{gun72} J.E. Gunn \& J.R. Gott III, \Journal{\APJ}{176}{1}{1972}.
\bibitem{gur71} H. Gursky, E. Kellogg, S. Murray, C. Leong, 
H. Tananbaum, \& R. Giacconi, \Journal{\APJ}{167}{L81}{1971}.
\bibitem{gur72} H. Gursky, R. Levinson, E. Kellogg, S. Murray, 
H. Tananbaum, R. Giacconi, \& A. Cavaliere, \Journal{\APJ}{173}{L99}{1972}.
\bibitem{han82a} R.J. Hanisch, \Journal{\AA}{111}{97}{1982a}.
\bibitem{han82b} R.J. Hanisch, \Journal{\AA}{116}{137}{1982b}.
\bibitem{han79} R.J. Hanisch, T.A. Matthews, \& M.M. Davis, 
\Journal{\AJ}{84}{946}{1979}.
\bibitem{har78} D.E. Harris \& G.K. Miley, \Journal{\AAS}{34}{117}{1978}.
\bibitem{hau73} M.G. Hauser \& P.J.E. Peebles, \Journal{\APJ}{185}{757}{1973}.
\bibitem{hee56} D.S. Heeschen, \Journal{\APJ}{124}{660}{1956}.
\bibitem{hei80} G.M. Heiligman \& E.L. Turner, \Journal{\APJ}{236}{745}{1980}.
\bibitem{hel79} G. Helou, E.E. Salpeter, \& N. Krumm, 
\Journal{\APJ}{228}{L1}{1979}.
\bibitem{hen64} M. H\'enon, \Journal{{\em Ann. d'Astroph.}}{27}{83}{1964}.
\bibitem{hen79} J.P. Henry, G. Branduardi, U. Briel, D. Fabricant,
E. Feigelson, S. Murray, A. Soltan, \& H. Tananbaum, 
\Journal{\APJ}{234}{L15}{1979}.
\bibitem{hen82} J.P. Henry, U. Briel, J.E. Gunn, \& A. Soltan, 
\Journal{\APJ}{262}{1}{1982}.
\bibitem{her85} F.W. Herschel, \Journal{{\em Phil. Trans.}}{75}{213}{1785}.
\bibitem{her11} F.W. Herschel, \Journal{{\em Phil. Trans.}}{101}{269}{1811}.
\bibitem{her57} E. Herzog, P. Wild, \& F. Zwicky \Journal{\PA}{69}{409}{1957}.
\bibitem{hic82} P. Hickson, \Journal{\APJ}{255}{382}{1982}.
\bibitem{him80} A. Himmes \& P. Biermann, \Journal{\AA}{86}{11}{1980}.
\bibitem{hin80} P. Hintzen, J.S. Scott, \& J.D. McKee, 
\Journal{\APJ}{242}{857}{1980}.
\bibitem{hod59} P.W. Hodge, \Journal{\PA}{71}{28}{1959}.
\bibitem{hod60} P.W. Hodge, \Journal{\PA}{72}{188}{1960}.
\bibitem{hof80} G.L. Hoffman, D.W. Olson, \& E.E. Salpeter, 
\Journal{\APJ}{242}{861}{1980}.
\bibitem{hol97} B.P. Holden, A.K. Romer, R.C. Nichol, \& M.P. Ulmer,
\Journal{\AJ}{114}{1701}{1997}.
\bibitem{hol37} E. Holmberg, \Journal{{\em Annals Obs. Lund}}{6}{1}{1937}.
\bibitem{hol40} E. Holmberg, \Journal{\APJ}{92}{200}{1940}.
\bibitem{hol41} E. Holmberg, \Journal{\APJ}{94}{385}{1941}.
\bibitem{hol50} E. Holmberg, 
\Journal{{\em Medd. Lund Astron. Obs. Ser.II}}{128}{1}{1950}.
\bibitem{hol58} E. Holmberg, 
\Journal{{\em Medd. Lund Astron. Obs. Ser.II}}{136}{1}{1958}.
\bibitem{hol61} E. Holmberg, \Journal{\AJ}{66}{620}{1961}.
\bibitem{hol74} E. Holmberg, \Journal{{\em Ark. Astron.}}{5}{305}{1974}.
\bibitem{hoy83} F. Hoyle, \Journal{{\em Ap. Space Sci.}}{93}{1}{1983}.
\bibitem{hub29} E. Hubble, \Journal{{\em Proc. Natl. Acad. Sci.}}{15}
{168}{1929}.
\bibitem{hub36} E. Hubble, {\em The Realm of the Nebul\ae~} (New Haven:
Yale Univ. Press, 1936)
\bibitem{hub31} E. Hubble \& M.L. Humason, \Journal{\APJ}{74}{43}{1931}.
\bibitem{huc82} J.P. Huchra \& M.J. Geller, \Journal{\APJ}{257}{423}{1982}.
\bibitem{hum34} M.L. Humason, \Journal{\PA}{46}{275}{1934}.
\bibitem{hum36} M.L. Humason, \Journal{\APJ}{83}{10}{1936}.
\bibitem{hum56} M.L. Humason, N.U. Mayall \& A.R. Sandage, 
\Journal{\AJ}{61}{97}{1956}.
\bibitem{hun72} R. Hunt \& D.W. Sciama, \Journal{\MN}{157}{335}{1972}.
\bibitem{jaf77} W.J. Jaffe, \Journal{\APJ}{212}{1}{1977}.
\bibitem{jaf79} W.J. Jaffe \& L. Rudnick, \Journal{\APJ}{233}{453}{1979}.
\bibitem{joe78} M. J\^oeveer, J. Einasto, \& E. Tago, 
\Journal{\MN}{185}{357}{1978}.
\bibitem{joh79} M.W. Johnson, R.G. Cruddace, G. Fritz, S. Shulman,
\& H. Friedman, \Journal{\APJ}{231}{L45}{1979}.
\bibitem{jon79} C. Jones, E. Mandel, J. Schwarz, W. Forman, S.S. Murray,
\& F.R. Harnden Jr., \Journal{\APJ}{234}{L21}{1979}.
\bibitem{jus59} K. Just, \Journal{\APJ}{129}{268}{1959}.
\bibitem{kah59} F.D. Kahn \& L. Woltjer, \Journal{\APJ}{130}{705}{1959}.
\bibitem{kar65} I.D. Karachentsev, \Journal{{\em Astrofiz.}}{1}{303}{1965}.
\bibitem{kar66} I.D. Karachentsev, \Journal{{\em Astrofiz.}}{2}{81}{1966}.
\bibitem{kar73} V.E. Karachentseva, 
\Journal{{\em Soob. Spets. Astrof. Obs.}}{8}{3}{1973}.
\bibitem{kel73} E. Kellogg, S. Murray, R. Giacconi, T. Tananbaum, 
\& H. Gursky, \Journal{\APJ}{185}{L13}{1973}.
\bibitem{ken83a} R.C. Kennicutt Jr., \Journal{\AJ}{88}{483}{1983}.
\bibitem{ken82} S.M. Kent \& J.E. Gunn, \Journal{\AJ}{87}{945}{1982}.
\bibitem{ken83b} S.M. Kent \& W.L.W. Sargent, \Journal{\AJ}{88}{697}{1983}.
\bibitem{kin62} I. King, \Journal{\AJ}{67}{471}{1962}.
\bibitem{kin66} I. King, \Journal{\AJ}{71}{64}{1966}.
\bibitem{kir77} R.P. Kirshner, \Journal{\APJ}{212}{319}{1977}.
\bibitem{kir81} R.P. Kirshner, A. Oemler Jr., P.L. Schechter, 
\& S.A. Shectman, \Journal{\APJ}{248}{L57}{1981}.
\bibitem{kle69} A.R. Klemola, \Journal{\AJ}{74}{804}{1969}.
\bibitem{kli81} F.R. Klinkhamer, C.A. Norman, \Journal{\APJ}{243}{L1}{1981}.
\bibitem{kly83} A.A. Klypin \& A.I. Kopylov, \Journal{\SAL}{9}{41}{1983}.
\bibitem{ko65} J.C. Ko, \Journal{\AJ}{70}{681}{1965}.
\bibitem{koo81} D.C. Koo, \Journal{\APJ}{251}{L75}{1981}.
\bibitem{kow69} C.T. Kowal, \Journal{\PA}{81}{608}{1969}.
\bibitem{kru74} E.C. Krupp, \Journal{\PA}{86}{385}{19}.
\bibitem{lak77} G. Lake \& R.B. Partridge, \Journal{\NAT}{270}{502}{1977}.
\bibitem{lak80} G. Lake \& R.B. Partridge, \Journal{\APJ}{237}{378}{1980}.
\bibitem{lar59} M.I. Large, D.S. Mathewson, C.G.T. Haslam, 
\Journal{\NAT}{183}{1663}{1959}.
\bibitem{lar72} R.B. Larson, \Journal{\NAT}{236}{21}{1972}.
\bibitem{lar75} R.B. Larson \& H.L. Dinerstein, \Journal{\PA}{87}{911}{1975}.
\bibitem{lar80} R.B. Larson, B.M. Tinsley, \& C.N. Caldwell, 
\Journal{\APJ}{237}{692}{1980}.
\bibitem{lay62} D. Layzer, \Journal{\APJ}{136}{138}{1962}.
\bibitem{lea75} S.M. Lea, \Journal{{\em Astrophys. Letters}}{16}{141}{1975}.
\bibitem{lea77} S.M. Lea, \Journal{{\em Highl. Astron.}}{4}{329}{1977}.
\bibitem{lea76} S.M. Lea \& D.S. De Young, \Journal{\APJ}{210}{647}{1976}.
\bibitem{lea82} S.M. Lea, R. Mushotzky, \& S.S. Holt, 
\Journal{\APJ}{262}{24}{1982}.
\bibitem{lea73} S.M. Lea, J. Silk, E. Kellogg, \& S. Murray 
\Journal{\APJ}{184}{L105}{1973}.
\bibitem{lec75} M. Lecar, \Journal{{\em IAU Symp.}}{69}{161}{1975}.
\bibitem{lew72} B.M. Lewis, \Journal{\AA}{16}{165}{1972}.
\bibitem{lew76} B.M. Lewis, \Journal{\NAT}{261}{302}{1976}.
\bibitem{lim59} D.N. Limber, \Journal{\APJ}{130}{414}{1959}.
\bibitem{lim62} D.N. Limber, \Journal{{\em IAU Symp.}}{15}{239}{1962}.
\bibitem{lob97} C. Lobo, A. Biviano, F. Durret, D. Gerbal, O. Le F\`evre, 
A. Mazure, \& E. Slezak, \Journal{\AA}{317}{385}{1997}.
\bibitem{lov00} J. Loveday, \Journal{\MN}{312}{557}{2000}.
\bibitem{luc83} J.R. Lucey, \Journal{\MN}{204}{33}{1983}.
\bibitem{luc80} J.R. Lucey, R.J. Dickens, \& J.A. Dawe, 
\Journal{\NAT}{285}{305}{1980}.
\bibitem{lum92} S.L. Lumsden, R.C. Nichol, C.A. Collins, \& L. Guzzo,
\Journal{\MN}{258}{L1}{1992}
\bibitem{lun27} K. Lundmark, \Journal{{\em Uppsala Medd.}}{30}{1}{1927}.
\bibitem{lyn67} D. Lynden-Bell, \Journal{\MN}{136}{101}{1967}.
\bibitem{lyn86} R. Lynds \& V. Petrosian, \Journal{{\em BAAS}}{18}{1014}{1986}.
\bibitem{mac76} H.T. MacGillivray, et al., \Journal{\MN}{176}{649}{1976}.
\Journal{{\em C. R. Acad. Sci. Paris. -- A}}{280}{1551}{1975}.
\bibitem{mar77} A.B. Marchant \& S.L. Shapiro, \Journal{\APJ}{215}{1}{1977}.
\bibitem{mat78a} W.G. Mathews, \Journal{\APJ}{219}{413}{1978}.
\bibitem{mat78b} W.G. Mathews \& J.N. Bregman, \Journal{\APJ}{224}{308}{1978}.
\bibitem{mat81} R.D. Mathieu \& H. Spinrad, \Journal{\APJ}{251}{485}{1981}.
\bibitem{mat64} T.A. Matthews, W.W. Morgan, \& M. Schmidt, 
\Journal{\APJ}{140}{35}{1964}.
\bibitem{mat77} K. Mattila, \Journal{\AA}{60}{425}{1977}.
\bibitem{may60} N.U. Mayall, \Journal{{\em Ann. d'Astroph.}}{23}{344}{1960}.
\bibitem{mck80} J.D. McKee, R.F. Mushotzky, E.A. Boldt, S.S. Holt,
F.E. Marshall, S.H. Pravdo, \& P.J. Serlemitsos, 
\Journal{\APJ}{242}{843}{1980}.
\bibitem{mee71} J.F. Meekins, G. Fritz, T.A. Chubb, \& H. Friedman, 
\Journal{\NAT}{231}{107}{1971}.
\bibitem{mel88} Y. Mellier, G. Mathez, A. Mazure, B. Chauvineau, \&
D. Proust, \Journal{\AA}{199}{67}{1988}.
\bibitem{mel75} J. Melnick \& H. Quintana, \Journal{\APJ}{198}{L97}{1975}.
\bibitem{mel77} J. Melnick \& W.L. Sargent, \Journal{\APJ}{215}{401}{1977}.
\bibitem{mes84} C. Messier, {\em Connaissance des Temps} (Paris: 1784).
\bibitem{mil72} G.K. Miley, G.C. Perola, P.C. van der Kruit, \&
H. van der Laan, \Journal{\NAT}{237}{269}{1972}.
\bibitem{mil83a} M. Milgrom, \Journal{\APJ}{270}{384}{1983}.
\bibitem{mil83b} R.H. Miller, \Journal{\APJ}{270}{390}{1983}.
\bibitem{min60} R. Minkowki, \Journal{\APJ}{132}{908}{1960}.
\bibitem{mit77a} R.J. Mitchell \& J.L. Culhane, \Journal{\MN}{178}{75P}{1977}.
\bibitem{mit76} R.J. Mitchell, J.L. Culhane, P.J.N. Davison, \& J.C. Ives,
\Journal{\MN}{175}{29P}{1976}.
\bibitem{mit77b} R.J. Mitchell, J.C. Ives, \& J.L. Culhane, 
\Journal{\MN}{181}{25P}{1977}.
\bibitem{mor75} W.W. Morgan, S. Kayser, \& R.A. White, 
\Journal{\AJ}{199}{545}{1975}.
\bibitem{mor65} W.W. Morgan \& J.R. Lesh, \Journal{\APJ}{142}{1364}{1965}.
\bibitem{mos77} C. Moss \& R.J. Dickens, \Journal{\MN}{178}{701}{1977}.
\bibitem{mul59} C.A. Muller, \Journal{{\em IAU Symp.}}{9}{465}{1959}.
\bibitem{mus78} R.F. Mushotzky, P.J. Serlemitsos, B.W. Smith, E.A. Boldt,
\& S.S. Holt, \Journal{\APJ}{225}{21}{1978}.
\bibitem{mus79} R.F. Mushotzky \& B.W. Smith, 
\Journal{{\em Highl. Astr.}}{5}{735}{1979}.
\bibitem{nap75} W.McD. Napier \& B.N.G. Guthrie, \Journal{\MN}{170}{7}{1975}.
\bibitem{nav96} J.F. Navarro, C.S. Frenk, \& S.D.M. White, 
\Journal{\APJ}{462}{563}{1996}.
\bibitem{ney61} J. Neyman, T.Page, \& E.L. Scott, \Journal{\AJ}{66}{633}{1961}.
\bibitem{ney52} J. Neyman \& E.L. Scott, \Journal{\APJ}{116}{144}{1952}.
\bibitem{ney53} J. Neyman, E.L. Scott, \& C.D. Shane, 
\Journal{\APJ}{117}{92}{1953}.
\bibitem{ney54} J. Neyman, E.L. Scott, \& C.D. Shane, 
\Journal{\APJS}{1}{269}{1954}.
\bibitem{ney62} J. Neyman, E.L. Scott, \& W. Zonn \Journal{\AJ}{67}{583}{1962}.
\bibitem{nez69} E.M. Nezhinskii \& L.P. Osipkov, \Journal{\SA}{13}{540}{1969}.
\bibitem{noe70} P.D. Noerdingler, \Journal{\NAT}{228}{845}{1970}.
\bibitem{noo61} T. Noonan, \Journal{\PA}{73}{212}{1961}.
\bibitem{noo73} T.W. Noonan, \Journal{\AJ}{78}{26}{1973}.
\bibitem{noo81} T.W. Noonan, \Journal{\APJS}{45}{613}{1981}.
\bibitem{nor79} C. Norman \& J. Silk, \Journal{\APJ}{233}{L1}{1979}.
\bibitem{nul82} P.E.J. Nulsen, \Journal{\MN}{198}{1007}{1982}.
\bibitem{oem74} A. Oemler Jr., \Journal{\APJ}{194}{1}{1974}.
\bibitem{oem76} A. Oemler Jr., \Journal{\APJ}{209}{693}{1976}.
\bibitem{oem77} A. Oemler Jr., \Journal{{\em Highl. Astron.}}{4}{253}{1977}.
\bibitem{oke71} J.B. Oke, \Journal{\APJ}{170}{193}{1971}.
\bibitem{ome65} G.C. Omer Jr., T.L. Page, \& A.G. Wilson 
\Journal{\AJ}{70}{440}{1965}.
\bibitem{oor81} J.H. Oort, \Journal{\AA}{94}{359}{1981}.
\bibitem{ost77a} J.P. Ostriker, in {\em ``The evolution of galaxies
and stellar populations,} B.M. Tinsley, R.B. Larson eds., p.369
(New Haven: Yale Univ. Obs., 1977).
\bibitem{ost77b} J.P. Ostriker \& M.A. Hausman \Journal{\APJ}{217}{L125}{1977}.
\bibitem{ost73} J.P. Ostriker \& P.J.E. Peebles, 
\Journal{\APJ}{186}{467}{1973}.
\bibitem{ost74} J.P. Ostriker, P.J.E. Peebles, \& A. Yahil, 
\Journal{\APJ}{193}{L1}{1974}.
\bibitem{ost76} J.P. Ostriker \& S.D. Tremaine, 
\Journal{\APJ}{202}{L113}{1976}.
\bibitem{oze78} L.M. Ozernoy \& M. Reinhardt, 
\Journal{{\em IAU Symp.}}{79}{98}{1978}.
\bibitem{oze79}  L.M. Ozernoy \& M. Reinhardt, 
\Journal{{\em Ap. Space Sci.}}{60}{267}{1979}.
\bibitem{paa64} G. Paal, \Journal{{\em Comm. Konkoly Obs.}}{54}{1}{1964}.
\bibitem{pag52} T. Page, \Journal{\APJ}{116}{63}{1952}.
\bibitem{pag82} H. Pagels \& J. Primack, \Journal{\PRL}{48}{223}{1982}.
\bibitem{par72} Yu. N. Parijsky, \Journal{\SA}{16}{1048}{1972}.
\bibitem{pay74} C. Payne-Gaposchkin, \Journal{\SKY}{47}{311}{1974}.
\bibitem{pee65} P.J.E. Peebles, \Journal{\APJ}{142}{1317}{1965}.
\bibitem{pee70} P.J.E. Peebles, \Journal{\AJ}{75}{13}{1970}.
\bibitem{pee71} P.J.E. Peebles, {\em Physical Cosmology} 
(Princeton: Princeton Univ. Press, 1971).
\bibitem{pee74} P.J.E. Peebles, \Journal{\AA}{32}{197}{1974}.
\bibitem{pee76a} P.J.E. Peebles, \Journal{\APJ}{205}{L109}{1976a}.
\bibitem{pee76b} P.J.E. Peebles, \Journal{\APJ}{205}{318}{1976b}.
\bibitem{pee68} P.J.E. Peebles \& R.H. Dicke, \Journal{\APJ}{154}{891}{1968}.
\bibitem{pen61} A.A. Penzias, \Journal{\AJ}{66}{293}{1961}.
\bibitem{per86} J. Perea, A. del Olmo, \& M. Moles, 
\Journal{\MN}{222}{49}{1986}.
\bibitem{per78} S.C. Perrenod, \Journal{\APJ}{224}{285}{1978}.
\bibitem{per80} S.C. Perrenod, \Journal{\APJ}{236}{373}{1980}.
\bibitem{per81} S.C. Perrenod \& J.P. Henry, \Journal{\APJ}{247}{L1}{1981}.
\bibitem{pog99} B.M. Poggianti, I. Smail, A. Dressler, J.W. Couch, A.J. Barger,
H. Butcher, R.S. Ellis, \& A. Oemler Jr., \Journal{\APJ}{518}{576}{1999}.
\bibitem{pos96} M. Postman, L.M. Lubin, J.E. Gunn, et al., 
\Journal{\AJ}{111}{615}{1996}.
\bibitem{pre74} W.H. Press \& P. Schechter, \Journal{\APJ}{187}{425}{1974}.
\bibitem{pro72} R.A. Proctor, \Journal{\MN}{33}{14}{1872}.
\bibitem{qui82} H. Quintana \& D.G. Lawrie, \Journal{\AJ}{87}{1}{1982}.
\bibitem{rea56} G. Reaves, \Journal{\AJ}{61}{69}{1956}.
\bibitem{rea62} G. Reaves, \Journal{\PA}{74}{392}{1962}.
\bibitem{rea66} G. Reaves, \Journal{\PA}{78}{407}{1966}.
\bibitem{rea68} G. Reaves, \Journal{\PA}{80}{564}{1968}.
\bibitem{rea74} G. Reaves, \Journal{\SA}{18}{307}{1974}.
\bibitem{rey23a} J.H. Reynolds, \Journal{\MN}{83}{147}{1923a}.
\bibitem{rey23b} J.H. Reynolds, \Journal{\MN}{84}{76}{1923b}.
\bibitem{ric76} D.O. Richstone, \Journal{\APJ}{204}{642}{1976}.
\bibitem{ric92} D.O. Richstone, A. Loeb, \& E.L. Turner, 
\Journal{\APJ}{393}{477}{1992}.
\bibitem{rob69} M.S. Roberts, \Journal{\AJ}{74}{859}{1969}.
\bibitem{rob73} M.S. Roberts \& A.H. Rots, \Journal{\AA}{26}{483}{1973}.
\bibitem{rog65} D.H. Rogstad, G.W. Rougoor, \& J.B. Whiteoak, 
\Journal{\APJ}{142}{1665}{1965}.
\bibitem{rol81b} J. Roland, \Journal{\AA}{93}{407}{1981}.
\bibitem{rol81a} J. Roland, H. Sol, I. Pauliny-Toth, \& A. Witzel,
\Journal{\AA}{100}{7}{1981}.
\bibitem{roo65} H.J. Rood, {\em PhD thesis,} (Univ. Michigan, 1965)
\bibitem{roo69} H.J. Rood, \Journal{\APJ}{158}{657}{1969}.
\bibitem{roo74a} H.J. Rood, \Journal{\APJ}{188}{451}{1974a}.
\bibitem{roo74b} H.J. Rood, \Journal{\APJ}{194}{27}{1974b}.
\bibitem{roo73} H.J. Rood \& G.O. Abell, 
\Journal{{\em Astroph. Letters}}{13}{69}{1973}.
\bibitem{roo72} H.J. Rood, T.L. Page, E.C. Kintner, \& I.K. King,
\Journal{\APJ}{175}{627}{1972}.
\bibitem{roo70} H.J. Rood, V.C.A. Rothman, \& B.E. Turnrose, 
\Journal{\APJ}{162}{411}{1970}.
\bibitem{roo71} H.J. Rood \& G.N. Sastry, \Journal{\PA}{83}{313}{1971}.
\bibitem{roo68} H.J. Rood \& B.E. Turnrose, \Journal{\APJ}{152}{1057}{1968}.
\bibitem{roo82} N. Roos \& S.J. Aarseth, \Journal{\AA}{114}{41}{1982}.
\bibitem{roo79} N. Roos \& C.A. Norman, \Journal{\AA}{76}{75}{1979}.
\bibitem{ros76} J.A. Rose, \Journal{\AAS}{23}{109}{1976}.
\bibitem{ros77} J.A. Rose, \Journal{\APJ}{211}{311}{1977}.
\bibitem{row72} M. Rowan-Robinson, \Journal{\AJ}{77}{543}{1972}.
\bibitem{roz72} M. R\'ozycka, \Journal{{\em Acta Astr.}}{22}{93}{1972}.
\bibitem{san85} A. Sandage, B. Binggeli, \& G.A. Tammann, 
\Journal{\AJ}{90}{1759}{1985}.
\bibitem{san76} A. Sandage \& G.A. Tammann, \Journal{\APJ}{207}{L1}{1976}.
\bibitem{san90} M. Sanrom\`a \& E. Salvador-Sol\'e, 
\Journal{\APJ}{360}{16}{1990}.
\bibitem{sar80} C.L. Sarazin, \Journal{\APJ}{236}{75}{1980}.
\bibitem{sar83} C.L. Sarazin \& R.W. O'Connell, \Journal{\APJ}{268}{552}{1983}.
\bibitem{sas68} G.N. Sastry, \Journal{\PA}{80}{252}{1968}.
\bibitem{sch76a} A. Scheepmaker, G.R. Ricker, K. Brecher, S.G. Ryckman, J.E.
Ballintine, J.P. Doty, P.M. Downey, \& W.H.G. Lewin, \Journal{\APJ}{205}{L65}{1976}.
\bibitem{sch76b} P. Schechter, \Journal{\APJ}{203}{297}{1976}.
\bibitem{sch78} L. Schipper \& I.R. King, \Journal{\APJ}{220}{798}{1978}.
\bibitem{sch80a} D.A. Schwartz et al., \Journal{\APJ}{238}{L53}{1980}.
\bibitem{sch80b} D.A. Schwartz, J. Schwarz, \& W. Tucker, 
\Journal{\APJ}{238}{L59}{1980}.
\bibitem{sch54} M. Schwarzschild, \Journal{\AJ}{59}{273}{1954}.
\bibitem{sci82} D.W. Sciama, \Journal{\MN}{198}{1P}{1982}.
\bibitem{sco54} E.L. Scott, C.D. Shane, \& M.D. Swanson,
\Journal{\APJ}{119}{91}{1954}.
\bibitem{sei65} G.A. Seielstad \& J.B. Whiteoak, 
\Journal{\APJ}{142}{616}{1965}.
\bibitem{ser77} P.J. Serlemitsos, B.W. Smith, E.A. Boldt, S.S. Holt,
\& J.H. Swank, \Journal{\APJ}{211}{L63}{1977}.
\bibitem{sha74} R.K. Shakhbazyan \& M.B. Petrosyan, 
\Journal{{\em Astrofiz.}}{10}{13}{1974}.
\bibitem{sha54} C.D. Shane \& C.A. Wirtanen, \Journal{\AJ}{59}{285}{1954}.
\bibitem{sha26} H. Shapley, \Journal{{\em Harvard Obs. Bull.}}{838}{3}{1926}.
\bibitem{sha30} H. Shapley, \Journal{{\em Harvard Obs. Bull.}}{874}{9}{1930}.
\bibitem{sha33} H. Shapley, \Journal{{\em Proc.Nat.Acad.Sci.Washington}}
{19}{591}{1933}.
\bibitem{sha34} H. Shapley, \Journal{{\em Harvard Obs. Bull.}}{896}{3}{1934}.
\bibitem{sha3?} H. Shapley, \Journal{{\em Harvard Obs. Bull.}}{903}{17}{1936}.
\bibitem{sha32} H. Shapley \& S. Ames, 
\Journal{{\em Harvard Ann.}}{88}{43}{1932}.
\bibitem{sha59} A.S. Sharov, \Journal{\SA}{36}{784}{1959}.
\bibitem{sha81} N.A. Sharp, \Journal{\MN}{195}{857}{1981}.
\bibitem{sil68} J. Silk, \Journal{\APJ}{151}{459}{1968}.
\bibitem{sil76} J. Silk, \Journal{\APJ}{208}{646}{1976}.
\bibitem{smi79b} B.W. Smith, R.F. Mushotzky, \& P.J. Serlemitsos, 
\Journal{\APJ}{227}{37}{1979}.
\bibitem{smi79a} R.W. Smith, \Journal{\JHA}{29}{133}{1979}.
\bibitem{smi36} S. Smith, \Journal{\APJ}{83}{23}{1936}.
\bibitem{sno70} T.P. Snow Jr., \Journal{\AJ}{75}{237}{1970}.
\bibitem{sol72} A.B. Solinger \& W.H. Tucker, \Journal{\APJ}{175}{L107}{1972}.
\bibitem{sou87} G. Soucail, B. Fort, Y. Mellier, \& J.-P. Picat, 
\Journal{\AA}{172}{L14}{1987}.
\bibitem{spi77} H. Spinrad, in {\em ``The evolution of galaxies
and stellar populations,} B.M. Tinsley, R.B. Larson eds., p.301
(New Haven: Yale Univ. Obs., 1977).
\bibitem{spi51} L. Spitzer Jr. \& W. Baade, \Journal{\APJ}{113}{413}{1951}.
\bibitem{sta98} S.A. Stanford, P.R. Eisenhardt, \& M. Dickinson, 
\Journal{\APJ}{492}{461}{1998}.
\bibitem{ste83} F.W. Stecker \& Q. Shafi, \Journal{\PRL}{50}{928}{1983}.
\bibitem{ste77} M. Stephan, \Journal{\MN}{37}{334}{1877}.
\bibitem{sto82} J. Stocke, J. Liebert, T. Maccacaro, I. Gioia, R. Griffiths,
\& J. Danziger, \Journal{\PA}{94}{759}{1982}.
\bibitem{sto80} A. Stockton, \Journal{{\em IAU Symp.}}{92}{89}{1980}.
\bibitem{sto55} S.N. Stone, \Journal{\PA}{67}{183}{1955}.
\bibitem{str79b} M.F. Struble, \Journal{\AJ}{84}{27}{1979}.
\bibitem{str79a} M.F. Struble \& S.A. Bludman, 
\Journal{{\em Ap. Space Sci.}}{64}{301}{1979}.
\bibitem{sul81} W.T. Sullivan III, G.D. Bothun, B. Bates, \& R.A. Schommer,
\Journal{\AJ}{86}{919}{1981}.
\bibitem{sul78} W.T. Sullivan III \& P.E. Johnson, 
\Journal{\APJ}{225}{751}{1978}.
\bibitem{sun70} R.A. Syunyaev \& Ya.B. Zel'dovich, 
\Journal{{\em Ap. Space Sci.}}{7}{20}{1970}.
\bibitem{sun72a} R.A. Syunyaev \& Ya.B. Zel'dovich, 
\Journal{{\em Comm. Ap. Space Phys.}}{4}{173}{1972a}.
\bibitem{sun72b} R.A. Syunyaev \& Ya.B. Zel'dovich, \Journal{\AA}{20}{189}{1972b}.
\bibitem{sza76} A.S. Szalay \& G. Marx, \Journal{\AA}{49}{437}{1976}.
\bibitem{tak72} B. Takase, \Journal{\PAJ}{24}{295}{1972}.
\bibitem{tam72} G.A. Tammann, \Journal{\AA}{21}{355}{1972}.
\bibitem{tan82} K.I. Tanaka, Y. Fujishima, \& M. Fujimoto, 
\Journal{\PAJ}{34}{147}{1982}.
\bibitem{tar80} M. Tarenghi, G. Chincarini, H.J. Rood, \& L.A. Thompson, 
\Journal{\APJ}{235}{724}{1980}.
\bibitem{tar78a} M. Tarenghi, W.G. Tifft, G. Chincarini, H.J. Rood, 
\& L.A. Thompson, \Journal{{\em IAU Symp.}}{79}{263}{1978}.
\bibitem{tar78b} J.C. Tarter, \Journal{\APJ}{220}{749}{1978}.
\bibitem{tar74} J.C. Tarter \& J. Silk, 
\Journal{{\em Q.Jl.R.Astr.Soc.}}{15}{122}{1974}.
\bibitem{the88} L.S. The \& S.D.M. White, \Journal{\AJ}{95}{15}{1988}.
\bibitem{tho01} T. Thomas \& P. Katgert, {\em in preparation} (2001).
\bibitem{tho78} L.A. Thompson \& S.A. Gregory, \Journal{\APJ}{220}{809}{1978}.
\bibitem{tho80} L.A. Thompson \& S.A. Gregory, \Journal{\APJ}{242}{1}{1980}.
\bibitem{tho93} L.A. Thompson \& S.A. Gregory, \Journal{\AJ}{106}{2197}{1993}.
\bibitem{thu77} T.X. Thuan \& J. Kormendy, \Journal{\PA}{89}{466}{1977}.
\bibitem{thu81} T.X. Thuan \& W. Romanishin, \Journal{\APJ}{248}{439}{1981}.
\bibitem{tif76} W.G. Tifft \& S.A. Gregory, \Journal{\APJ}{205}{696}{1976}.
\bibitem{tin70} B.M. Tinsley, \Journal{{\em Ap. Space Sci.}}{6}{344}{1970}.
\bibitem{tom37} C.W. Tombaugh, \Journal{\PA}{49}{259}{1937}.
\bibitem{too72} A. Toomre \& J. Toomre, \Journal{\APJ}{178}{623}{1972}.
\bibitem{tub43} M. Tuberg, \Journal{\APJ}{98}{501}{1943}.
\bibitem{tul80} R.B. Tully, \Journal{\APJ}{237}{390}{1980}.
\bibitem{tul77} R.B. Tully \& J.R. Fischer, \Journal{\AA}{54}{661}{1977}.
\bibitem{tur76a} E.L. Turner \& J.R. Gott III, \Journal{\APJS}{32}{409}{1976a}.
\bibitem{tur76b} E.L. Turner \& J.R. Gott III, \Journal{\APJ}{209}{6}{1976b}.
\bibitem{tur74} E.L. Turner \& W.L.W. Sargent, \Journal{\APJ}{194}{587}{1974}.
\bibitem{tur70} B.E. Turnrose \& H.J. Rood, \Journal{\APJ}{159}{773}{1970}.
\bibitem{ulm81} M.P. Ulmer et al., \Journal{\APJ}{243}{681}{1981}.
\bibitem{val79} M.J. Valtonen \& G.G. Byrd, \Journal{\APJ}{230}{655}{1979}.
\bibitem{val61} G.B. van Albada, \Journal{\AJ}{66}{590}{1961}.
\bibitem{vdb60a} S. van den Bergh, \Journal{\APJ}{131}{558}{1960a}.
\bibitem{vdb60b} S. van den Bergh, \Journal{\MN}{121}{387}{1960b}.
\bibitem{vdb60c} S. van den Bergh, \Journal{\PA}{72}{312}{1960c}.
\bibitem{vdb61} S. van den Bergh, \Journal{\PA}{73}{46}{1961}.
\bibitem{vdb69} S. van den Bergh, \Journal{\NAT}{224}{891}{1969}.
\bibitem{vdb75} S. van den Bergh, \Journal{\ARAA}{13}{217}{1975}.
\bibitem{vdb76} S. van den Bergh, \Journal{\APJ}{206}{883}{1976}.
\bibitem{vdb99a} S. van den Bergh, \Journal{\AAR}{9}{273}{1999a}.
\bibitem{vdb99b} S. van den Bergh, {\em astro-ph/9904251} (1999b).
\bibitem{vdb00} S. van den Bergh, {\em astro-ph/0005314} (2000).
\bibitem{ves82} W.T. Vestrand, \Journal{\AJ}{87}{1266}{1982}.
\bibitem{vig77} L. Vigroux, \Journal{\AA}{56}{473}{1977}.
\bibitem{vik94} A. Vikhlinin, W. Forman, \& C. Jones, 
\Journal{\APJ}{435}{162}{1994}.
\bibitem{vil81} A. Vilenkin, \Journal{{\em Phys. Rev. D}}{24}{2082}{1981}.
\bibitem{vis77} N. Visvanathan \& A. Sandage, \Journal{\APJ}{216}{214}{1977}.
\bibitem{vho60} S. von Hoerner, \Journal{{\em Z. Astroph.}}{50}{184}{1960}.
\bibitem{vor69} B.A. Vorontsov-Velyaminov, \Journal{\SA}{13}{235}{1969}.
\bibitem{wel71} G.A. Welch \& G.N. Sastry, \Journal{\APJ}{169}{L3}{1971}.
\bibitem{whi76a} S.D.M. White, \Journal{\MN}{177}{717}{1976a}.
\bibitem{whi76b} S.D.M. White, \Journal{\MN}{174}{19}{1976b}.
\bibitem{whi77} S.D.M. White, \Journal{\MN}{179}{33}{1977}.
\bibitem{whi78} S.D.M. White, \Journal{\MN}{184}{185}{1978}.
\bibitem{whi93} S.D.M. White, U.G. Briel, \& J.P. Henry, 
\Journal{\MN}{261}{L8}{1993}.
\bibitem{whi81} S.D.M. White \& J. Silk \Journal{\APJ}{241}{864}{1980}.
\bibitem{whi80} S.D.M. White, J. Silk \& J.P. Henry, 
\Journal{\APJ}{251}{L65}{1980}.
\bibitem{wil70} M.A.G. Willson, \Journal{\MN}{151}{1}{1970}.
\bibitem{wil97} G. Wilson, I. Smail, R.S. Ellis, \& W.J. Couch, 
\Journal{\MN}{284}{915}{1997}.
\bibitem{wol01} M. Wolf, \Journal{\AN}{155}{127}{1901}.
\bibitem{wol02} M. Wolf, \Journal{{\em Pub. Astr. Obs. 
K\"onigstuhl-Heidelberg}}{I}{127}{1902}. 
\bibitem{wol05} M. Wolf, \Journal{\AN}{170}{211}{1905}.
\bibitem{wol74} R.S. Wolff, H. Helava, T. Kifune, \& M.C. Weisskopff, 
\Journal{\APJ}{193}{L53}{1974}.
\bibitem{woo67} N.J. Woolf, \Journal{\APJ}{148}{287}{1967}.
\bibitem{wri74} M. Wright, J. Tarter, \& J. Silk, \Journal{\AA}{36}{441}{1974}.
\bibitem{yah73} A. Yahil \& J.P. Ostriker, \Journal{\APJ}{185}{787}{1973}.
\bibitem{yah77} A. Yahil \& N.V. Vidal, \Journal{\APJ}{214}{347}{1977}.
\bibitem{yup69} J.T. Yu \& P.J.E. Peebles, \Journal{\APJ}{158}{103}{1969}.
\bibitem{zel70} Ya.B. Zel'dovich, \Journal{\AA}{5}{84}{1970}.
\bibitem{zuc93} E. Zucca, G. Zamorani, R. Scaramella, \& G. Vettolani, 
\Journal{\APJ}{407}{470}{1993}.
\bibitem{zuc97} E. Zucca et al., \Journal{\AA}{326}{477}{1997}.
\bibitem{zwi33} F. Zwicky, \Journal{{\em Helv. Phys. Acta}}{6}{110}{1933}.
\bibitem{zwi37} F. Zwicky, \Journal{\APJ}{86}{217}{1937}.
\bibitem{zwi38} F. Zwicky, \Journal{\PA}{50}{210}{1938}.
\bibitem{zwi42a} F. Zwicky, \Journal{\APJ}{95}{555}{1942a}.
\bibitem{zwi42b} F. Zwicky, \Journal{\PA}{54}{185}{1942b}.
\bibitem{zwi50a} F. Zwicky, \Journal{\PA}{62}{196}{1950a}.
\bibitem{zwi50b} F. Zwicky, \Journal{\PA}{62}{256}{1950b}.
\bibitem{zwi51} F. Zwicky, \Journal{\PA}{63}{61}{1951}.
\bibitem{zwi52a} F. Zwicky, \Journal{\PA}{64}{247}{1952a}.
\bibitem{zwi52b} F. Zwicky, \Journal{\PA}{64}{242}{1952b}.
\bibitem{zwi53} F. Zwicky, \Journal{\PA}{65}{215}{1953}.
\bibitem{zwi56} F. Zwicky, \Journal{\PA}{68}{331}{1956}.
\bibitem{zwi57a} F. Zwicky, \Journal{\PA}{69}{518}{1957a}.
\bibitem{zwi57b} F. Zwicky, {\em Morphological Astronomy} 
(Berlin: Springer--Verlag, 1957b)
\bibitem{zwi63} F. Zwicky, \Journal{\PA}{75}{373}{1963}.
\bibitem{zwi61} F. Zwicky, E. Herzog, P. Wild, M. Karpowicz, \& 
C.T. Kowal {\em Catalogue of Galaxies and Clusters of Galaxies} 
(Pasadena: Calif. Inst. Technol., 1961--68).
\bibitem{zwi60} F. Zwicky \& M.L. Humason, \Journal{\APJ}{132}{627}{1960}.
\bibitem{zwi64} F. Zwicky \& M.L. Humason, \Journal{\APJ}{139}{269}{1964}.
\end{thebibliography}
\end{document}